\documentclass[12pt,a4paper]{article}

\usepackage[hidelinks]{hyperref}

\usepackage{afterpage, amsfonts, amsmath, amssymb, amsthm, bm, bbm, caption, color, enumitem, float, graphicx, epstopdf, kantlipsum, lipsum, longtable, lscape, mathrsfs, multirow, nccmath, nonfloat, pdflscape, rotating, tocloft}

\usepackage[flushleft]{threeparttable}

\usepackage[margin=1in]{geometry}

\usepackage[round]{natbib}
\usepackage[doublespacing]{setspace}
\usepackage[table]{xcolor}

\def\be{\begin{equation}}
\def\ee{\end{equation}}
\def\bea{\begin{eqnarray}}
\def\eea{\end{eqnarray}}

\author{}
\title{}

\DeclareMathOperator*{\argmin}{\arg\!\min}

\DeclareMathOperator*{\diag}{\normalfont\textrm{diag}}

\begin{document}
\newcommand\blfootnote[1]{
\begingroup
\renewcommand\thefootnote{}\footnote{#1}
\addtocounter{footnote}{-1}
\endgroup
}

\newtheorem{corollary}{Corollary}
\newtheorem{definition}{Definition}
\newtheorem{lemma}{Lemma}
\newtheorem{proposition}{Proposition}
\newtheorem{remark}{Remark}
\newtheorem{theorem}{Theorem}
\newtheorem{assumption}{Assumption}
\newtheorem{example}{Example}

\numberwithin{corollary}{section}
\numberwithin{definition}{section}
\numberwithin{equation}{section}
\numberwithin{lemma}{section}
\numberwithin{proposition}{section}
\numberwithin{remark}{section}
\numberwithin{theorem}{section}

\allowdisplaybreaks[4]

\begin{titlepage}

{\small

\begin{center}
{\Large \bf Decomposition of Bilateral Trade Flows Using a Three-Dimensional Panel Data Model\footnote{\noindent  Peng acknowledges the Australian Research Council Discovery Grants Program for its financial support under Grant Number DP210100476. 

\textit{Correspondence}: Yufeng Mao, Department of Econometrics and Business Statistics, Monash University, Caulfield East, VIC 3145, Australia. Email: Yufeng.Mao1@monash.edu
}}
 
\medskip

{\sc Yufeng Mao$^\sharp$, Bin Peng$^\sharp$, Mervyn Silvapulle$^\sharp$, Param Silvapulle$^\sharp$ 

and Yanrong Yang$^*$}
\medskip

$^\sharp$Monash University and $^*$Australian National University

\bigskip
\bigskip 

\today

\bigskip

\begin{abstract}
This study decomposes the bilateral trade flows using a three-dimensional panel data model. Under the scenario that all three dimensions diverge to infinity, we propose an estimation approach to identify the number of global shocks and country-specific shocks sequentially, and establish the asymptotic theories accordingly. From the practical point of view, being able to separate the pervasive and nonpervasive shocks in a multi-dimensional panel data is crucial for a range of applications, such as, international financial linkages, migration flows, etc. In the numerical studies, we first conduct intensive simulations to examine the theoretical findings, and then use the proposed approach to investigate the international trade flows from two major trading groups (APEC and EU) over 1982-2019, and quantify the network of bilateral trade.
\end{abstract}

\end{center}

\noindent{\em Keywords}: Three-Dimensional Panel Data, Bilateral Trade, Asymptotic Theory

\medskip

\noindent{\em JEL classification}: C23, P45

}

\end{titlepage}

\section{Introduction} \label{Section1}

All countries of the world are nowadays connected with each other more or less through varieties of bilateral trade. Getting reliable and up-to-date statistics on exports and imports of different countries is thus crucial in order to provide a detailed insight into the most recent trading patterns. Given an increasing interest in understanding such a complex network, we see the rising popularity of multi-dimensional models over the past decade, e.g., \cite{MNP2013}, \cite{BHM2016}, \cite{andreou2019inference}, \cite{choi2020canonical}, \cite{kapetanios2020estimation}, just to name a few. Excellent reviews on the applications and theoretical developments of multi-dimensional panel data models can be respectively seen in \cite{BEP2016} and \cite{Jorg2016} for instance.

Despite a vast amount of research on two dimensional factor models (see \citealp{BN2008} for an excellent review), it seems that the literature of multi-dimensional models has not even settled on how to effectively distinguish pervasive and nonpervasive economic shocks, where pervasive and nonpervasive shocks refer to those affecting the entire network and those affecting only a part of the network respectively (e.g., \citealp{Wang2008, ER2017}). In this regard, the presentation \eqref{Eq2.2} of Section \ref{Section2} provides a clear visualization using matrix form. 

To solve the aforementioned issue, different algorithms have been proposed (e.g., \citealp{Jorg2016} and references therein), but from the theoretical point of view the progress has not been pushed forward much since \cite{Wang2008}.  We now comment on the relevant literature. \cite{ER2017} extend the study of \cite{Wang2008} to allow for long run dependence, and both papers numerically rely on some initial estimates on the pervasive and nonpervasive factors. In our view, the requirement on initial estimates is due to the fact that both studies aim to estimate the pervasive and nonpervasive factors in one objective function, which as a consequence leads to a complex minimization problem. Thus, the numerical implementation often becomes complex, and is hard to be justified. In another two works, both \cite{CKKK2018} and \cite{andreou2019inference} propose sequential procedures to identify and estimate pervasive and nonpervasive shocks, in which canonical correlation analysis (CCA) are adopted. However, only two of the three dimensions are allowed to diverge in both studies. \cite{Han2019} considers a shrinkage estimation approach to explore the group effects of the factor structure, which can be computationally expensive, as the choice of tuning parameter often plays an important role in practice.

From the practical point of view, being able to separate the pervasive and nonpervasive shocks in a multi-dimensional panel data is crucial for a range of applications. First, as mentioned in the beginning of the paper, accounting for pervasive and nonpervasive shocks reveals a detailed network structure of the international trade. We will come back to it in the empirical study section. A second example is better understanding business-cycle fluctuations across countries and regions (\citealp{KOW2003}). Along this line of research, identifying the common fluctuations across macroeconomic aggregates worldwide has always been one of the priorities (e.g., \citealp{GHR1997}). The emergence of multi-dimensional panel data models provides an excellent framework to facilitate the investigation. Another field which urgently calls for development on multi-dimensional panel data models is associated with migration flows. As well understood, the rate of migration between two countries does not depend solely on their relative attractiveness, but also on the one of alternative destinations (\citealp{BERTOLI201379}). Given the increasing mobility of the entire population, how to better capture the bilateral flows therefore becomes vital now more than ever. Other examples requiring multi-dimensional panel data models can also be found in \cite{CKKK2018}, \cite{kapetanios2020estimation}, etc.

Having presented the above challenges and necessities, in this study, we specifically consider a three-dimensional panel data model with unobserved global (pervasive) and country-specific (nonpervasive) factors, which has been exposed in the literature but has not been fully solved to the best of the authors' knowledge. On theory, our contributions are the following three-fold: (1). under the scenario that all three dimensions can diverge to infinity, we propose an estimation approach to identify the number of global shocks and country-specific shocks sequentially; (2). the newly proposed approach is easy to implement, and the asymptotic theories are established accordingly; (3). we further conduct intensive numerical studies to examine the finite sample performance of the newly proposed approach using both simulated and real datasets. In the empirical study, we then apply the approach to decompose the network of bilateral trade using country level data from two major trading groups (APEC and EU) over the period 1982-2019. 
We find that the country-specific shocks become more volatile in recent years, which may indicate the increasing instability of the inward and outward bilateral trade costs over the past couple of decades. In addition, we show that the trade flows involving China mainland, Germany and the United States show relatively strong sensitivity to global shocks, which reflects the fact that, in general, they are leading export and import countries worldwide. We note that the relationship among Canada, Mexico, and the United States is also highly sensitive to different shocks, which somewhat reflects the fact that all three of them are highly economically related through North American Free Trade Agreement (NAFTA) that eliminates some trade barriers and promotes the trading activities. 

The structure of this paper is as follows. Section \ref{Section2} presents the model with the estimation approach, and establishes the asymptotic properties accordingly. In Section \ref{Section3}, we conduct intensive simulations to examine the finite sample performance of the newly proposed approach. Section \ref{Section4} provides an empirical study using country level bilateral trade data. Section \ref{Section5} concludes. Due to the limit of space, the preliminary lemmas and the proofs  are given in the online supplementary appendices.

Before proceeding further, it is convenient to introduce some notation: $\| \cdot\|_{F}$ denotes the Euclidean norm of a vector or the Frobenius norm of a matrix; for a matrix $\textbf{A}$, its spectral norm is defined as $\| \textbf{A} \|_2 = \sqrt{\lambda_{\max} \{ \textbf{A}^{\prime} \textbf{A} \} }$, where $\lambda_{\max} \{ \cdot \}$ denotes the maximum eigenvalue; $\textbf{M}_{\textbf{A}} = \textbf{I} - \textbf{P}_{\textbf{A}} $ denotes the orthogonal projection matrix generated by matrix $\textbf{A}$, where $ \textbf{P}_{\textbf{A}} = \textbf{A}(\textbf{A}' \textbf{A})^{-1} \textbf{A}'$ and $\textbf{A}$ is a matrix with full column rank;  let $\to_P$ and $\to_D$ denote convergence in probability and in distribution, respectively; we write $a \asymp b$ if $a=O_P(b)$ and $b=O_P(a)$; let $\diag(\textbf{A}, \textbf{B})$ denotes the block-diagonal matrix that takes $\textbf{A}$ and $\textbf{B}$ as the upper left and lower right blocks; $\mbox{vec}(\textbf{A})$ stands for the vectorization operation; $\mathbb{I}(\cdot)$ stands for the indicator function.

\section{Model \& Methodology} \label{Section2}

In this section, we first present the model, then provide the estimation approach, and finally establish the asymptotic theories accordingly.

\subsection{The Setup} \label{Section2.1}

Having presented our motivations in Section \ref{Section1}, we specifically consider the next model in this study.

\begin{eqnarray} \label{Eq2.1}
y_{ijt} = \bm{\gamma}_{ij}' \bm{g}_t + \bm{\lambda}_{E, ij}' \bm{f}_{E, it} + \bm{\lambda}_{I,ij}' \bm{f}_{I,j t} + u_{ijt},
\end{eqnarray}
where  $i=1,\ldots,M$ index the exporters,  $j=1,\ldots,N$ index the importers, and $t=1,\ldots,T$ index the time periods. We observe $y_{ijt}$'s only, and $u_{ijt}$'s are the idiosyncratic error terms. $\bm{g}_t$ is an $r_g\times 1$ unobservable global factor, which is regarded as global shocks and may capture the globalisation trends. Some detailed explanation on the globalisation trends can be found in \cite{kapetanios2020estimation}, and we shall be more specific on this so-called ``trend" in the empirical study of Section \ref{Section4}. $\bm{f}_{E,it}$ and $\bm{f}_{I,jt}$ represent the unobservable $r_{E,i}\times 1$ and $r_{I,j}\times 1$ country-specific factors. Specifically, $\bm{f}_{E,it}$ is referred to as an exporter factor which affects all import partners associated with export country $i$ and $\bm{f}_{I,jt}$ is referred to as an importer factor which affects all export partners associated with import country $j$. The country-specific factors may capture the unobservable multilateral trade resistances (MTRs) that are different for exporters and importers. Loosely speaking, MTRs refer to the barriers which each of exporter and importer face in their trade with all their trading partners. We refer interested readers to \cite{anderson2003gravity} for a comprehensive discussion on MTR. $\bm{\gamma}_{ij}$, $\bm{\lambda}_{E,ij}$ and $\bm{\lambda}_{I,ij}$ are the corresponding factor loadings. Throughout this paper, we always use the subscript $_g$ to denote the variables associated with the global factors, and use the subscripts $_E$ and $_I$ to denote the variables associated with the exporters and importers respectively.

The model \eqref{Eq2.1} is in fact not new, and has been mentioned in \citet[eq. 18]{Jorg2016}, \citet[eq. 1]{CKKK2018}, and \citet*[eq. 2]{kapetanios2020estimation} among others for different purposes. In what follows, we propose an easily implemented methodology to recover the structure of the right hand side of \eqref{Eq2.1}, when all three dimensions are allowed to diverge to infinity. Precisely, we first estimate the numbers of global and country-specific factors (i.e., the values of $r_g$, $r_{E,i}$'s and $r_{I,j}$'s), and then establish inferences for global and country-specific shocks.

\begin{remark}\label{Remark0}
Before proceeding further, we comment on an important identification issue. For simplicity, we suppose that $r_g =1$ and $g_t \equiv 1$, and suppose further that  

\begin{eqnarray*}
&&\bm{f}_{E, it} =\bm{f}_E +\bm{\eta}_{E, it}\quad\text{with}\quad E[\bm{\eta}_{E, it} ]=0,\nonumber \\
&&\bm{f}_{I, it} =\bm{f}_I +\bm{\eta}_{I, it}\quad\text{with}\quad E[\bm{\eta}_{I, it} ]=0.
\end{eqnarray*}
Then, the model \eqref{Eq2.1} becomes

\begin{eqnarray}
y_{ijt} = \bm{\gamma}_{ij}^* + \bm{\lambda}_{E, ij}' \bm{\eta}_{E, it} + \bm{\lambda}_{I,ij}' \bm{\eta}_{I,j t} + u_{ijt},
\end{eqnarray}
where $ \bm{\gamma}_{ij}^*  = \bm{\gamma}_{ij}+\bm{\lambda}_{E,ij}' \bm{f}_{E} + \bm{\lambda}_{I,ij}' \bm{f}_{I} $. It then infers that for a model having a multi-layer factor structure, only one layer can have non-zero mean factors. 
\end{remark}
\noindent Having said Remark \ref{Remark0}, without loss of generality, we assume that

\begin{eqnarray}\label{ID1}
 E [\bm{f}_{E, it}]=\bm{0}  \quad \text{and}\quad E[ \bm{f}_{I,j t}]=\bm{0}
\end{eqnarray}
for country-specific factors throughout this study.

\medskip

As repeatedly pointed out in the literature (e.g., \citealp{Wang2008, Jorg2016, CKKK2018}), investigating \eqref{Eq2.1} relies on how to utilize the sparse structure of the next presentation.

\begin{eqnarray}\label{Eq2.2}
\begin{pmatrix}
y_{11t}\\
\vdots\\
y_{M1t}\\
\vdots\\
y_{1Nt}\\
\vdots\\
y_{MNt}\\
\end{pmatrix}
=
\begin{pmatrix}
\bm{\gamma}_{11}' & \bm{\lambda}_{E,11}' &  \cdots & \textbf{0} & \bm{\lambda}_{I,11}' &  \cdots & \textbf{0} \\
\vdots       & \vdots & \ddots & \vdots & \vdots & \ddots & \vdots\\ 
\bm{\gamma}_{M1}' & \textbf{0}  & \cdots & \bm{\lambda}_{E,M1}' & \bm{\lambda}_{I,M1}' & \cdots & \textbf{0} \\
\vdots &  &  \vdots &  &   & \vdots & \\
\bm{\gamma}_{1N}' & \bm{\lambda}_{E,1N}' & \cdots & \textbf{0} & \textbf{0}  & \cdots & \bm{\lambda}_{I,1N}'\\
\vdots       & \vdots & \ddots & \vdots & \vdots  & \ddots & \vdots\\ 
\bm{\gamma}_{MN}' & \textbf{0} & \cdots & \bm{\lambda}_{E,MN}' & \textbf{0} & \cdots & \bm{\lambda}_{I,MN}'\\
\end{pmatrix}
\begin{pmatrix}
\bm{g}_{t}\\
\bm{f}_{E, 1  t}\\
\vdots\\
\bm{f}_{E, Mt}\\
\bm{f}_{I, 1 t}\\
\vdots\\
\bm{f}_{I, N t}\\
\end{pmatrix}
+
\begin{pmatrix}
u_{11t}\\
\vdots\\
u_{M1t}\\
\vdots\\
u_{1Nt}\\
\vdots\\
u_{MNt}\\
\end{pmatrix}\nonumber \\
\end{eqnarray}
In view of \eqref{Eq2.2}, a few facts emerge:

\begin{enumerate}
\item In order to estimate \eqref{Eq2.2}, one needs to identify the number of factors for each $\bm{g}_t$, $\bm{f}_{E,it}$ and $\bm{f}_{I,jt}$. Traditional PCA usually requires a low rank setting. However, having $\bm{g}_t$, $\bm{f}_{E,it}$'s and $\bm{f}_{I,jt}$'s in one column as in \eqref{Eq2.2} yields a factor with a diverging dimension, which suggests that recovering all factors and loadings in one goal seems to be challenging. Thus, it motivates us to consider a multiple steps approach below.

\item The country-specific factors associated with exporters and importers are interchangeable, as the sparse structure associated with the corresponding factor loadings depends on how we rank $y_{ijt}$ with respect to $i$ and $j$ only. Thus, we would expect to recover the exporter and importer factors in a parallel manner.

\item As clearly seen in \eqref{Eq2.2}, $\bm{g}_t$ has an impact on every single $y_{ijt}$, although the magnitude depends on the value of $\bm{\gamma}_{ij}$. However, $\bm{f}_{E,it}$ or $\bm{f}_{I,jt}$ affects only an asymptotically negligible subset of $y_{ijt}$'s due to the sparse structure. From the signal-to-noise ratio point of view, we expect that the global factors are easier to be identified. Intuitively speaking, they can be estimated first if principal component analysis (PCA) is employed. As the country-specific factors contain the second tier of signal, they should be recovered after removing the dominating ones. 

\end{enumerate}
In Section \ref{Section2.1} below, we propose a multi-step estimation approach based on the aforementioned points.

\subsection{The Estimation Approach}\label{Section2.2}

We are now ready to present the estimation approach, which is a  procedure involving multiple steps. The outline is as follows. 

\begin{enumerate}
\item[\textbf{Step 1}] Conduct PCA to identify the number of global factors $r_g$, and estimate the global factor, which contains the strongest ``signal" as explained under \eqref{Eq2.2}.

\item[\textbf{Step 2}] Remove the estimated global factor, then simultaneously conduct multiple PCA to estimate the number of country-specific factors $r_{E,i}$' and $r_{I,j}$'s, and recover the country-specific factors, which contain ``signals" weaker than the global factor but stronger than the error terms. 
\end{enumerate}

First, we write \eqref{Eq2.1} in matrix form to facilitate the development. Throughout, the subscript $_\bullet$ always stands for including all available sample in the corresponding dimension for notational simplicity. 

\begin{eqnarray} \label{Eq2.3}
\textbf{Y}=\bm{\Gamma}\textbf{G}'+ \bm{\Lambda}_E\textbf{F}_E'+ \bm{\Lambda}_I\textbf{F}_I' + \textbf{U},
\end{eqnarray}
where the response variables and error terms are defined by

\begin{eqnarray}\label{Eq2.4}
&&\textbf{Y} =(\textbf{Y}_{\bullet \bullet 1},\ldots, \textbf{Y}_{\bullet \bullet T}),\quad\textbf{Y}_{\bullet \bullet t} = (y_{1 1t},\ldots, y_{M1t},\ldots, y_{1 Nt},\ldots, y_{MNt})',\nonumber \\
&&\textbf{U} =(\textbf{U}_{\bullet \bullet 1},\ldots, \textbf{U}_{\bullet \bullet T}) ,\quad\textbf{U}_{\bullet \bullet t} = (u_{1 1t},\ldots, u_{M1t},\ldots, u_{1 Nt},\ldots, u_{MNt})';
\end{eqnarray}
the global factors and loadings are defined by

\begin{eqnarray}\label{Eq2.5}
\textbf{G} = (\bm{g}_1,\ldots, \bm{g}_T)',\quad\bm{\Gamma}= (\bm{\gamma}_{11}, \ldots, \bm{\gamma}_{M1},\ldots,\bm{\gamma}_{1N}, \ldots, \bm{\gamma}_{MN})';
\end{eqnarray}
and the country-specific factors and loadings are defined by

\begin{eqnarray}\label{Eq2.6}
\textbf{F}_E &=& (\textbf{F}_{E,1},\ldots,\textbf{F}_{E,M}),\quad\textbf{F}_{E,i} = (\bm{f}_{E,i1},\ldots, \bm{f}_{E,iT})', \nonumber \\
\textbf{F}_I &=& (\textbf{F}_{I,1},\ldots,\textbf{F}_{I,N}),\quad\textbf{F}_{I,j} = (\bm{f}_{I,j1},\ldots, \bm{f}_{I,jT})',\nonumber \\
\bm{\Lambda}_E &=& (  \diag \{ \bm{\Lambda}_{E,\bullet 1}' \}, \ldots,  \diag \{ \bm{\Lambda}_{E,\bullet N}' \})',\quad\bm{\Lambda}_{E,\bullet j} = (\bm{\lambda}_{E,1j}, \ldots, \bm{\lambda}_{E,Mj})',\nonumber \\
\bm{\Lambda}_I &=& \diag\{\bm{\Lambda}_{I, \bullet 1}, \ldots, \bm{\Lambda}_{I, \bullet N}\},\quad\bm{\Lambda}_{I,\bullet j} = (\bm{\lambda}_{I,1j}, \ldots, \bm{\lambda}_{I,Mj})'.
\end{eqnarray}

\medskip

With the above notations in hand, we are ready to present the details of each step with necessary discussions.

\textbf{Step 1} --- Conduct PCA on $\frac{1}{MNT}\textbf{Y}'\textbf{Y} $ as follows.

\begin{eqnarray}\label{Eq2.7}
\widehat{\textbf{G}}\textbf{V}_{g} =\frac{1}{MNT} \textbf{Y}'\textbf{Y} \widehat{\textbf{G}},
\end{eqnarray}
in which $\frac{1}{T}\widehat{\textbf{G}}'\widehat{\textbf{G}} =\textbf{I}_{k_{\max}}$, $\textbf{V}_{g} =\diag\{ \widehat{\rho}_{g,1},\ldots, \widehat{\rho}_{g,k_{\max}} \}$ with $ \widehat{\rho}_{g,1}\ge \cdots \ge \widehat{\rho}_{g,k_{\max}}$ being the largest $k_{\max}$ eigenvalues, $k_{\max} \ (> r_g)$ is a user-specified fixed large integer. By \eqref{Eq2.7}, we implement the following two sub-steps.

\begin{enumerate}
\item[\textbf{Step 1.1}] Estimate the number of global factors $r_g$ by

\begin{equation}\label{Eq2.8}
\widehat{r}_g = \argmin_{0 \le k \le k_{\max}} \left\{ \frac{ \widehat{\rho}_{g,k+1}}{ \widehat{\rho}_{g,k}}  \cdot \mathbb{I} \left( \widehat{\rho}_{g,k}\ge\omega_{MNT}  \right) + \mathbb{I} \left( \widehat{\rho}_{g,k}< \omega_{MNT}  \right) \right\},
\end{equation}
where $\omega_{MNT} = 1/\ln (\max\{M, N, T\})$, and $\widehat{\rho}_{g,0} =1$ is a mock eigenvalue. 

\item[\textbf{Step 1.2}] Estimate $\textbf{G}$ by letting $\widehat{\textbf{G}}$ include the first $\widehat{r}_g$ columns only, where we have slightly abused the notation $\widehat{\textbf{G}}$. The loading matrix is estimated by $\widehat{\bm{\Gamma}} =\frac{1}{T} \textbf{Y} \widehat{\textbf{G}} $. 
\end{enumerate}

\textbf{Step 2} includes two parallel sections: \textbf{Part 1} and \textbf{Part 2}. 

\textbf{Part 1} --- For each $j=1,\ldots,N$, conduct PCA:

\begin{eqnarray}\label{Eq2.9}
\widehat{\textbf{F}}_{I,j}\textbf{V}_{I,j}=\frac{1}{MT}(\textbf{Y}_{I, j} - \widehat{\bm{\Gamma}}_{I, j} \widehat{\textbf{G}}' )' (\textbf{Y}_{I, j} - \widehat{\bm{\Gamma}}_{I, j} \widehat{\textbf{G}}' )\widehat{\textbf{F}}_{I,j},
\end{eqnarray}
where $ \textbf{Y}_{I,j} = (\textbf{Y}_{\bullet j1},\ldots,  \textbf{Y}_{\bullet jT}  )$ with $  \textbf{Y}_{\bullet jt} =(y_{1jt},\ldots, y_{Mjt})'$, $\widehat{\bm{\Gamma}}_{I, j}$ includes the $M$ rows of $\widehat{\bm{\Gamma}}$ corresponding the $j^{th}$ importer, $\frac{1}{T}\widehat{\textbf{F}}_{I,j}' \widehat{\textbf{F}}_{I,j}=\textbf{I}_{k_{\max}}$, and $\textbf{V}_{I,j} =\diag\{\widehat{\rho}_{Ij,1},\ldots, \widehat{\rho}_{Ij,k_{\max}} \}$ with $\widehat{\rho}_{Ij,1}\ge \cdots \ge \widehat{\rho}_{Ij,k_{\max}}$ being the largest $k_{\max}\ (> r_{I,j})$ eigenvalues. By \eqref{Eq2.9}, implement the followings.

\begin{enumerate}
\item[\textbf{Part 1.1}] Estimate $r_{I,j}$ by

\begin{equation}\label{Eq2.10}
\widehat{r}_{I,j} = \argmin_{0 \le k \le k_{\max}} \left\{ \frac{ \widehat{\rho}_{Ij,k+1}}{ \widehat{\rho}_{Ij,k}}  \cdot \mathbb{I} \left( \widehat{\rho}_{Ij,k}\ge \omega_{MNT} \right) + \mathbb{I} \left( \widehat{\rho}_{Ij,k} < \omega_{MNT} \right) \right\},
\end{equation}
where $\widehat{\rho}_{Ij,0} =1$ is a mock eigenvalue. 

\item[\textbf{Part 1.2}] Estimate $\textbf{F}_{I,j} = ( \bm{f}_{I,j1},\ldots, \bm{f}_{I,jT})'$ by letting $\widehat{\textbf{F}}_{I,j}$ include the first $\widehat{r}_{I,j}$ columns only.  The loading matrix $\bm{\Lambda}_{I,\bullet j}$ defined in \eqref{Eq2.6} is estimated by $\widehat{\bm{\Lambda}}_{I, \bullet j} = \frac{1}{T}(\textbf{Y}_{I, j} - \widehat{\bm{\Gamma}}_{I, j} \widehat{\textbf{G}}' ) \widehat{\textbf{F}}_{I,j} $. 
\end{enumerate}

\textbf{Part 2} --- For each $i=1,\ldots,M$, conduct PCA:

\begin{eqnarray}\label{Eq2.11}
\widehat{\textbf{F}}_{E,i}\textbf{V}_{E,i}=\frac{1}{NT}(\textbf{Y}_{E, i} - \widehat{\bm{\Gamma}}_{E, i} \widehat{\textbf{G}}' )' (\textbf{Y}_{E, i} - \widehat{\bm{\Gamma}}_{E, i} \widehat{\textbf{G}}' ) \widehat{\textbf{F}}_{E,i},
\end{eqnarray}
where $ \textbf{Y}_{E,i} = (\textbf{Y}_{i \bullet 1},\ldots,  \textbf{Y}_{i\bullet T}  )$ with $\textbf{Y}_{i\bullet t} =(y_{i1t},\ldots, y_{iNt})'$, $\widehat{\bm{\Gamma}}_{E, i}$ includes the $N$ rows of $\widehat{\bm{\Gamma}}$ corresponding to the $i^{th}$ exporter, $\frac{1}{T}\widehat{\textbf{F}}_{E,i}' \widehat{\textbf{F}}_{E,i}=\textbf{I}_{k_{\max}}$, and $\textbf{V}_{E,i} =\diag\{\widehat{\rho}_{Ei,1},\ldots, \widehat{\rho}_{Ei,k_{\max}} \}$ with $\widehat{\rho}_{Ei,1}\ge \cdots \ge \widehat{\rho}_{Ei,k_{\max}}$ being the largest $k_{\max}\ (> r_{E,i})$ eigenvalues. By \eqref{Eq2.11}, we conduct the followings.

\begin{enumerate}
\item[\textbf{Part 2.1}] Estimate $r_{E,i}$ by

\begin{equation}\label{Eq2.12}
\widehat{r}_{E,i} = \argmin_{0 \le k \le k_{\max}} \left\{ \frac{\widehat{\rho}_{Ei,k+1}}{\widehat{\rho}_{Ei,k}}  \cdot \mathbb{I} \left( \widehat{\rho}_{Ei,k}\ge \omega_{MNT} \right) + \mathbb{I} \left( \widehat{\rho}_{Ei,k} < \omega_{MNT} \right) \right\},
\end{equation}
where $\widehat{\rho}_{Ei,0} =1$ is a mock eigenvalue.  

\item[\textbf{Part 2.2}] Estimate $\textbf{F}_{E,i} = (\bm{f}_{E,i1},\ldots, \bm{f}_{E,iT})'$ by letting $\widehat{\textbf{F}}_{E,i}$ include the first $\widehat{r}_{E,i}$ columns only. The loading matrix $\bm{\Lambda}_{E,i\bullet} =(\bm{\lambda}_{E,i1}, \ldots, \bm{\lambda}_{E,iN})'$ is estimated by $\widehat{\bm{\Lambda}}_{E,i \bullet} = \frac{1}{T}(\textbf{Y}_{E, i} - \widehat{\bm{\Gamma}}_{E, i} \widehat{\textbf{G}}' ) \widehat{\textbf{F}}_{E,i} $. 
\end{enumerate}

\medskip

\begin{remark}
We make a few comments on the estimation approach. (1). The use of eigenvalue ratio in \eqref{Eq2.8}, \eqref{Eq2.10} and \eqref{Eq2.12} is in the same spirit of \cite{LamYao2012} and \cite{AhnHorenstein2013}. (2). The threshold $\omega_{MNT} $ is to bypass a technical challenge raised in \citet[eq. 3.3]{LamYao2012}, and the mock eigenvalues $\widehat{\rho}_{g,0}$, $\widehat{\rho}_{Ij,0}$'s and $\widehat{\rho}_{Ei,0}$'s are designed to capture the cases where there are no global factors, or some of the country-specific factors do not exist. From the dimension reduction point of view, it is crucial to have a procedure which accounts for zero factors under the three dimensional panel data framework. (3). $k_{\max}$ is a user-defined fixed integer. Practically, one can adopt any reasonable large value which suits the empirical study (e.g., \citealp{FLM13, PX2019}). 
\end{remark}

\subsection{Consistency}

In this subsection, we show that the number of factors can be identified consistently in each step with necessary conditions. The asymptotic distributions are established in the next subsection. 

To facilitate the development, we impose the following conditions.

\begin{assumption} \label{ass1}
\item

\begin{enumerate}
\item As $T\to \infty$, $\normalfont \frac{1}{T}\textbf{G}'  \textbf{G}\to_P \bm{\Sigma}_{\textbf{G}}$, where $\normalfont \bm{\Sigma}_{\textbf{G}}$ is a deterministic positive definite matrix. Also, $\max_{t\ge 1}E \| \bm{g}_t \|_{F}^4 < \infty$.

\item Suppose that \eqref{ID1} holds. Moreover, $\max_{i\ge 1,t\ge 1}E\| \bm{f}_{E, it} \|_F^4 < \infty$ and $\|\normalfont\textbf{F}_{E}\|_2 = O_P(\sqrt{T} \vee \sqrt{M})$. Also, $\max_{j\ge 1,t\ge 1}E\| \bm{f}_{I, jt} \|_F^4 < \infty$ and $\|\normalfont\textbf{F}_{I}\|_2 = O_P(\sqrt{T} \vee \sqrt{N})$.
\end{enumerate}
\end{assumption}

\begin{assumption} \label{ass2}
\item 

\begin{enumerate}
\item As $(M, N) \to (\infty,\infty)$, $\frac{1}{MN}\bm{\Gamma}' \bm{\Gamma}  \to_P \bm{\Sigma}_{\bm{\Gamma}}$, where $\bm{\Sigma}_{\bm{\Gamma}}$ is a deterministic positive definite matrix. Also, $\max_{i\ge1, j\ge 1}E \| \bm{\gamma}_{ij} \|_{F}^4 < \infty$.

\item Suppose that $\max_{i\ge1,j\ge 1}E\| \bm{\lambda}_{E, ij}\|_F^4<\infty$ and $\max_{i\ge1,j\ge 1}\| \bm{\lambda}_{E, ij}\|_F = O_P(\sqrt{\ln (MN)})$. Also, $\max_{i\ge1,j\ge 1}E\| \bm{\lambda}_{I, ij}\|_F^4<\infty$ and $\max_{i\ge1,j\ge 1}\| \bm{\lambda}_{I, ij}\|_F = O_P(\sqrt{\ln (MN)})$.

\end{enumerate}
\end{assumption}

\begin{assumption}  \label{ass3}
\item

\begin{enumerate}
\item Let $\{u_{ijt}\ |\ i\ge 1,j\ge 1, t\ge 1\}$ be independent of the other variables. Let $\mathcal{F}_{-\infty}^0$ and $\mathcal{F}_\tau^\infty$ denote the $\sigma$-algebras generated by $\normalfont\{\textbf{U}_{\bullet \bullet t} \ | \ t \le  0\}$ and $\normalfont\{\textbf{U}_{\bullet \bullet t} \ | \ t \ge  \tau\}$ respectively, where $\normalfont \textbf{U}_{\bullet \bullet t} = (u_{11t},\ldots, u_{M1t},\ldots,u_{1Nt},\ldots, u_{MNt})'$. Define the mixing coefficient $\alpha(\tau) = \sup_{A\in \mathcal{F}_{-\infty}^0, B\in \mathcal{F}_\tau^\infty} \left|\Pr(A)\Pr(B) -\Pr(AB) \right|$.

\begin{enumerate}
\item Let $\normalfont\{\textbf{U}_{\bullet \bullet t} \ | \ t\ge 1\}$ be strictly stationary and $\alpha$-mixing such that for some $\nu>0$, $ \max_{i\ge 1,j\ge 1} E|u_{ijt}|^{4+\nu} <\infty$, and the mixing coefficient satisfies $ \sum_{t=1}^\infty  [\alpha(t)]^{\nu/(2+\nu)}$ $< \infty$.  

\item $E[u_{ijt}]=0$, $ \max_{i\ge 1, j\ge 1}\sigma_{ij}^2<\infty$ and $ \sum_{(i,j) \neq (m,n)} |\sigma_{ijmn}|=O(MN)$, where $\sigma_{ij}^2 = E[u_{ijt}^2]$ and $\sigma_{ijmn} =E[u_{ijt}u_{mnt}]$ for $t\ge 1$. In addition, suppose that\\ $\sum_{i,m=1}^M \sum_{j,n=1}^N \sum_{t,s=1}^T |E[ u_{ijt}u_{mns}] |=O(MNT)$.
\end{enumerate}

\item Suppose that $r_g<\infty$, $ \max_{i\ge 1}r_{E,i}<\infty$, and $ \max_{j\ge 1}r_{I,j}<\infty$.
\end{enumerate}
\end{assumption}

Assumption \ref{ass1} imposes restrictions on the global and country-specific factors, which are not more restrictive than Assumption 1.i of \cite{CKKK2018}. The conditions on the spectral norm of $\textbf{F}_E$ and $\textbf{F}_I$ are widely adopted in the literature (e.g., \citealp*[Assumption A.1.iii]{LiQianSu} and \citealp[Assumption A.1.v]{LuSu}). Extensive discussions with examples on this type of assumption can be found in \cite{Moon}.

Assumption \ref{ass2} puts restrictions on the loadings associated with the global and country-specific factors. The bounds on $\max_{i\ge1,j\ge 1}\| \bm{\lambda}_{E, ij}\|_F$ and $\max_{i\ge1,j\ge 1}\| \bm{\lambda}_{I, ij}\|_F$ are fairly standard. See Assumption A7 of \cite{CHL2012} for example.

Assumption \ref{ass3}.1 assumes that the error terms $u_{ijt}$'s follow stationary time series process over $t$, and simultaneously allow for weak cross-sectional dependence over $i$ and $j$. Assumption \ref{ass3}.2 requires $r_g$, $r_{E,i}$'s and $r_{I,j}$'s to be bounded, which nests $r_g=0$, $r_{E,i}=0$ and $r_{I,j}=0$ as special cases.

Under these conditions, we present the first theorem of this paper below.

\begin{theorem} \label{theorem1}
Under Assumptions \ref{ass1}-\ref{ass3}, as $(M,N,T) \to (\infty,\infty,\infty)$, 

\begin{enumerate}
\item in {\normalfont \textbf{Step 1.1}}, $\Pr(\widehat{r}_g = r_g) \to 1$;

\item in {\normalfont \textbf{Step 1.2}}, $\normalfont\frac{1}{\sqrt{T}} \| \widehat{\textbf{G}} - \textbf{G} \textbf{H} \|_{F} = O_P \left( \frac{\sqrt{ \ln (MN)}}{\min \{\sqrt{M}, \sqrt{N}, \sqrt{T}\}} \right)$, where $\normalfont\textbf{H} = \frac{1}{MN}\bm{\Gamma}' \bm{\Gamma}\cdot \frac{1}{T}\textbf{G}' \widehat{\textbf{G}}  \cdot(\textbf{V}_{g}^{\dag})^{-1}$, and $\normalfont\textbf{V}_{g}^{\dag}$ is the $r_g\times r_g$ leading principal submatrix of $\normalfont \textbf{V}_g$.
\end{enumerate}
\end{theorem}

Theorem \ref{theorem1}.1 shows that $r_g$ can be estimated consistently, while Theorem \ref{theorem1}.2 indicates that we can only recover $\textbf{G}$ up to a rotation matrix. From the signal-to-noise ratio point of view, only the space spanned by the global factors can be recovered in \textbf{Step 1}. 

\begin{remark}
It is noteworthy that when establishing Theorem \ref{theorem1}, no harsh conditions are imposed between the global factor structure and the country-specific ones. In this sense, although the rate of Theorem \ref{theorem1} is slow, we show that the global factors can be identified from the data first with minimum cost. In the traditional literature, the fact has barely been mentioned. To the best of the authors' knowledge, the only exception is Remark 4 of \cite{Han2019}. In Appendix A of the online supplementary file, we provide a sharper rate for the estimation of the global factor when more structures are adopted. The details are summarized in Lemma \ref{lemma2_improved}.
\end{remark}

\medskip

Having presented the results associated with the global factors, we investigate the country-specific ones, and further impose the following conditions.

\begin{assumption}\label{ass4}
\item 

\begin{enumerate}
\item  For $i=1,\ldots,M$  and $j=1,\ldots,N$, suppose that the following conditions hold:

\begin{enumerate}
\item $\normalfont \frac{1}{T} \| \textbf{G}'\textbf{F}_{E, i} \|_F  = O_P(T^{a_{E,i}})$ and $\normalfont\frac{1}{T} \| \textbf{G}'\textbf{F}_{I,j} \|_F = O_P(T^{a_{I,j}})$, where $\normalfont\textbf{F}_{E, i}$ and $\normalfont\textbf{F}_{I,j}$ are defined in \eqref{Eq2.6}, $\max_{i\ge 1}a_{E,i}< 0$, and $\max_{j\ge 1}a_{I,j}< 0$;

\item $\max_{i\ge 1}\|\normalfont\frac{1}{T}\textbf{F}_{E,i}' \textbf{F}_{E, i}  - \bm{\Sigma}_{\textbf{F}_{E,i}} \|_F =o_P(1)$ and $\max_{j\ge 1}\|\normalfont\frac{1}{T}\textbf{F}_{I, j}' \textbf{F}_{I, j}  - \bm{\Sigma}_{\textbf{F}_{I,j}} \|_F =o_P(1)$, where $\bm{\Sigma}_{\textbf{F}_{E,i}}$ and $\bm{\Sigma}_{\textbf{F}_{I,j}}$ are deterministic positive definite matrices;

\item $\normalfont \max_{i\ge 1}\|\frac{1}{N}\bm{\Lambda}_{E, i\bullet }'  \bm{\Lambda}_{E, i\bullet }  - \bm{\Sigma}_{\bm{\Lambda}_{E,i\bullet }}\|_F =o_P(1)$ and $\normalfont \max_{j\ge 1}\|\frac{1}{M}\bm{\Lambda}_{I,\bullet j}'  \bm{\Lambda}_{I,\bullet j}  - \bm{\Sigma}_{\bm{\Lambda}_{I,\bullet j}}\|_F =o_P(1)$, where $\normalfont\bm{\Lambda}_{E,i\bullet } = (\bm{\lambda}_{E,i1}, \ldots, \bm{\lambda}_{E,iN})'$, $\normalfont\bm{\Lambda}_{I,\bullet j}$ is defined under \eqref{Eq2.6}, and $\bm{\Sigma}_{\bm{\Lambda}_{E,i\bullet }}$ and $\bm{\Sigma}_{\bm{\Lambda}_{I,\bullet j}}$ are deterministic positive definite matrices.
\end{enumerate}

\item Suppose that $\max_{j\ge 1} \sum_{i \ne m} \sigma_{ijmj}=O(M)$, and $\max_{i\ge 1} \sum_{j\ne n} \sigma_{ijin}=O(N)$, where $\sigma_{ijmn}$ is defined in Assumption \ref{ass3}.
\end{enumerate}
\end{assumption}

Assumption \ref{ass4}.1.(a) requires certain orthogonality between the global factors and country-specific factors. Specifically, the values of $a_{E,i}$ and $a_{I,j}$ measure the degree of orthogonality between the global and country-specific factors. If $a_{E,i}=a_{I,j} =-\infty$, this condition essentially reduces to Assumption A of \cite{Ando}, where they show the necessity of orthogonality in order to identify the common and group-specific factors under a two-dimensional panel data framework. Similar discussions on orthogonality can also be seen in \cite{andreou2019inference}. Assumptions \ref{ass4}.1.(b) and \ref{ass4}.1.(c) impose more conditions on the blocks of factors and loadings associated with exporters and importers, which are fairly standard. Assumption \ref{ass4}.2 further regulates the weak cross-sectional dependence of the error terms.

With Assumption \ref{ass4} in hand, the country-specific factor structures can be successfully recovered in \textbf{Step 2}. The details are summarized in the next theorem.

\begin{theorem} \label{theorem2}
Under Assumptions \ref{ass1}-\ref{ass4}, as $(M,N,T) \to (\infty,\infty,\infty)$,

\begin{enumerate}
\item For $j=1,\ldots,N$, 

\begin{enumerate}
\item in {\normalfont \textbf{Part 1.1} of \textbf{Step 2}}, $\Pr(\widehat{r}_{I,j}= r_{I,j}) \to 1$;

\item in {\normalfont \textbf{Part 1.2} of \textbf{Step 2}}, $\normalfont\frac{1}{\sqrt{T}} \| \widehat{\textbf{F}}_{I,j} - \textbf{F}_{I,j} \textbf{H}_{I,j} \|_{F} = O_P \left( \frac{\sqrt{ \ln (MN)}}{\min \{\sqrt{M}, \sqrt{N}, \sqrt{T}\}} + T^{a_{I,j}} \right)$, where $\normalfont\textbf{H}_{I,j} = \frac{1}{M}\bm{\Lambda}_{I, \bullet j}' \bm{\Lambda}_{I, \bullet j}\cdot \frac{1}{T}\textbf{F}_{I,j}' \widehat{\textbf{F}}_{I,j} \cdot(\textbf{V}_{I,j}^{\dag})^{-1}$, and $\normalfont\textbf{V}_{I,j}^{\dag}$ is the $r_{I,j}\times r_{I,j}$ leading principal submatrix of $\normalfont \textbf{V}_{I,j}$.
\end{enumerate}

\item For $i=1,\ldots,M$, 

\begin{enumerate}
\item in {\normalfont \textbf{Part 2.1} of \textbf{Step 2}}, $\Pr(\widehat{r}_{E,i}= r_{E,i}) \to 1$;

\item in {\normalfont \textbf{Part 2.2} of \textbf{Step 2}}, $\normalfont\frac{1}{\sqrt{T}} \| \widehat{\textbf{F}}_{E,i} - \textbf{F}_{E,i} \textbf{H}_{E,i} \|_{F} = O_P \left( \frac{\sqrt{ \ln (MN)}}{\min \{\sqrt{M}, \sqrt{N}, \sqrt{T}\}} +T^{a_{E,i}} \right)$, where $\normalfont\textbf{H}_{E,i} = \frac{1}{N}\bm{\Lambda}_{E, i\bullet}' \bm{\Lambda}_{E,i \bullet }\cdot \frac{1}{T}\textbf{F}_{E,i}' \widehat{\textbf{F}}_{E,i} \cdot(\textbf{V}_{E,i}^{\dag})^{-1}$, and $\normalfont\textbf{V}_{E,i}^{\dag}$ is the $r_{E,i}\times r_{E,i}$ leading principal submatrix of $\normalfont \textbf{V}_{E,i}$.
\end{enumerate}
\end{enumerate}
\end{theorem}

Theorem \ref{theorem2} shows that $r_{E,i}$ and $r_{I,j}$ can be estimated consistently. Moreover, $\widehat{\textbf{F}}_{I,j}$ and $\widehat{\textbf{F}}_{E,i}$ respectively recover $\textbf{F}_{I,j}$ and $\textbf{F}_{E,i}$ up to rotation matrices. 

\medskip

Till now, we conclude that we have successfully recovered the network presented by \eqref{Eq2.2}. To establish inferences for the estimation approach, we study the asymptotic distributions associated with \textbf{Step 1} and \textbf{Step 2} in the next subsection.

\subsection{Asymptotic Distribution}

In order to establish the asymptotic distributions, the following assumptions are necessary to facilitate the development.

\begin{assumption}\label{ass5}
\item 
\begin{enumerate}
\item Let $\normalfont \frac{1}{\sqrt{MN}} \| \bm{\Gamma}'\bm{\Lambda}_{E} \|_F  = O_P(1)$ and $\normalfont\frac{1}{\sqrt{MN}} \| \bm{\Gamma}'\bm{\Lambda}_{I} \|_F = O_P(1)$.

\item $\normalfont \frac{1}{T} \textbf{G}'\textbf{G} = \textbf{I}_{r_g}$ and $\normalfont \bm{\Gamma}' \bm{\Gamma}$ is a diagonal matrix with distinct entries.

\item Suppose that $\normalfont \frac{1}{\sqrt{MN}} \sum_{i=1}^{M} \sum_{j=1}^{N} \bm{\gamma}_{ij} \upsilon_{ijt} \to_{D} N(\textbf{0}, \bm{\Phi}_{t})$ for $t=1,\ldots, T$, where $\upsilon_{ijt} = \bm{\lambda}_{E, ij}' \bm{f}_{E, it} + \bm{\lambda}_{I,ij}' \bm{f}_{I,j t} + u_{ijt}$.
\end{enumerate}
\end{assumption}

 \begin{assumption}\label{ass6}
\item 

\begin{enumerate}

\item Suppose that $\normalfont \frac{1}{T} \| \textbf{F}_{E, i}'\textbf{F}_{I, j} \|_F  = O_P(T^{b_{EI, ij}})$, where $\max_{i\ge 1,j\ge1} b_{EI,ij} < 0$.

\item \begin{enumerate}
\item $\normalfont \frac{1}{T} \textbf{F}_{I,j}'\textbf{F}_{I,j} = \textbf{I}_{r_{I,j}}$ and $\normalfont \bm{\Lambda}_{I,\bullet j}' \bm{\Lambda}_{I,\bullet j}$ is a diagonal matrix with distinct entries;

\item $\normalfont \frac{1}{T} \textbf{F}_{E,i}'\textbf{F}_{E,i} = \textbf{I}_{r_{E,i}}$ and $\normalfont \bm{\Lambda}_{E, i \bullet}' \bm{\Lambda}_{E, i \bullet}$ is a diagonal matrix with distinct entries.
\end{enumerate}

\item \begin{enumerate}
\item $\normalfont \frac{1}{\sqrt{M}} \sum_{i=1}^{M} \bm{\lambda}_{I,ij} (\bm{\lambda}_{E, ij}' \bm{f}_{E, it} + u_{ijt}) \to_{D} N(\textbf{0}, \bm{\Omega}_{I,jt})$ for each pair of $(j,t)$;

\item $\normalfont \frac{1}{\sqrt{N}} \sum_{j=1}^{N} \bm{\lambda}_{E,ij} (\bm{\lambda}_{I, ij}' \bm{f}_{I, jt} + u_{ijt}) \to_{D} N(\textbf{0}, \bm{\Omega}_{E,it})$ for each pair of $(i,t)$.
\end{enumerate}
\end{enumerate}
\end{assumption}

Assumption \ref{ass5}.1 requires certain orthogonality between global factor loadings and country-specific factor loadings, which is not unusual in the literature. For instance, \cite{LamYao2012} explain the rational behind such a setting at length. Assumption \ref{ass5}.2 further imposes conditions for the purpose of identification, which has been extensively discussed in \cite{BN2013} and \cite{FanLiaoWang}. In view of Remark \ref{Remark0}, Assumption \ref{ass5}.3 is fairly standard. We further explain Assumption \ref{ass5}.3 together with Assumption \ref{ass6}.3 below.
 
Similar to Assumption \ref{ass5}.1, Assumption \ref{ass6}.1 requires certain orthogonality but focusing on the export factors and importer factors, while Assumption \ref{ass6}.2 is for the purpose of identification. Assumption \ref{ass6}.3 is somewhat interesting. Take  
\begin{equation*}
     \frac{1}{\sqrt{N}} \sum_{j=1}^{N} \bm{\lambda}_{E,ij} (\bm{\lambda}_{I, ij}' \bm{f}_{I, jt} + u_{ijt}) \to_{D} N(\textbf{0}, \bm{\Omega}_{E,it})
\end{equation*}
as an example, which says the asymptotic distribution associated with the $i$-th exporter factor at time $t$ is not only driven by the error component, but also is driven by its entire importer network. The same argument applies to the importer factor. In this way, the networks of export and import are entangled with each other. Mathematically, it requires country-specific shocks to have mean 0, which is ensured by \eqref{ID1}. See Assumption 1.ii of \cite{CKKK2018} and Assumption 1.a of \cite{Han2019} for similar settings.
 
To close our theoretical investigation, we summarize the asymptotic distributions associated with the global and country-specific factors in the next theorem.

\begin{theorem} \label{theorem_global_clt}
Under Assumptions \ref{ass1}-\ref{ass5}, Let $(M,N,T) \to (\infty,\infty,\infty)$.

\begin{enumerate}
\item[1.]  If $\sqrt{MN} ( \frac{1}{T} +\Delta_{g, MNT}^*) \to 0$, then $\normalfont \sqrt{MN} (\widehat{\bm{g}}_t - \bm{g}_t) \to_{D} N (\textbf{0}, \bm{\Sigma}_{\bm{\Gamma}}^{-1} \bm{\Phi}_{t} \bm{\Sigma}_{\bm{\Gamma}}^{-1} )$ for each $t $.
\end{enumerate}
In addition, let Assumption \ref{ass6} also hold.

\begin{enumerate}
\item[2.] If $\sqrt{M} ( \frac{1}{\sqrt{T}} + \Delta_{Ij, MNT}^{*}) \to 0$, then $\normalfont \sqrt{M} (\widehat{\bm{f}}_{I,jt} - \bm{f}_{I,jt}) \to_{D} N (\textbf{0}, \bm{\Sigma}_{\bm{\Lambda}_{I,\bullet j}}^{-1} \bm{\Omega}_{I, jt} \bm{\Sigma}_{\bm{\Lambda}_{I,\bullet j}}^{-1} ) $ for each $(j,t)$;

\item[3.] If $\sqrt{N} (\frac{1}{\sqrt{T}} + \Delta_{Ei, MNT}^{*}) \to 0$, then $\normalfont \sqrt{N} (\widehat{\bm{f}}_{E,it} - \bm{f}_{E,it}) \to_{D} N (\textbf{0}, \bm{\Sigma}_{\bm{\Lambda}_{E, i \bullet}}^{-1} \bm{\Omega}_{E, it} \bm{\Sigma}_{\bm{\Lambda}_{E, i \bullet}}^{-1} )$ for each $(i,t)$.
\end{enumerate}
In the above, $\Delta_{g, MNT}^{*} $, $\Delta_{Ij, MNT}^{*} $ and $\Delta_{Ei, MNT}^{*} $ are defined as follows.

\begin{eqnarray*}
\Delta_{g, MNT}^{*} &=& \frac{T^{\max_{i}a_{E,i}}}{\sqrt{N}} + \frac{T^{\max_{j}a_{I,j}}}{\sqrt{M}} + \frac{\sqrt{\ln(MN)} \cdot \left( T^{\max_{i}a_{E,i}} + T^{\max_{j}a_{I,j}} \right) }{\sqrt{T}} \\
&& + \ln(MN) \cdot ( T^{2\max_{i}a_{E,i}} + T^{2\max_{j}a_{I,j}}); \\
\Delta_{Ij, MNT}^{*} &=& \frac{\ln(MN) \cdot T^{\max_{j}a_{I,j}}}{ \min\{\sqrt{M}, \sqrt{N}, \sqrt{T}\} } + \sqrt{\ln(MN)} \cdot ( T^{\max_{i}a_{E,i}} + T^{\max_i b_{EI, ij}} ) + T^{a_{I,j}}; \\
\Delta_{Ei, MNT}^{*} &=& \frac{\ln(MN) \cdot T^{\max_{i}a_{E,i}}}{ \min\{\sqrt{M}, \sqrt{N}, \sqrt{T}\} } + \sqrt{\ln(MN)} \cdot ( T^{\max_{j}a_{I,j}} + T^{\max_j b_{EI, ij}} ) + T^{a_{E,i}}.
\end{eqnarray*}
\end{theorem}

The condition $\frac{\sqrt{MN}}{T}\to 0$ in the first result of Theorem \ref{theorem_global_clt} is equivalent to $\frac{\sqrt{N}}{T}\to 0$ in Theorem 1 of \cite{BN2013} in which a two dimension model is considered. The condition $\sqrt{MN} \cdot \Delta_{g, MNT}^*\to 0$ requires the orthogonality between the global and country-specific factor structures are strong enough in order to achieve the optimal rate $\sqrt{MN}$. If we adopt the orthogonality as in \cite{Ando} and \cite{andreou2019inference}, then this condition will completely vanish.

In order to achieve asymptotic normality for the country-specific factors, slightly stronger restrictions (such as $\frac{M}{T}\to 0$ and $\frac{N}{T}\to 0$) are imposed in the body of this theorem on top of Assumption \ref{ass6}, which is due to the fact that we need to account for the estimation bias caused by Step 1 of the estimation approach. It is noteworthy that $\frac{M}{T}\to 0$ and $\frac{N}{T}\to 0$ imply $\frac{MN}{T^2}\to 0$, which has been discussed above. Therefore, we claim the newly imposed conditions are reasonable, and are only slightly stronger than those used in traditional two dimensional analysis. The conditions $\sqrt{M} \cdot \Delta_{Ij, MNT}^*\to 0$ and $\sqrt{N} \cdot \Delta_{Ei, MNT}^*\to 0$ require the orthogonality between the global and country-specific factor structures are strong enough in order to achieve the optimal rates $\sqrt{M}$ and $\sqrt{N}$. Again, if orthogonality is adopted, these conditions will disappear automatically.

\section{Simulation}\label{Section3}

In this section, we examine the finite sample performance of the methodology proposed in Section \ref{Section2}. Specifically, the data generating process (DGP) is as follows.

\begin{eqnarray}
y_{ijt} = \bm{\gamma}_{ij}' \bm{g}_t + \bm{\lambda}_{E, ij}' \bm{f}_{E, it} + \bm{\lambda}_{I,ij}' \bm{f}_{I,j t} + u_{ijt},
\end{eqnarray}
where  $i=1,\ldots,M$,  $j=1,\ldots,N$, and $t=1,\ldots,T$. The global factors, country-specific factors and idiosyncratic errors are generated by the following AR(1) processes

\begin{eqnarray*}
\bm{g}_t &=& \phi_{g} \bm{g}_{t-1} + \bm{v}_{g,t} \quad \mbox{with}\quad \bm{v}_{g, t} \sim i.i.d. \ N(\textbf{0}, \textbf{I}_{r_g}), \\
 \bm{f}_{E, it}&=& \phi_{E,i}  \bm{f}_{E, i,t-1} + \bm{v}_{E, it} \quad \mbox{with}\quad  \bm{v}_{E, it} \sim i.i.d. \ N(\textbf{0}, \textbf{I}_{r_{E,i}}), \\
 \bm{f}_{I,jt} &=& \phi_{I,j} \bm{f}_{I,j,t-1} + \bm{v}_{I,jt} \quad \mbox{with}\quad  \bm{v}_{I,jt} \sim i.i.d.\ N(\textbf{0}, \textbf{I}_{r_{I,j}}) , \\
  u_{ijt} &=& \phi_{u} u_{ij, t-1} + e_{ijt} \quad \mbox{with} \quad e_{ijt} \sim i.i.d.\  N(0, 1),
\end{eqnarray*}
where $i.i.d.$ stands for independent and identically distributed. The factor loadings are generated as: $\bm{\gamma}_{ij} \sim i.i.d. \ N(\textbf{0}, \textbf{I}_{r_g})$, $\bm{\lambda}_{E,ij} \sim i.i.d. \ N(\textbf{0}, \textbf{I}_{r_{E,i}})$, and $\bm{\lambda}_{I,ij}  \sim i.i.d. N(\textbf{0}, \textbf{I}_{r_{I,j}}).$

We consider the following two cases.

\begin{enumerate}
\item[DGP 1:] Let $\phi_g = \phi_{E,i} = \phi_{I,j} = \phi_{u} = 0$, $r_{g} = 3$, $r_{E,i} = 2$ for $i=1,\ldots,M$, and $r_{I,j} = 1$ for $j=1,\ldots,N$;
\item[DGP 2:] Let $\phi_g = \phi_{E,i} = \phi_{I,j} = \phi_{u} = 0.5$, and the rest values are the same as those in DGP 1.
\end{enumerate}
For each DGP, we conduct the estimation approach of Section \ref{Section2} by letting $M,N,T \in \{20,40,60,80\}$, and implement 1000 replications for each given sample size. 

To measure the performance of the proposed estimation approach, we define a few criteria below. First, we measure the detection on different factors, and start from the global factor structure.

\begin{eqnarray*}
P_{g,c} = \frac{1}{1000}\sum_{\ell =1}^{1000} \mathbb{I}(\widehat{r}_g^\ell =r_g), \quad P_{g,u} = \frac{1}{1000}\sum_{\ell =1}^{1000} \mathbb{I}(\widehat{r}_g^\ell < r_g),\quad P_{g,o} = \frac{1}{1000}\sum_{\ell =1}^{1000} \mathbb{I}(\widehat{r}_g^\ell >r_g), 
\end{eqnarray*}
where $\widehat{r}_g^\ell$ defines the estimated of $r_g$ at the $\ell^{th}$ replication. It is clear that $P_{g,c}$, $P_{g,u}$ and $P_{g,o}$ define the probabilities of correctly, under and over select the number of global factors. For the export factors, we define

\begin{eqnarray*}
&&P_{E,c} = \frac{1}{1000}\sum_{\ell =1}^{1000} \frac{1}{M}\sum_{i=1}^M\mathbb{I}(\widehat{r}_{E,i}^\ell =r_{E,i}), \quad P_{E,u} = \frac{1}{1000}\sum_{\ell =1}^{1000} \frac{1}{M}\sum_{i=1}^M\mathbb{I}(\widehat{r}_{E,i}^\ell < r_{E,i}),\\
&&P_{E,o} = \frac{1}{1000}\sum_{\ell =1}^{1000} \frac{1}{M}\sum_{i=1}^M\mathbb{I}(\widehat{r}_{E,i}^\ell > r_{E,i}) ,
\end{eqnarray*}
where $\widehat{r}_{E,i}^\ell $ stands for the estimated $\widehat{r}_{E,i}$ at the $\ell^{th}$ replication. Also, it is obvious that $P_{E,c}$, $P_{E,u}$ and $P_{E,o}$ define the probabilities of correctly, under and over select the number of export factors. Similarly, we can define $P_{I,c}$, $P_{I,u}$ and $P_{I,o}$ for the import factors. The details are omitted for the sake of conciseness.

Second, we measure the estimation on different factors. Recall that we have defined $\textbf{G}$, $\textbf{F}_{E,i}$ and $\textbf{F}_{I,j}$ under \eqref{Eq2.3}, and then further define

\begin{eqnarray*}
&& \mbox{RMSE}_{\textbf{G}} 
= \sqrt{ \frac{1}{1000} \sum_{\ell=1}^{1000} \| \textbf{P}_{\widehat{\textbf{G}}^\ell} - \textbf{P}_{\textbf{G}^\ell}  \|_F^2 },\\
&& \mbox{RMSE}_{ \textbf{E}} = \sqrt{ \frac{1}{1000} \sum_{\ell =1}^{1000}  \frac{1}{M} \sum_{i=1}^{M} \| \textbf{P}_{\widehat{\textbf{F}}_{E,i}^\ell} - \textbf{P}_{\textbf{F}_{E,i}^\ell} \|_F^2 }, \\
&& \mbox{RMSE}_{ \textbf{I}} = \sqrt{ \frac{1}{1000} \sum_{\ell =1}^{1000}  \frac{1}{N} \sum_{j=1}^{N} \| \textbf{P}_{\widehat{\textbf{F}}_{I,j}^\ell} - \textbf{P}_{\textbf{F}_{I,j}^\ell} \|_F^2 }.
\end{eqnarray*}
In the above formulas, we let $\widehat{\textbf{G}}^\ell$ and $\textbf{G}^\ell $ include the estimated and true global factors from the $\ell^{th}$ replication. Similarly, we define $\widehat{\textbf{F}}_{E,i}^\ell$ and $\textbf{F}_{E,i}^\ell$ for the exporter factors, and define $\widehat{\textbf{F}}_{I,j}^\ell$ and $\textbf{F}_{I,j}^\ell$ for the importer factors.

\medskip

We summarize the simulation results in Table \ref{Table1} to Table \ref{Table3}. Note that due to the limit of space, the results of some combinations of $(M,N,T)$ are dropped in all tables. In Table \ref{Table1}, it is clear that as the sample size goes up, the values of $P_{g,c}$, $P_{E,c}$ and $P_{I,c}$ converge to 1. When the sample size is relatively small, it seems that we tend to under select the number of factors. Once all $M$, $N$, $T$ are greater than and equal to 40, the selection on the factors is quite accurate. In Table \ref{Table2}, we consider a DGP with more time series correlation, and the pattern is almost identical to those presented in Table \ref{Table1}. Table \ref{Table3} reports the results of $\mbox{RMSE}_{\textbf{G}} $, $\mbox{RMSE}_{\textbf{E}}$ and $\mbox{RMSE}_{\textbf{I}} $. It is not surprising that all values of $\mbox{RMSE}$ converge to 0, as the sample size goes up. Moreover, the values of $\mbox{RMSE}_{\textbf{E}} $ and $\mbox{RMSE}_{\textbf{I}} $ are larger than $\mbox{RMSE}_{\textbf{G}}$ in general, which should be expected. The reason is that Step 2 includes the estimation bias associated with Step 1, although the bias is negligible in the asymptotic sense under certain restrictions. It is noteworthy that the values of $\mbox{RMSE}_{\textbf{E}}$ are larger than those of $\mbox{RMSE}_{\textbf{I}} $, which is due to the fact that more unobservable factors are included for the exporters.

Having justified the validity of the proposed estimation approach through simulations, we are now ready to move on to the empirical study in the next section.

\section{Empirical Study}\label{Section4}

In this section, we use the proposed methodology to investigate the international trade flows.

\subsection{The Data}

We use monthly bilateral export volumes of commodity goods among 23 countries/region over the period of 1982-2019. The export flows data are collected from the Direction of Trade Statistics (DOTS) of International Monetary Fund (IMF) available at \url{https://www.imf.org/external/index.htm}. We use the FOB (free on board) value of exports of goods denominated in U.S. dollars and restrict the sample to 506 country-pairs of 23 countries/regions from two major trading groups over a 456-month period from January, 1982 to December, 2019. 

\begin{itemize}
\item Asia-Pacific Economic Cooperation (APEC): Australia (AUS), China Mainland  (CHN), Hong Kong (HKG), Indonesia (IDN), Japan (JPN), Korea (KOR), Malaysia (MYS), New Zealand (NZL), Singapore (SGP), Thailand (THA), Canada (CAN), Mexico (MEX), United States (USA)

\item European Union (EU): Denmark (DNK), Finland (FIN), France (FRA), Germany (DEU), Ireland (IRL), Italy (ITA), Netherlands (NLD), Spain (ESP), Sweden (SWE), United Kingdom (GBR)
\end{itemize}
Canada, Mexico and United States are also the members of North American Free Trade Agreement (NAFTA). As they are already included in APEC, we no longer specifically mention NAFTA in this study. It is worth pointing out that a similar dataset is considered in \cite{chen2019modeling} to investigate the patterns in the dynamic network of international trade. The difference between their study and our paper lies on the setting of factor structure. While we consider multiple layers of the factor structure, their study focuses on one layer only with a different presentation. As a consequence, the two models and the corresponding estimation approaches are not directly comparable.  

In what follows, the combination of an export country/region and one of its import partner is referred to as a country pair. For example, the export flow from the United States to Australia and the export flow from Australia to the United States are the bilateral export flows for two different country pairs.

\subsection{Estimation Results}

We first report the estimated numbers of global and country-specific factors. Specifically, only one global factor is identified from the sample. The estimated numbers of exporter factors and importer factors are summarized in Table \ref{Table4}. As shown in the table, majorities have only 1 or 2 factors with the importer factors of IDN being the only exception.

\begin{figure} [H]
\includegraphics[scale=0.75]{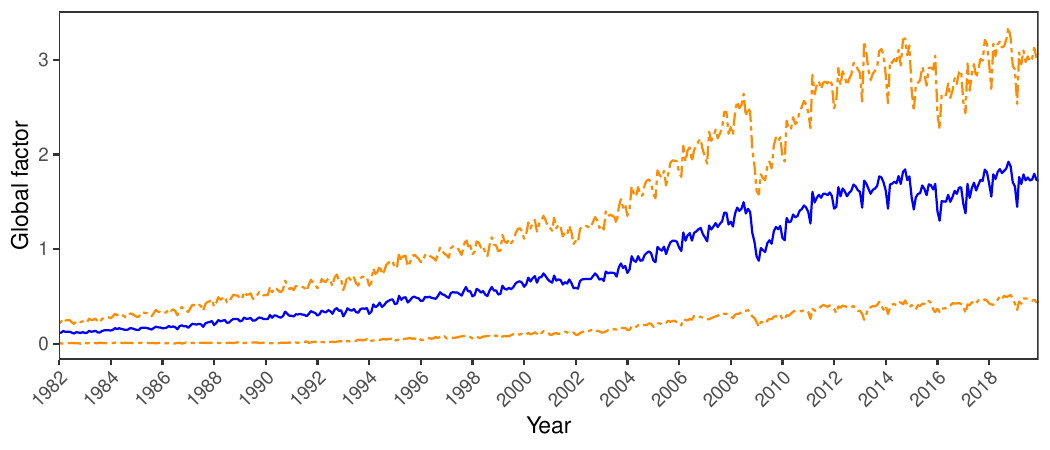}
\caption{Global factor for bilateral export flows from January, 1982 to December, 2019. The blue solid line represents the estimated global factor and the orange dashed lines provide a 95\% confidence interval.}
\centering
\label{Global_factor}
\end{figure}

Figure \ref{Global_factor} shows the estimated global factor which has a clear upward trend. First, let's explain why such a behaviour can be captured under the proposed framework. Note that Assumption \ref{ass1} requires $\frac{1}{T}\textbf{G}'\textbf{G}\to_P \bm{\Sigma}_{\textbf{G}}$ only. As a special case, it may possess a form like

\begin{eqnarray}\label{trend}
\frac{1}{T}\sum_{t=1}^T [g(\tau_t)]^2 \to \int_0^1 [g(w)]^2 dw,
\end{eqnarray}
where $\tau_t=t/T$, and $g(\cdot)$ can be functions such as $g(w)=w$, $g(w)=w^2$, etc. Therefore, the upward trending is obviously included. Detailed discussions on trending behaviour like \eqref{trend} can be seen in \cite{YGP2020}. As explained in \cite{Wang2008} and \cite{Jorg2016}, the global factor may be interpreted as global shocks on the entire network of international trade, e.g., the Global Financial Crisis. Our finding is somewhat consistent with their arguments. For example, there is a sudden and severe drop around 2009 which captures the so-called ``great trade collapse", a consequence of the 2008 financial crisis, occurred between the third quarter of 2008 and the second quarter of 2009. We refer interested readers to \cite{BJY2012} for more details on great trade collapse. In addition, we note that the global factor becomes more volatile over the sample period, which may indicate the increasing vulnerability of countries to shocks on trade due to globalization over the past couple of decades.

\begin{figure} [H]
\includegraphics[scale=0.6]{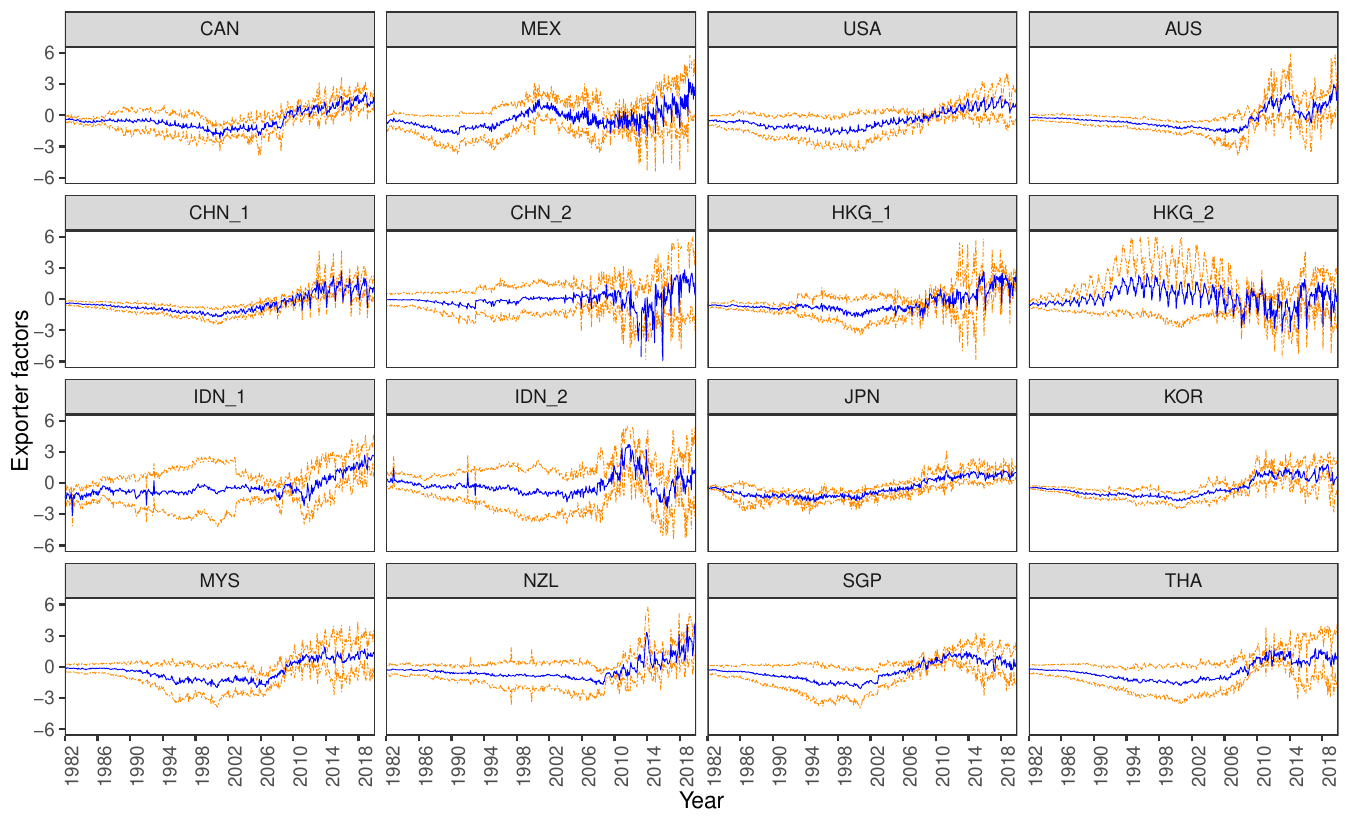}
\caption{Exporter factors for 13 APEC countries/region in the sample from January, 1982 to December, 2019. The blue solid line represents the estimated global factor and the orange dashed lines provide a 95\% confidence interval.}
\centering
\label{Exporter_factors1}
\end{figure}

\begin{figure} [H]
\includegraphics[scale=0.6]{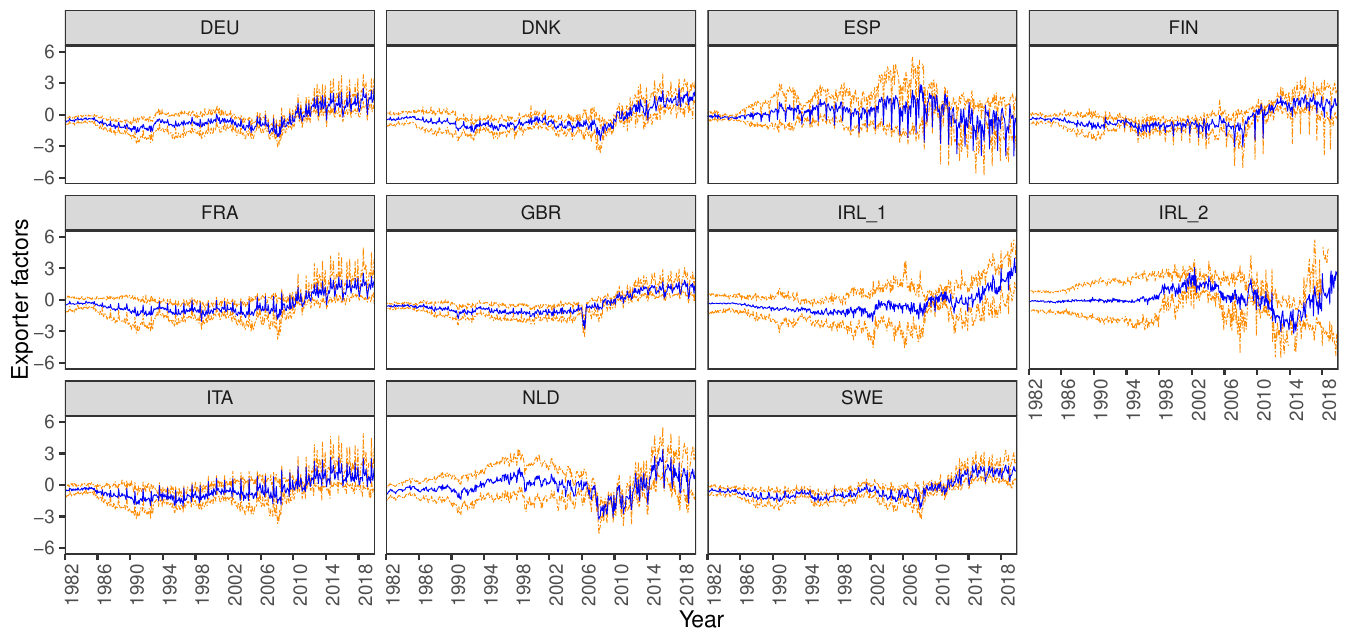}
\caption{Exporter factors for 10 EU countries in the sample from January, 1982 to December, 2019. The blue solid line represents the estimated global factor and the orange dashed lines provide a 95\% confidence interval.}
\centering
\label{Exporter_factors2}
\end{figure}

\begin{figure} [H]
\includegraphics[scale=0.6]{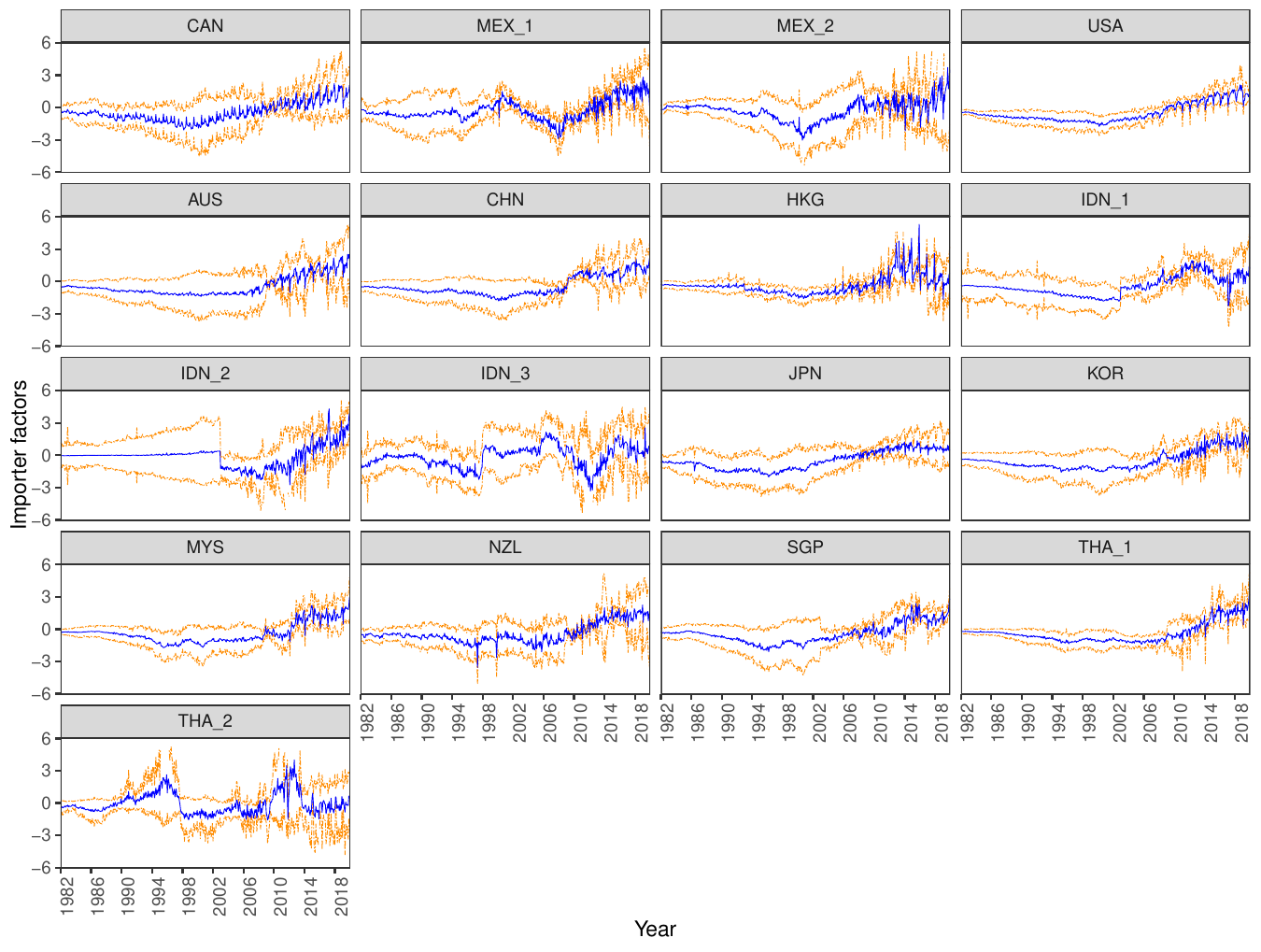}
\caption{Importer factors for 13 APEC countries/region in the sample from January, 1982 to December, 2019. The blue solid line represents the estimated global factor and the orange dashed lines provide a 95\% confidence interval.}
\centering
\label{Importer_factors1}
\end{figure}

\begin{figure} [H]
\includegraphics[scale=0.6]{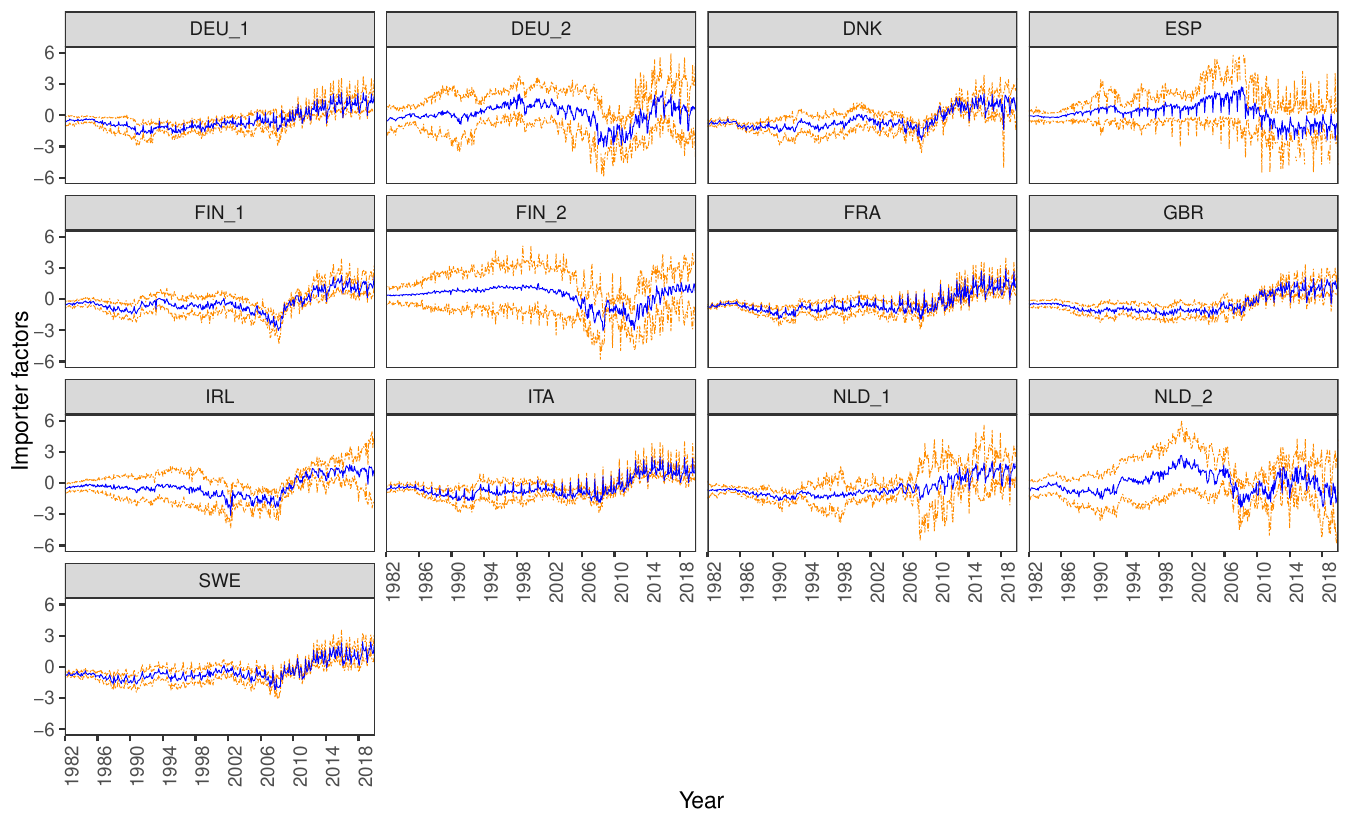}
\caption{Importer factors for 10 EU countries in the sample from January, 1982 to December, 2019. The blue solid line represents the estimated global factor and the orange dashed lines provide a 95\% confidence interval.}
\centering
\label{Importer_factors2}
\end{figure}

Figures \ref{Exporter_factors1} - \ref{Importer_factors2} show the estimated exporter factors and importer factors. Specifically, Figure \ref{Exporter_factors1} and Figure \ref{Exporter_factors2} present he exporter factors associated with the countries of APEC and EU respectively. Figure \ref{Importer_factors1} and Figure \ref{Importer_factors2} show the importer factors associated with the countries of APEC and EU respectively. The exporter factors can be interpreted as country-specific shocks of export countries which affect the trade volumes from the exporters to the import partners. Similarly, the importer factors can be interpreted as country-specific shocks of import countries which affect the trade volumes from the importers to the export partners. As mentioned in Section \ref{Section2.1}, the exporter and importer factors may capture the unobservable outward and inward multilateral trade resistances (MTRs) for different exporters and importers respectively, which can be seen as measures of outward and inward bilateral trade costs for different exporters and importers. The detailed discussions on the connection between multilateral resistances and country-specific factors can be found in \cite{kapetanios2020estimation}, where the exporter and importer factors are always referred to as source and destination country factors. 
For almost all country-specific factors, we can observe the increase of the volatility, especially from the beginning of the 21st century, indicating the increasing instability of the inward and outward bilateral trade costs for most of the countries in our sample. Under the assumption of bilateral trade costs symmetry, it follows that the inward and outward multilateral resistances are the same for the same country \citep{anderson2003gravity}. By comparing the estimated exporter and importer factors for the same country, it can be seen that this symmetry in the multilateral resistances is partially supported by the data. For example, the exporter and importer factors for USA share the similar trend.

\begin{figure} [H]
\includegraphics[scale=0.7]{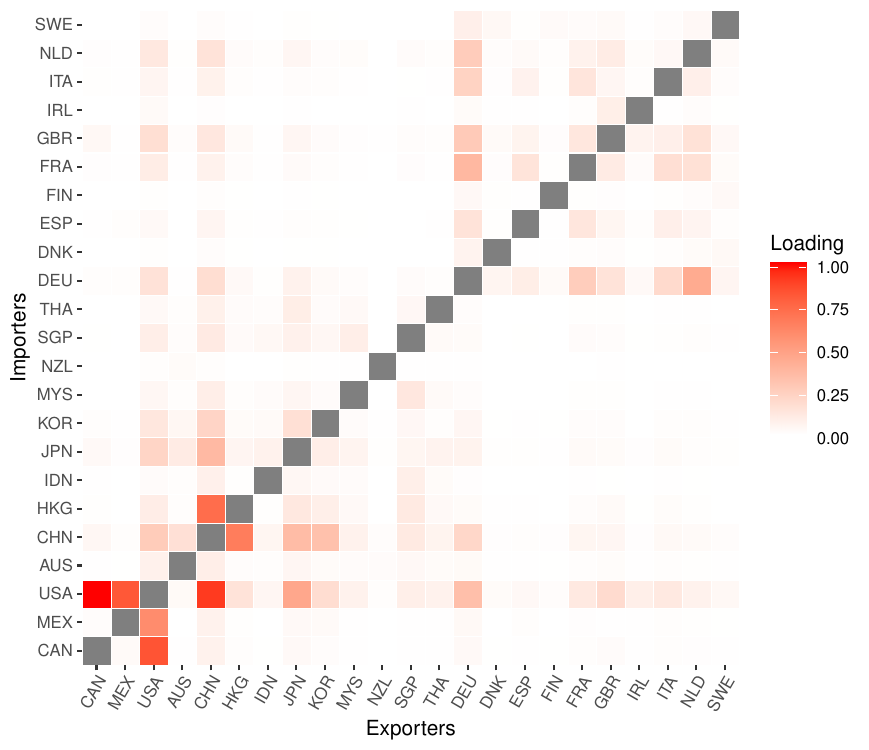}
\caption{Global factor loadings for bilateral export flows. The corresponding figures are rescaled to have values between $0$ and $1$.}
\centering
\label{Global_factor_loading}
\end{figure}

Figure \ref{Global_factor_loading} presents the heat map of the global factor loadings for different country pairs. Since the global factor loadings are positive for all country pairs, we rescale them to $[0,1]$ for better presentation. The global factor loading can be interpreted as the responses of the trade volumes for different country pairs to the global shocks. The colour of each cell reflects the sensitivity of the trade volume between two countries to the global shocks. For example, in Figure \ref{Global_factor_loading}, the darkest cell corresponding to the export flow from CAN to USA indicates that the export volume from CAN to USA is the most sensitive relationship among all bilateral export flows in the sample. Also, the country pairs like CHN and HKG, CHN and USA, MEX and USA also show strong sensitivity to the global shocks. The relationship among USA, MEX and CAN partially can be explained by the fact that all three of them are the members of NAFTA, which eliminates some trade barriers among the three parties and promotes the trading activities. The similar patterns can also be observed among countries from EU and Asia respectively. Overall, by comparing the values in different rows and columns of the plot, it can be seen that the trade flows involving USA, CHN and DEU show relatively strong sensitivity to global shocks, which indicates that, in general, they are leading export and import countries worldwide.

\begin{figure} [H]
\includegraphics[scale=0.7]{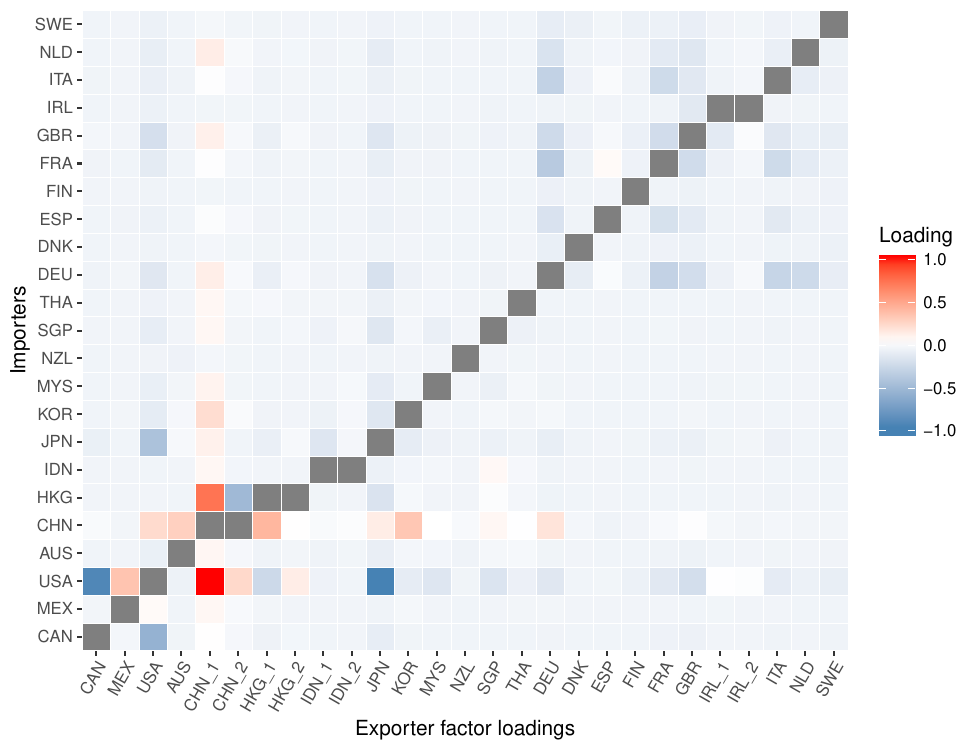}
\caption{Exporter factor loadings for bilateral export flows. The corresponding figures are rescaled to have values between $-1$ and $1$.}
\centering
\label{Exporter_factor_loadings}
\end{figure}

\begin{figure} [H]
\includegraphics[scale=0.7]{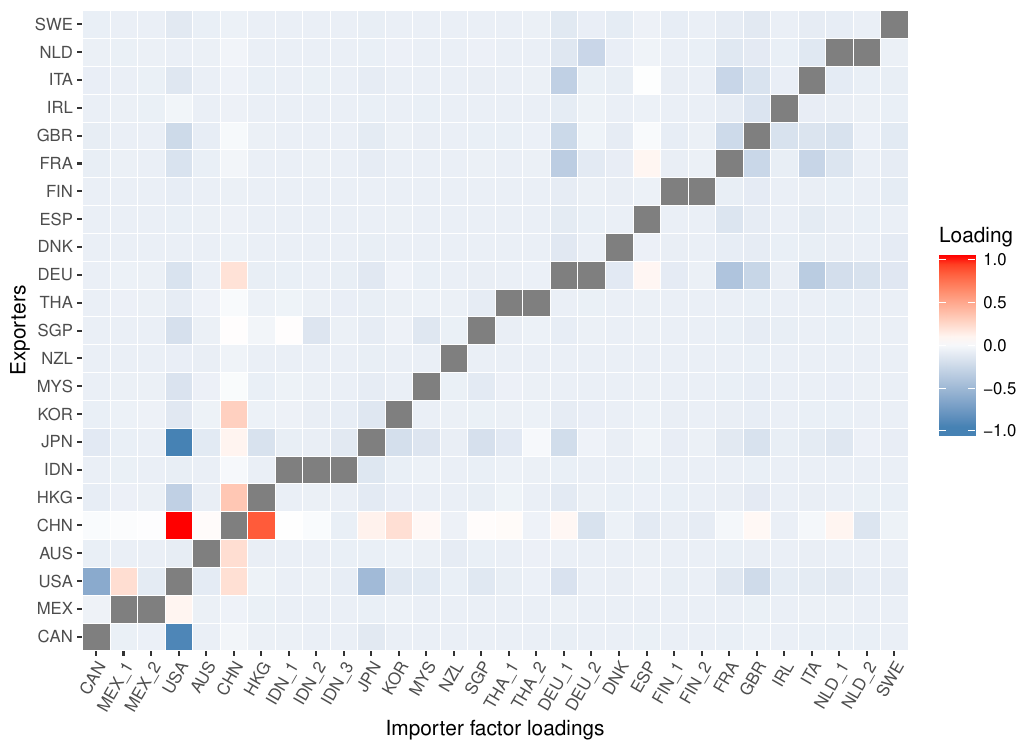}
\caption{Importer factor loadings for bilateral export flows. The corresponding figures are rescaled to have values between $-1$ and $1$.}
\centering
\label{Importer_factor_loadings}
\end{figure}

Figure \ref{Exporter_factor_loadings} and Figure \ref{Importer_factor_loadings} present the heat maps of the exporter factor loadings and the importer factor loadings, respectively. Each column in the plots represents a country-specific factor loading corresponding to an exporter or importer factor. Similar to the global factor loading, the exporter and importer factor loadings are rescaled to have values between $-1$ and $1$. The exporter factor loadings corresponding to different export countries measure the responses of their import partners to the shocks on those export countries. The importer factor loadings can be interpreted in the same manner. As shown in Figure \ref{Exporter_factor_loadings}, the export flow from CAN to USA is relatively sensitive to the exporter shocks of CAN. This is also the case for the export flow from JPN to USA which is shown to be sensitive to the exporter shocks of JPN. On the other hand, Figure \ref{Importer_factor_loadings} shows that the export flows from both CAN and JPN to USA are also sensitive to the country-specific importer shocks of USA. The country-specific factor loadings for other countries can be interpreted similarly.

\section{Conclusion}\label{Section5}
In this study, we specifically consider a three-dimensional panel data model, which has been exposed in the literature but has not been fully solved to the best of the authors' knowledge. On theory, our contributions are the following three-fold: (1). under the scenario that all three dimensions can diverge to infinity, we propose an estimation approach to identify the number of global shocks and country-specific shocks sequentially; (2). the newly proposed approach is easy to implement, and the asymptotic theories are established accordingly; (3). we further conduct intensive numerical studies to examine the finite sample performance of the newly proposed approach using both simulated and real datasets. In the empirical study, we then apply the approach to decompose the network of bilateral trade using country level data from two major trading groups (APEC and EU) over the period 1982-2019. 
We find that the country-specific shocks become more volatile in recent years, which may indicate the increasing instability of the inward and outward bilateral trade costs over the past couple of decades. In addition, we show that the trade flows involving China mainland, Germany and the United States show relatively strong sensitivity to global shocks, which reflects the fact that, in general, they are leading export and import countries worldwide. We note that the relationship among Canada, Mexico, and the United States is also highly sensitive to different shocks, which somewhat reflects the fact that all three of them are highly economically related through NAFTA that eliminates some trade barriers and promotes the trading activities.

{\footnotesize
\bibliography{mybibliography}
}

\newpage

\small

\begin{table}[htbp]\footnotesize
\begin{threeparttable}
  \centering
  \caption{DGP 1 -- The percentages of correctly, under and over selecting factors. Specifically, $P_{g,c}$, $P_{g,u}$ and $P_{g,o}$ are for the global factors; $P_{E,c}$, $P_{E,u}$ and $P_{E,o}$ are for the export factors; $P_{I,c}$, $P_{I,u}$ and $P_{I,o}$ are for the import factors. For the sake of space, some combinations of $(M,N,T)$ are omitted in the table.} \label{Table1}
    \begin{tabular}{lllcccccccccccc}
    \hline \hline                  
    $M$ & $N$ & $T$ & $P_{g,c}$ & $P_{g,u}$ & $P_{g,o}$ && $P_{E,c}$ & $P_{E,u}$ & $P_{E,o}$ && $P_{I,c}$ & $P_{I,u}$ & $P_{I,o}$ \\
   \cline{4-6} \cline{8-10} \cline{12-14}
    20 & 20 & 20   & 0.650 & 0.349 & 0.001 && 0.439 & 0.418 & 0.144 && 0.682 & 0.049 & 0.269 \\
       &    & 40   & 0.978 & 0.022 & 0.000 && 0.655 & 0.287 & 0.059 && 0.844 & 0.046 & 0.110  \\
       &    & 60   & 0.995 & 0.005 & 0.000 && 0.710 & 0.238 & 0.051 && 0.878 & 0.035 & 0.087  \\
       &    & 80   & 1.000 & 0.000 & 0.000 && 0.734 & 0.211 & 0.055 && 0.890 & 0.033 & 0.077 \\  
    \cline{4-14}
    40 & 40 & 20   & 0.836 & 0.164 & 0.000 && 0.629 & 0.306 & 0.065 && 0.854 & 0.062 & 0.084 \\
       &    & 40   & 0.997 & 0.003 & 0.000 && 0.883 & 0.107 & 0.010 && 0.963 & 0.022 & 0.015 \\
       &    & 60   & 0.999 & 0.001 & 0.000 && 0.936 & 0.054 & 0.010 && 0.980 & 0.010 & 0.009 \\
       &    & 80   & 1.000 & 0.000 & 0.000 && 0.958 & 0.033 & 0.010 && 0.986 & 0.006 & 0.008 \\       
    \cline{4-14}
    60 & 60 & 20   & 0.884 & 0.116 & 0.000 && 0.731 & 0.227 & 0.042 && 0.906 & 0.053 & 0.041  \\
       &    & 40   & 0.998 & 0.002 & 0.000 && 0.955 & 0.041 & 0.003 && 0.988 & 0.009 & 0.002 \\
       &    & 60   & 1.000 & 0.000 & 0.000 && 0.985 & 0.013 & 0.002 && 0.996 & 0.003 & 0.002 \\
       &    & 80   & 1.000 & 0.000 & 0.000 && 0.993 & 0.005 & 0.002 && 0.998 & 0.001 & 0.001 \\       
    \cline{4-14}
    80 & 80 & 20   & 0.865 & 0.135 & 0.000 && 0.759 & 0.185 & 0.056 && 0.915 & 0.041 & 0.044 \\
       &    & 40   & 1.000 & 0.000 & 0.000 && 0.980 & 0.019 & 0.001 && 0.995 & 0.004 & 0.001 \\
       &    & 60   & 1.000 & 0.000 & 0.000 && 0.996 & 0.003 & 0.001 && 0.999 & 0.001 & 0.000 \\
       &    & 80   & 1.000 & 0.000 & 0.000 && 0.999 & 0.001 & 0.000 && 1.000 & 0.000 & 0.000 \\ 
    \cline{4-14}
    20 & 80 & 20   & 0.759 & 0.241 & 0.000 && 0.691 & 0.193 & 0.116 && 0.711 & 0.043 & 0.246 \\
       &    & 40   & 0.990 & 0.010 & 0.000 && 0.970 & 0.022 & 0.008 && 0.853 & 0.041 & 0.106 \\
       &    & 60   & 1.000 & 0.000 & 0.000 && 0.996 & 0.003 & 0.001 && 0.884 & 0.033 & 0.084 \\
       &    & 80   & 1.000 & 0.000 & 0.000 && 0.999 & 0.001 & 0.000 && 0.899 & 0.027 & 0.074 \\
    \cline{4-14}
    80 & 20 & 20   & 0.819 & 0.181 & 0.000 && 0.473 & 0.420 & 0.107 && 0.886 & 0.044 & 0.070 \\
       &    & 40   & 0.998 & 0.002 & 0.000 && 0.650 & 0.294 & 0.056 && 0.995 & 0.005 & 0.000 \\
       &    & 60   & 1.000 & 0.000 & 0.000 && 0.714 & 0.233 & 0.053 && 0.999 & 0.001 & 0.000 \\
       &    & 80   & 1.000 & 0.000 & 0.000 && 0.743 & 0.205 & 0.051 && 1.000 & 0.000 & 0.000 \\ 
    \hline \hline
    \end{tabular}
\end{threeparttable}
\end{table}

\begin{table}[htbp]\footnotesize
\begin{threeparttable}
  \centering
  \caption{DGP 2 -- The percentages of correctly, under and over selecting factors. Specifically, $P_{g,c}$, $P_{g,u}$ and $P_{g,o}$ are for the global factors; $P_{E,c}$, $P_{E,u}$ and $P_{E,o}$ are for the export factors; $P_{I,c}$, $P_{I,u}$ and $P_{I,o}$ are for the import factors. For the sake of space, some combinations of $(M,N,T)$ are omitted in the table.} \label{Table2}
    \begin{tabular}{lllcccccccccccc}
    \hline \hline
    $M$ & $N$ & $T$ & $P_{g,c}$ & $P_{g,u}$ & $P_{g,o}$ && $P_{E,c}$ & $P_{E,u}$ & $P_{E,o}$ && $P_{I,c}$ & $P_{I,u}$ & $P_{I,o}$ \\
   \cline{4-6} \cline{8-10} \cline{12-14}
    20 & 20 & 20   & 0.240 & 0.705 & 0.055 && 0.301 & 0.400 & 0.299 && 0.505 & 0.014 & 0.481 \\
       &    & 40   & 0.700 & 0.300 & 0.000 && 0.467 & 0.356 & 0.178 && 0.694 & 0.005 & 0.301 \\
       &    & 60   & 0.920 & 0.080 & 0.000 && 0.615 & 0.300 & 0.085 && 0.814 & 0.010 & 0.176  \\
       &    & 80   & 0.990 & 0.010 & 0.000 && 0.700 & 0.242 & 0.058 && 0.870 & 0.008 & 0.123 \\   
    \cline{4-14}
    40 & 40 & 20   & 0.280 & 0.660 & 0.060 && 0.339 & 0.380 & 0.280 && 0.543 & 0.023 & 0.435 \\
       &    & 40   & 0.820 & 0.180 & 0.000 && 0.668 & 0.240 & 0.092 && 0.881 & 0.010 & 0.109 \\
       &    & 60   & 0.990 & 0.010 & 0.000 && 0.858 & 0.126 & 0.016 && 0.970 & 0.006 & 0.024 \\
       &    & 80   & 1.000 & 0.000 & 0.000 && 0.916 & 0.073 & 0.011 && 0.988 & 0.001 & 0.011 \\              
    \cline{4-14}
    60 & 60 & 20   & 0.220 & 0.750 & 0.030 && 0.329 & 0.342 & 0.329 && 0.562 & 0.010 & 0.428 \\
       &    & 40   & 0.800 & 0.200 & 0.000 && 0.738 & 0.148 & 0.114 && 0.895 & 0.011 & 0.094 \\
       &    & 60   & 0.990 & 0.010 & 0.000 && 0.935 & 0.054 & 0.011 && 0.985 & 0.003 & 0.012 \\
       &    & 80   & 1.000 & 0.000 & 0.000 && 0.974 & 0.025 & 0.001 && 0.997 & 0.001 & 0.002 \\       
    \cline{4-14}
    80 & 80 & 20   & 0.290 & 0.700 & 0.010 && 0.342 & 0.347 & 0.311 && 0.613 & 0.009 & 0.378 \\
       &    & 40   & 0.850 & 0.150 & 0.000 && 0.794 & 0.110 & 0.096 && 0.919 & 0.006 & 0.075 \\
       &    & 60   & 0.990 & 0.010 & 0.000 && 0.964 & 0.028 & 0.008 && 0.994 & 0.001 & 0.005 \\
       &    & 80   & 1.000 & 0.000 & 0.000 && 0.991 & 0.008 & 0.001 && 1.000 & 0.000 & 0.000 \\     
    \cline{4-14}
    20 & 80 & 20   & 0.239 & 0.698 & 0.063 && 0.340 & 0.344 & 0.316 && 0.494 & 0.013 & 0.493 \\
       &    & 40   & 0.720 & 0.280 & 0.000 && 0.676 & 0.117 & 0.207 && 0.716 & 0.006 & 0.278 \\
       &    & 60   & 0.990 & 0.010 & 0.000 && 0.965 & 0.026 & 0.009 && 0.845 & 0.007 & 0.149 \\
       &    & 80   & 1.000 & 0.000 & 0.000 && 0.990 & 0.010 & 0.000 && 0.878 & 0.005 & 0.117 \\   
    \cline{4-14}
    80 & 20 & 20   & 0.270 & 0.670 & 0.060 && 0.305 & 0.394 & 0.301 && 0.582 & 0.025 & 0.393 \\
       &    & 40   & 0.750 & 0.250 & 0.000 && 0.485 & 0.366 & 0.149 && 0.851 & 0.008 & 0.141 \\
       &    & 60   & 0.970 & 0.030 & 0.000 && 0.625 & 0.307 & 0.068 && 0.981 & 0.002 & 0.017 \\
       &    & 80   & 1.000 & 0.000 & 0.000 && 0.691 & 0.254 & 0.055 && 0.999 & 0.001 & 0.000 \\
    \hline \hline
    \end{tabular}
\end{threeparttable}
\end{table}

\begin{table}[htbp]\footnotesize
\begin{threeparttable}
  \centering
  \caption{The results of $\mbox{RMSE}_{\textbf{G}} $, $\mbox{RMSE}_{\textbf{E}} $ and $\mbox{RMSE}_{\textbf{I}} $. Specifically, $\mbox{RMSE}_{\textbf{G}} $, $\mbox{RMSE}_{\textbf{E}} $ and $\mbox{RMSE}_{\textbf{I}} $ measure the estimation on global factors, export factors and importer factors, respectively. For the sake of space, some combinations of $(M,N,T)$ are omitted in the table.} \label{Table3}
    \begin{tabular}{lllccccccccc}
    \hline \hline
        &     &     & \multicolumn{3}{c}{DGP 1} && \multicolumn{3}{c}{DGP 2} \\
    $M$ & $N$ & $T$ & $\mbox{RMSE}_{\textbf{G}} $ & $\mbox{RMSE}_{\textbf{E}} $ & $\mbox{RMSE}_{\textbf{I}} $ && $\mbox{RMSE}_{\textbf{G}}$  & $\mbox{RMSE}_{\textbf{E}} $ & $\mbox{RMSE}_{\textbf{I}} $ \\
    \cline{4-6} \cline{8-10}
    20 & 20 & 20   & 0.778 & 1.172 & 1.034 && 1.189 & 1.370 & 1.321 \\
       &    & 40   & 0.368 & 0.989 & 0.793 && 0.805 & 1.174 & 1.146 \\
       &    & 60   & 0.314 & 0.910 & 0.733 && 0.496 & 1.038 & 0.961 \\
       &    & 80   & 0.292 & 0.884 & 0.695 && 0.359 & 0.947 & 0.891 \\
    \cline{4-10}
    40 & 40 & 20   & 0.519 & 1.015 & 0.761 && 1.164 & 1.266 & 1.165 \\
       &    & 40   & 0.191 & 0.778 & 0.574 && 0.581 & 0.960 & 0.790 \\
       &    & 60   & 0.168 & 0.683 & 0.525 && 0.275 & 0.784 & 0.610 \\
       &    & 80   & 0.154 & 0.637 & 0.499 && 0.222 & 0.716 & 0.549 \\
    \cline{4-10}
    60 & 60 & 20   & 0.423 & 0.941 & 0.680 && 1.144 & 1.234 & 1.189 \\
       &    & 40   & 0.141 & 0.687 & 0.512 && 0.582 & 0.953 & 0.718 \\
       &    & 60   & 0.118 & 0.595 & 0.454 && 0.261 & 0.715 & 0.546 \\
       &    & 80   & 0.108 & 0.546 & 0.423 && 0.177 & 0.652 & 0.500 \\
    \cline{4-10}
    80 & 80 & 20   & 0.440 & 0.915 & 0.664 && 1.110 & 1.243 & 1.130 \\
       &    & 40   & 0.110 & 0.655 & 0.470 && 0.539 & 0.888 & 0.679 \\
       &    & 60   & 0.094 & 0.557 & 0.421 && 0.231 & 0.698 & 0.500 \\
       &    & 80   & 0.086 & 0.504 & 0.394 && 0.163 & 0.603 & 0.458 \\ 
    \cline{4-10}
    20 & 80 & 20   & 0.618 & 0.978 & 1.009 && 1.186 & 1.251 & 1.337 \\
       &    & 40   & 0.232 & 0.671 & 0.792 && 0.711 & 0.956 & 1.062 \\
       &    & 60   & 0.180 & 0.571 & 0.719 && 0.295 & 0.678 & 0.818 \\
       &    & 80   & 0.168 & 0.519 & 0.689 && 0.236 & 0.612 & 0.739 \\    
    \cline{4-10}
    80 & 20 & 20   & 0.560 & 1.131 & 0.694 && 1.136 & 1.427 & 1.088 \\
       &    & 40   & 0.191 & 0.961 & 0.492 && 0.657 & 1.131 & 0.723 \\
       &    & 60   & 0.164 & 0.890 & 0.432 && 0.307 & 0.958 & 0.534 \\
       &    & 80   & 0.154 & 0.866 & 0.403 && 0.217 & 0.925 & 0.468 \\    
    \hline \hline
    \end{tabular}
    \end{threeparttable} 
\end{table}

\begin{table}[htbp]\footnotesize
\begin{threeparttable}
\centering
\caption{Estimated number of exporter factors and importer factors.} \label{Table4}
\begin{tabular}{lcccccccc}
    \hline \hline
  &  \multicolumn{8}{c}{APEC}\\
             & CAN & MEX & USA & AUS & CHN & HKG & IDN & JPN \\
\cline{2-9}
Exporter  & 1   & 1   & 1   & 1   & 2   & 2  & 2   & 1 \\
Importer  & 1   & 2   & 1   & 1   & 1   & 1  & 3   & 1 \\
            & KOR & MYS & NZL & SGP & THA \\
\cline{2-6}
Exporter  & 1   & 1   & 1   & 1  & 1    \\
Importer  & 1   & 1   & 1   & 1  & 2  \\ \\
  &  \multicolumn{8}{c}{EU}\\
                & DEU & DNK & ESP & FIN & FRA & GBR & IRL & ITA    \\
\cline{2-9}
Exporter & 1   & 1   & 1   & 1   & 1   & 1   & 2   & 1  \\
Importer & 2   & 1   & 1   & 2   & 1   & 1   & 1   & 1  \\
               & NLD & SWE    \\
\cline{2-3}
Exporter & 1   & 1     \\
Importer & 2   & 1     \\
\hline \hline
\end{tabular}
\end{threeparttable}
\end{table}

\newpage

\vspace*{1 cm}

\newpage

\begin{center}
{\large \bf Online Supplementary Appendices to \\``Decomposition of Bilateral Trade Flows Using a Three-Dimensional Panel Data Model"}

\bigskip
\bigskip

{\sc Yufeng Mao$^\sharp$, Bin Peng$^\sharp$, Mervyn Silvapulle$^\sharp$, Param Silvapulle$^\sharp$

and Yanrong Yang$^*$}

\medskip

$^\sharp$Monash University and $^*$Australian National University

\end{center}

\small

This file includes two appendices. Appendix A presents the preliminary lemmas, and the proofs of the main results. We relegate the secondary results and the associated proofs to Appendix B. Specifically, Appendix \ref{SecA.1} scratches the outline of the proofs, while Appendix \ref{SecA.2} presents the preliminary lemmas, which facilitate the development of the main results. Appendix \ref{SecA.3} summaries the proofs for each step. In Appendix B, Appendix \ref{AppendixB.2} states the secondary lemmas, while Appendix \ref{ProofofLB} includes all the corresponding proofs.

\setcounter{page}{1}

\section*{Appendix A}

\renewcommand{\theequation}{A.\arabic{equation}}
\renewcommand{\thesection}{A.\arabic{section}}
\renewcommand{\thefigure}{A.\arabic{figure}}
\renewcommand{\thetable}{A.\arabic{table}}
\renewcommand{\thelemma}{A.\arabic{lemma}}
\renewcommand{\theassumption}{A.\arabic{assumption}}
\renewcommand{\thetheorem}{A.\arabic{theorem}}

\setcounter{equation}{0}
\setcounter{lemma}{0}
\setcounter{section}{0}
\setcounter{table}{0}
\setcounter{figure}{0}
\setcounter{assumption}{0}

\section{Outline of the Proofs}\label{SecA.1}

In this section, we provide the outline of the proofs. Lemma \ref{lemma_nof1} presents the relevant results about the eigenvalues associated with $\frac{1}{MNT} \textbf{Y}'\textbf{Y}$ of (\ref{Eq2.7}), which yields the estimation of $r_g$, and thus leads to Theorem \ref{theorem1} of the main text. After establishing Theorem \ref{theorem1}, we take the value of $r_g$ as granted and focus on the estimation of country-specific factors. Specifically, we derive Lemma \ref{lemma_nof1_local} and Lemma \ref{lemma_nof2_local} in the same manner as Lemma \ref{lemma_nof1} by making use of the newly imposed Assumption \ref{ass4}, which allows us to further estimate the number of exporter factors and importer factors, and thus establish Theorem \ref{theorem2} of the main text. 

After identifying the number of factors successfully, Lemma \ref{lemma2_improved}-\ref{lemma_clt0_local} provide the further results in order to establish the asymptotic distributions associated with the global and country-specific factors. Specifically, Lemma \ref{lemma2_improved} improve the convergence rate of the global factor presented in Theorem \ref{theorem1} by making use of Assumption \ref{ass4} and Assumption \ref{ass5}. Then we establish the first result of Theorem \ref{theorem_global_clt} in the main text which provides the asymptotic distribution for the global factor. Based on the results in Lemma \ref{lemma2_improved}.1, Lemma \ref{lemma6_improved} updates the convergence rate of the country-specific factors presented in Theorem \ref{theorem2} of the main text. Similar to the development of the asymptotic distribution for the global factor, we then establish the asymptotic distributions for the exporter and importer factors in the second and third results of Theorem \ref{theorem_global_clt} in the main text after imposing Assumption \ref{ass6}.

\section{Preliminary Lemmas}\label{SecA.2}

\begin{lemma} \label{lemma_nof1}
Let Assumptions \ref{ass1}-\ref{ass3} hold, and let $(M,N,T) \to (\infty,\infty,\infty)$.

\begin{enumerate}

\item $\normalfont\frac{1}{\sqrt{T}} \| \widehat{\textbf{G}}^{\dag} - \textbf{G} \textbf{H} \|_{F} = O_P \left( \frac{\sqrt{ \ln (MN)}}{\min \{\sqrt{M}, \sqrt{N}, \sqrt{T}\}} \right)$, where $\normalfont \widehat{\textbf{G}}^\dag$ includes the first $r_g$ columns of $\normalfont\widehat{\textbf{G}}$ only, $\normalfont\textbf{H} = \frac{1}{MN}\bm{\Gamma}' \bm{\Gamma}\cdot \frac{1}{T}\textbf{G}' \widehat{\textbf{G}}^{\dag} \cdot(\textbf{V}_{g}^{\dag})^{-1}$ is an $r_g \times r_g$ invertible matrix, $\normalfont\textbf{V}_{g}^{\dag}$ is the $r_g\times r_g$ leading principal submatrix of $\normalfont \textbf{V}_g$ and $\normalfont\| \textbf{H} \|_F = O_P(1)$;

\item $ |\widehat{\rho}_{g,k} - \rho_{g,k}^{\triangledown}| =  O_P \left( \frac{\sqrt{ \ln (MN)}}{\min \{\sqrt{M}, \sqrt{N}, \sqrt{T}\}} \right)$ for $k=1,\ldots, r_g$, where $\normalfont \widehat{\rho}_{g,k} = \frac{1}{T} \widehat{\textbf{G}}_{k}^{\dag \prime} \widehat{\bm{\Sigma}} \widehat{\textbf{G}}_{k}^{\dag}$ with $\normalfont\widehat{\textbf{G}}_{k}^{\dag}$ being the $k$-th column of $\normalfont\widehat{\textbf{G}}^{\dag}$, $\normalfont\rho_{g,k}^{\triangledown} =\frac{1}{T} \textbf{h}_k^{\prime} \textbf{G}^{\prime} \bm{\Sigma} \textbf{G}\textbf{h}_k$, $\widehat{\bm{\Sigma}} =\normalfont\frac{1}{MNT}\textbf{Y}'\textbf{Y}$, $\normalfont\bm{\Sigma} = \frac{1}{MNT}\textbf{G} \bm{\Gamma}' \bm{\Gamma} \textbf{G}' $, and $\normalfont\textbf{h}_k$ is the $k$-th column of $\normalfont\textbf{H}$; 
\item $ |\widehat{\rho}_{g,k} | = O_P \left(\frac{ \ln (MN)}{\min \{M,N,T\}} \right)$ for $k=r_g+1,\ldots,k_{\max}$.
\end{enumerate}
\end{lemma}

\begin{lemma} \label{lemma_nof1_local}
Let Assumptions \ref{ass1}-\ref{ass4} hold, and let $(M,N,T) \to (\infty,\infty,\infty)$. For $j= 1,\ldots, N$,

\begin{enumerate}

\item $\normalfont\frac{1}{\sqrt{T}} \| \widehat{\textbf{F}}_{I,j}^{\dag} - \textbf{F}_{I, j} \textbf{H}_{I,j} \|_{F} = O_P \left( \frac{\sqrt{ \ln (MN)}}{\min \{\sqrt{M}, \sqrt{N}, \sqrt{T}\}} + T^{a_{I,j}}\right)$, where $\normalfont \widehat{\textbf{F}}_{I,j}^\dag$ includes the first $r_{I,j}$ columns of $\normalfont\widehat{\textbf{F}}_{I,j}$ only, $\normalfont\textbf{H}_{I,j} = \frac{1}{M}\bm{\Lambda}_{I, \bullet j}' \bm{\Lambda}_{I, \bullet j}\cdot \frac{1}{T}\textbf{F}_{I,j}' \widehat{\textbf{F}}_{I,j}^{\dag} \cdot(\textbf{V}_{I,j}^{\dag})^{-1}$ is an $r_{I,j} \times r_{I,j}$ invertible matrix, $\normalfont\textbf{V}_{I,j}^{\dag}$ is the $r_{I,j}\times r_{I,j}$ leading principal submatrix of $\normalfont \textbf{V}_{I,j}$ and $\normalfont\| \textbf{H}_{I,j} \|_F = O_P(1)$;

\item $ |\widehat{\rho}_{Ij,k} - \rho_{Ij,k}^{\triangledown}| =  O_P \left( \frac{\sqrt{ \ln (MN)}}{\min \{\sqrt{M}, \sqrt{N}, \sqrt{T}\}} +T^{a_{I,j}} \right)$ for $k=1,\ldots, r_{I,j}$, where $\normalfont \widehat{\rho}_{Ij,k} = \frac{1}{T} \widehat{\textbf{F}}_{Ij,k}^{\dag \prime} \widehat{\bm{\Sigma}}_{I,j} \widehat{\textbf{F}}_{Ij,k}^{\dag}$ with $\normalfont\widehat{\textbf{F}}_{Ij,k}^{\dag}$ being the $k$-th column of $\normalfont\widehat{\textbf{F}}_{I,j}^{\dag}$, $\normalfont\rho_{Ij,k}^{\triangledown} =\frac{1}{T} \textbf{h}_{Ij,k}^{\prime} \textbf{F}_{I,j}^{\prime} \bm{\Sigma}_{I,j} \textbf{F}_{I,j}\textbf{h}_{Ij,k}$, $\normalfont\widehat{\bm{\Sigma}}_{I,j} = \frac{1}{MT}(\textbf{Y}_{I, j} - \widehat{\bm{\Gamma}}_{I, j} \widehat{\textbf{G}}' )' (\textbf{Y}_{I, j} - \widehat{\bm{\Gamma}}_{I, j} \widehat{\textbf{G}}' )$, $\normalfont\bm{\Sigma}_{I,j} = \frac{1}{MT}\textbf{F}_{I,j} \bm{\Lambda}_{I,\bullet j}' \bm{\Lambda}_{I,\bullet j} \textbf{F}_{I,j}' $, and $\normalfont\textbf{h}_{Ij,k}$ is the $k$-th column of $\normalfont\textbf{H}_{I,j}$; 

\item $ |\widehat{\rho}_{Ij,k} | = O_P \left(\frac{ \ln (MN)}{\min \{M,N,T\}} +T^{2a_{I,j}}\right)$ for $k=r_{I,j}+1,\ldots,k_{\max}$.
\end{enumerate}

\end{lemma}

\begin{lemma} \label{lemma_nof2_local}
Let Assumptions \ref{ass1}-\ref{ass4} hold, and let $(M,N,T) \to (\infty,\infty,\infty)$. For $i= 1,\ldots, M$,

\begin{enumerate}

\item $\normalfont\frac{1}{\sqrt{T}} \| \widehat{\textbf{F}}_{E,i}^{\dag} - \textbf{F}_{E, i} \textbf{H}_{E,i} \|_{F} = O_P \left( \frac{\sqrt{ \ln (MN)}}{\min \{\sqrt{M}, \sqrt{N}, \sqrt{T}\}} +T^{a_{E,i}}\right)$, where $\normalfont \widehat{\textbf{F}}_{E,i}^\dag$ includes the first $r_{E,i}$ columns of $\normalfont\widehat{\textbf{F}}_{E,i}$ only, $\normalfont\textbf{H}_{E,i} = \frac{1}{N}\bm{\Lambda}_{E, i\bullet }' \bm{\Lambda}_{E, i\bullet }\cdot \frac{1}{T}\textbf{F}_{E,i}' \widehat{\textbf{F}}_{E,i}^{\dag} \cdot(\textbf{V}_{E,i}^{\dag})^{-1}$ is an $r_{E,i} \times r_{E,i}$ invertible matrix, $\normalfont\textbf{V}_{E,i}^{\dag}$ is the $r_{E,i}\times r_{E,i}$ leading principal submatrix of $\normalfont \textbf{V}_{E,i}$ and $\normalfont\| \textbf{H}_{E,i} \|_F = O_P(1)$;

\item $ |\widehat{\rho}_{Ei,k} - \rho_{Ei,k}^{\triangledown}| =  O_P \left( \frac{\sqrt{ \ln (MN)}}{\min \{\sqrt{M}, \sqrt{N}, \sqrt{T}\}} +T^{a_{E,i}} \right)$ for $k=1,\ldots, r_{E,i}$,  where $\normalfont \widehat{\rho}_{Ei,k} = \frac{1}{T} \widehat{\textbf{F}}_{Ei,k}^{\dag \prime} \widehat{\bm{\Sigma}}_{E,i} \widehat{\textbf{F}}_{Ei,k}^{\dag}$ with $\normalfont\widehat{\textbf{F}}_{Ei,k}^{\dag}$ being the $k$-th column of $\normalfont\widehat{\textbf{F}}_{E,i}^{\dag}$, $\normalfont\rho_{Ei,k}^{\triangledown} =\frac{1}{T} \textbf{h}_{Ei,k}^{\prime} \textbf{F}_{E,i}^{\prime} \bm{\Sigma}_{E,i} \textbf{F}_{E,i}\textbf{h}_{Ei,k}$, $\normalfont\widehat{\bm{\Sigma}}_{E,i} = \frac{1}{NT}(\textbf{Y}_{E, i} - \widehat{\bm{\Gamma}}_{E, i} \widehat{\textbf{G}}' )' (\textbf{Y}_{E, i} - \widehat{\bm{\Gamma}}_{E, i} \widehat{\textbf{G}}' )$, $\normalfont\bm{\Sigma}_{E,i} = \frac{1}{NT}\textbf{F}_{E,i} \bm{\Lambda}_{E, i\bullet}' \bm{\Lambda}_{E,i\bullet} \textbf{F}_{E,i}' $, and $\normalfont\textbf{h}_{Ei,k}$ is the $k$-th column of $\normalfont\textbf{H}_{E,i}$; 

\item $ |\widehat{\rho}_{Ei,k} | = O_P \left(\frac{ \ln (MN)}{\min \{M,N,T\}} +T^{2a_{E,i}}\right)$ for $k=r_{E,i}+1,\ldots,k_{\max}$.
\end{enumerate}
\end{lemma}

\begin{lemma} \label{lemma2_improved}
Let Assumptions \ref{ass1}-\ref{ass3} and Assumption \ref{ass5}.1 hold. Denote that 

\begin{eqnarray*}
&&\Delta_{g, MNT} = \frac{\ln(MN) \cdot (T^{\max_{i}a_{E,i}} + T^{\max_{j}a_{I,j}})}{ \min\{\sqrt{M}, \sqrt{N}, \sqrt{T}\} } \\
\mbox{and } &&\Delta_{g, MNT}^{*} = \frac{T^{\max_{i}a_{E,i}}}{\sqrt{N}} + \frac{T^{\max_{j}a_{I,j}}}{\sqrt{M}} + \frac{\sqrt{\ln(MN)} \cdot \left(T^{\max_{i}a_{E,i}} + T^{\max_{j}a_{I,j}} \right) }{\sqrt{T}} \\
&&\;\;\;\;\;\; + \ln(MN) \cdot ( T^{2\max_{i}a_{E,i}} + T^{2\max_{j}a_{I,j}})
\end{eqnarray*}
for notational simplicity. As $(M,N,T) \to (\infty,\infty,\infty)$,

\begin{enumerate}
\item $\normalfont\frac{1}{\sqrt{T}} \| \widehat{\textbf{G}} - \textbf{G} \textbf{H} \|_{F} = O_P\left( \frac{\sqrt{\ln(MN)}}{\min \{T, \sqrt{MT}, \sqrt{NT} \}} + \frac{1}{\sqrt{MN}} + \Delta_{g, MNT} \right)$;

\item $\normalfont \frac{1}{T} \| \textbf{G}' (\widehat{\textbf{G}}  - \textbf{G} \textbf{H}) \|_{F} = O_P \left( \frac{1}{MN} + \frac{1}{T} + \Delta_{g, MNT}^{*} \right)$.
\end{enumerate}
In addition, suppose that Assumption \ref{ass5}.2 holds. Then

\begin{enumerate}
\item[3.] $\normalfont \textbf{H} = \textbf{I}_{r_g} + O_P \left( \frac{1}{MN} + \frac{1}{T} + \Delta_{g, MNT}^{*} \right)$.
\end{enumerate}
\end{lemma}

\begin{lemma} \label{lemma6_improved}
Let Assumptions \ref{ass1}-\ref{ass5} hold. Denote that
\begin{eqnarray*}
&&\Delta_{Ij, MNT} = \Delta_{g, MNT} + \sqrt{\ln(MN)} \cdot T^{\max_{i}a_{E,i}} + T^{a_{I,j}} \\ 
\mbox{and } &&\Delta_{Ei, MNT} = \Delta_{g, MNT} + \sqrt{\ln(MN)} \cdot T^{\max_{j}a_{I,j}} + T^{a_{E,i}}
\end{eqnarray*}
for notational simplicity, where $\Delta_{g, MNT}$ is defined in Lemma \ref{lemma2_improved}. As $(M,N,T) \to (\infty,\infty,\infty)$,

\begin{enumerate}
\item for $j=1,\ldots, N$, $\normalfont\frac{1}{\sqrt{MT}}\| \bm{\Gamma}_{I, j} \textbf{G}' - \widehat{\bm{\Gamma}}_{I, j}\widehat{\textbf{G}}' \|_F = O_P \left( \frac{1}{\sqrt{MN}} + \frac{1}{\sqrt{T}} + \Delta_{Ij, MNT} \right)$, where $\bm{\Gamma}_{I, j}$ is defined in the same way as $\widehat{\bm{\Gamma}}_{I, j}$;

\item for $i=1,\ldots, M$, $\normalfont\frac{1}{\sqrt{NT}}\| \bm{\Gamma}_{E,i} \textbf{G}' - \widehat{\bm{\Gamma}}_{E,i}\widehat{\textbf{G}}' \|_F = O_P \left( \frac{1}{\sqrt{MN}} + \frac{1}{\sqrt{T}} +  \Delta_{Ei, MNT} \right)$, where $\bm{\Gamma}_{E,i}$ is defined in the same way as $\widehat{\bm{\Gamma}}_{E,i}$.

\item[3.] for $j=1,\ldots, N$, $\normalfont \frac{1}{\sqrt{T}} \| \widehat{\textbf{F}}_{I,j} - \textbf{F}_{I, j} \textbf{H}_{I,j} \|_{F} = O_P\left( \frac{\sqrt{\ln (MN)}}{ \min\{\sqrt{M}, \sqrt{T}\} } +  \Delta_{Ij, MNT} \right)$;

\item[4.] for $i=1,\ldots,M$, $\normalfont \frac{1}{\sqrt{T}} \| \widehat{\textbf{F}}_{E,i} - \textbf{F}_{E, i} \textbf{H}_{E,i} \|_{F} = O_P\left( \frac{\sqrt{\ln (MN)}}{ \min\{ \sqrt{N}, \sqrt{T}\} } +  \Delta_{Ei, MNT} \right)$.
\end{enumerate}
\end{lemma}

\begin{lemma} \label{lemma_clt0_local}
Let Assumptions \ref{ass1}-\ref{ass5} and Assumption \ref{ass6}.1 hold, as $(M,N,T) \to (\infty,\infty,\infty)$,
\begin{enumerate}
\item $\normalfont \frac{1}{T} \| \textbf{F}_{I, j}' ( \widehat{\textbf{F}}_{I,j} - \textbf{F}_{I, j} \textbf{H}_{I,j} ) \|_{F} = O_P\left( \frac{\ln(MN)}{ M } + \frac{1}{\sqrt{MN}} + \frac{1}{\sqrt{T}} + \Delta_{Ij, MNT} + \sqrt{\ln(MN)}\cdot T^{\max_i b_{EI, ij}} \right)$;

\item $\normalfont \frac{1}{T} \| \textbf{F}_{E, i}' ( \widehat{\textbf{F}}_{E,i} - \textbf{F}_{E, i} \textbf{H}_{E,i} ) \|_{F} = O_P\left( \frac{\ln(MN)}{ N } + \frac{1}{\sqrt{MN}} + \frac{1}{\sqrt{T}} + \Delta_{Ei, MNT} + \sqrt{\ln(MN)} \cdot T^{\max_j b_{EI, ij}} \right)$.
\end{enumerate}
In addition, let Assumption \ref{ass6}.2 hold.

\begin{enumerate}
\item[3.] $\normalfont \textbf{H}_{I,j} = \textbf{I}_{r_{I,j}} + O_P\left( \frac{\ln(MN)}{ M } + \frac{1}{\sqrt{MN}} + \frac{1}{\sqrt{T}} + \Delta_{Ij, MNT} + \sqrt{\ln(MN)} \cdot T^{\max_i b_{EI, ij}} \right)$;

\item[4.] $\normalfont \textbf{H}_{E,j} = \textbf{I}_{r_{E,j}} + O_P\left( \frac{\ln(MN)}{ N } + \frac{1}{\sqrt{MN}} + \frac{1}{\sqrt{T}} + \Delta_{Ei, MNT} + \sqrt{\ln(MN)} \cdot T^{\max_j b_{EI, ij}} \right)$.
\end{enumerate}
\end{lemma}

\section{Proofs}\label{SecA.3}

We now start presenting the proofs..

\subsection{On Consistency} \label{AppendixA.2}

\noindent \textbf{Proof of Lemma \ref{lemma_nof1}}: 

(1). Expanding \eqref{Eq2.7}, we write

\begin{eqnarray*}
\widehat{\textbf{G}} \textbf{V}_{g}&=& \frac{1}{MNT} (\bm{\Gamma}\textbf{G}'+ \bm{\Lambda}_E\textbf{F}_E'+ \bm{\Lambda}_I\textbf{F}_I' + \textbf{U})' (\bm{\Gamma}\textbf{G}'+ \bm{\Lambda}_E\textbf{F}_E'+ \bm{\Lambda}_I\textbf{F}_I' + \textbf{U}) \widehat{\textbf{G}}\\
&=& (\textbf{A}_1 + \cdots + \textbf{A}_{16}) \widehat{\textbf{G}}
\end{eqnarray*}
where

\begin{eqnarray*}
&&\textbf{A}_1 = \frac{1}{MNT} \textbf{G} \bm{\Gamma}' \bm{\Gamma} \textbf{G}',\quad
\textbf{A}_2 = \frac{1}{MNT} \textbf{F}_E \bm{\Lambda}_E' \bm{\Lambda}_E \textbf{F}_E',\quad
\textbf{A}_3 = \frac{1}{MNT} \textbf{F}_{I} \bm{\Lambda}_I' \bm{\Lambda}_I \textbf{F}_{I}', \\
&&\textbf{A}_4 = \frac{1}{MNT} \textbf{U}'\textbf{U}, \quad \textbf{A}_5 = \frac{1}{MNT} \textbf{G} \bm{\Gamma}' \bm{\Lambda}_E \textbf{F}_E', \quad \textbf{A}_6 = \frac{1}{MNT} \textbf{G} \bm{\Gamma}' \bm{\Lambda}_I \textbf{F}_I', \\
&&\textbf{A}_7 = \frac{1}{MNT} \textbf{G} \bm{\Gamma}' \textbf{U}, \quad \textbf{A}_8 = \frac{1}{MNT} \textbf{F}_{E} \bm{\Lambda}_E' \bm{\Gamma} \textbf{G}', \quad \textbf{A}_9 = \frac{1}{MNT} \textbf{F}_{E} \bm{\Lambda}_E' \bm{\Lambda}_I \textbf{F}_{I}', \\ 
&&\textbf{A}_{10} = \frac{1}{MNT} \textbf{F}_{E} \bm{\Lambda}_E'  \textbf{U}, \quad \textbf{A}_{11} = \frac{1}{MNT} \textbf{F}_{I} \bm{\Lambda}_I' \bm{\Gamma} \textbf{G}', \quad \textbf{A}_{12} = \frac{1}{MNT} \textbf{F}_{I} \bm{\Lambda}_I'  \bm{\Lambda}_E \textbf{F}_{E}', \\
&&\textbf{A}_{13} = \frac{1}{MNT} \textbf{F}_{I} \bm{\Lambda}_I'  \textbf{U}, \quad \textbf{A}_{14} = \frac{1}{MNT} \textbf{U}' \bm{\Gamma} \textbf{G}', \quad \textbf{A}_{15} = \frac{1}{MNT} \textbf{U}' \bm{\Lambda}_E \textbf{F}_{E}',  \\
&&\textbf{A}_{16} = \frac{1}{MNT} \textbf{U}' \bm{\Lambda}_I \textbf{F}_{I}'.
\end{eqnarray*}
Thus, the following equality holds immediately.

\begin{equation} \label{eq3.2}
\widehat{\textbf{G}} \textbf{V}_{g} - \textbf{G}\cdot \frac{1}{MN}\bm{\Gamma}' \bm{\Gamma} \cdot \frac{1}{T}\textbf{G}' \widehat{\textbf{G}} = (\textbf{A}_2 + \cdots + \textbf{A}_{16}) \widehat{\textbf{G}},
\end{equation}
in which  

\begin{equation*}
\| \textbf{A}_2 + \cdots + \textbf{A}_{16} \|_2 \leq \| \textbf{A}_2 \|_2 + \cdots + \| \textbf{A}_{16} \|_2 = O_P \left( \frac{\sqrt{ \ln (MN)}}{\min \{\sqrt{M}, \sqrt{N}, \sqrt{T}\}} \right) 
\end{equation*}
by Lemma \ref{lemma2}.

We left multiply \eqref{eq3.2} by $\frac{1}{T}\textbf{G}' $ to yield

\begin{equation*}
\frac{1}{T}\textbf{G}' \widehat{\textbf{G}} \cdot \textbf{V}_{g} - \frac{1}{T}\textbf{G}' \textbf{G} \cdot \frac{1}{MN}\bm{\Gamma}' \bm{\Gamma} \cdot \frac{1}{T}\textbf{G}' \widehat{\textbf{G}}  = \frac{1}{T}\textbf{G}' ( \textbf{A}_2 + \cdots + \textbf{A}_{16}) \widehat{\textbf{G}} = o_P(1),
\end{equation*}
because $ \frac{1}{\sqrt{T}} \| \textbf{G} \|_F = O_P(1)$, $ \| \textbf{A}_2 + \cdots + \textbf{A}_{16} \|_2 = o_P(1)$ and $ \frac{1}{\sqrt{T}} \| \widehat{\textbf{G}} \|_F = O_P(1)$. Furthermore, $\frac{1}{T}\textbf{G}' \textbf{G} \cdot \frac{1}{MN}\bm{\Gamma}' \bm{\Gamma}  = \bm{\Sigma}_{\textbf{G}} \bm{\Sigma}_{\bm{\Gamma}} + o_P(1)$ by Assumptions \ref{ass1} and \ref{ass2}. It follows that

\begin{equation} \label{limit_eigen}
\bm{\Sigma}_{\textbf{G}} \bm{\Sigma}_{\bm{\Gamma}} \cdot \frac{1}{T}\textbf{G}' \widehat{\textbf{G}} = \frac{1}{T} \textbf{G}' \widehat{\textbf{G}} \cdot \textbf{V}_{g} + o_P(1).
\end{equation}

Note that $\frac{1}{T}\textbf{G}' \widehat{\textbf{G}}$ is of rank $r_g$, which then further indicates that $\textbf{V}_{g}$ has at least $r_g$ non-zero elements on the main diagonal which converge to the eigenvalues of $\bm{\Sigma}_{\textbf{G}} \bm{\Sigma}_{\bm{\Gamma}}$. We now can conclude that $\textbf{V}_{g}$ is of rank $r_g$ in the limit.

We are now ready to investigate $\widehat{\rho}_{g,k}$ with $k \leq r_g$. Because here $k \leq r_g$, we then focus on $\widehat{\textbf{G}}^{\dag}$ and $\textbf{V}_g^{\dag}$, which are defined in this Lemma. In particular, by the argument under (\ref{limit_eigen}), we can conclude that $\textbf{V}_{g}^{\dag} \to_P \textbf{V}_{0g}$, where $\textbf{V}_{0g} = \diag \{\rho_{g,1}, \ldots, \rho_{g,r_g}\}$ consisting of the eigenvalues of $\normalfont\bm{\Sigma}_{\textbf{G}} \bm{\Sigma}_{\bm{\Gamma}}$. By \eqref{eq3.2}, we can write

\begin{equation} \label{eq3.2b}
\widehat{\textbf{G}}^{\dag} \textbf{V}_g^{\dag} - \textbf{G}\cdot \frac{1}{MN} \bm{\Gamma}' \bm{\Gamma} \cdot \frac{1}{T}\textbf{G}' \widehat{\textbf{G}}^{\dag}  = (\textbf{A}_2 + \cdots + \textbf{A}_{16}) \widehat{\textbf{G}}^{\dag}.
\end{equation}

Left multiplying (\ref{eq3.2b}) by $\frac{1}{T}\widehat{\textbf{G}}^{\dag \prime}$ and using the fact that $\frac{1}{T}\widehat{\textbf{G}}^{\dag \prime} \widehat{\textbf{G}}^{\dag} = \textbf{I}_{r_g}$, we have
 
\begin{equation} \label{eq3.3}
\textbf{V}_{g}^{\dag} - \frac{1}{T}\widehat{\textbf{G}}^{\dag \prime}\textbf{G} \cdot \frac{1}{MN}\bm{\Gamma}' \bm{\Gamma}\cdot \frac{1}{T} \textbf{G}' \widehat{\textbf{G}}^{\dag}  = \frac{1}{T} \widehat{\textbf{G}}^{\dag \prime} (\textbf{A}_2 + \cdots + \textbf{A}_{16}) \widehat{\textbf{G}}^{\dag} = o_P(1),
\end{equation}
because $\frac{1}{\sqrt{T}}\| \widehat{\textbf{G}}^{\dag} \|_{F}  = O(1)$ and $ \| \textbf{A}_2 + \cdots + \textbf{A}_{16} \|_{2} = o_P(1)$ by Lemma \ref{lemma2}. By (\ref{eq3.3}) and $\textbf{V}_{g}^{\dag} \to_P \textbf{V}_{0g}$, it follows that

\begin{equation} \label{equivalence}
\frac{1}{T}\widehat{\textbf{G}}^{\dag \prime}\textbf{G} \cdot \frac{1}{MN}\bm{\Gamma}' \bm{\Gamma}\cdot \frac{1}{T} \textbf{G}' \widehat{\textbf{G}}^{\dag}\to_P \textbf{V}_{0g}.
\end{equation}
Hence, $\frac{1}{T}\widehat{\textbf{G}}^{\dag \prime}\textbf{G} \cdot \frac{1}{MN}\bm{\Gamma}' \bm{\Gamma}\cdot \frac{1}{T} \textbf{G}' \widehat{\textbf{G}}^{\dag}$ is invertible with probability approaching one (w.p.a.1), which implies that $\frac{1}{T}\textbf{G}' \widehat{\textbf{G}}^{\dag} $ is invertible w.p.a.1.

Right multiplying $ (\frac{1}{T}\textbf{G}' \widehat{\textbf{G}}^{\dag} )^{-1} (\frac{1}{MN}\bm{\Gamma}' \bm{\Gamma} )^{-1}$ on each side of (\ref{eq3.2b}) yields that

\begin{eqnarray} \label{eq3.4}
&&\widehat{\textbf{G}}^{\dag} \textbf{V}_{g}^{\dag} \left(\frac{1}{T}\textbf{G}' \widehat{\textbf{G}}^{\dag} \right)^{-1} \left(\frac{1}{MN}\bm{\Gamma}' \bm{\Gamma} \right)^{-1} - \textbf{G} \nonumber\\
&=& (\textbf{A}_2 + \cdots + \textbf{A}_{16}) \widehat{\textbf{G}}^{\dag} \left(\frac{1}{T}\textbf{G}' \widehat{\textbf{G}}^{\dag} \right)^{-1} \left(\frac{1}{MN}\bm{\Gamma}' \bm{\Gamma} \right)^{-1}.
\end{eqnarray}
By examining each term on the right hand side of \eqref{eq3.4}, we obtain that

\begin{eqnarray*}
&&\frac{1}{\sqrt{T}} \left\| \widehat{\textbf{G}}^{\dag} \textbf{V}_{g}^{\dag} \left(\frac{1}{T}\textbf{G}' \widehat{\textbf{G}}^{\dag} \right)^{-1} \left(\frac{1}{MN}\bm{\Gamma}' \bm{\Gamma} \right)^{-1} - \textbf{G} \right\|_{F} \\
&\leq& O(1) \cdot ( \| \textbf{A}_2 \|_{2} + \cdots + \| \textbf{A}_{16} \|_{2} ) \cdot \left\| \frac{1}{\sqrt{T}}\widehat{\textbf{G}}^{\dag} \cdot \left(\frac{1}{T}\textbf{G}' \widehat{\textbf{G}}^{\dag} \right)^{-1} \left( \frac{1}{MN}\bm{\Gamma}' \bm{\Gamma} \right)^{-1} \right\|_{F} \\
&\leq & O_P(1) \cdot ( \| \textbf{A}_2 \|_{2} + \cdots + \| \textbf{A}_{16} \|_{2} ) =O_P \left( \frac{\sqrt{ \ln (MN)}}{\min \{\sqrt{M}, \sqrt{N}, \sqrt{T}\}} \right).
\end{eqnarray*}
where the first inequality follows from (\ref{eq3.4}) and $\|\bm{\Omega}\|_F \le \sqrt{\text{rank}(\bm{\Omega})} \cdot \|\bm{\Omega} \|_2$, the second inequality follows from $\frac{1}{\sqrt{T}}\| \widehat{\textbf{G}}^{\dag} \|_{F}  = O_P(1)$ and $\| (\frac{1}{T}\textbf{G}' \widehat{\textbf{G}}^{\dag} )^{-1} (\frac{1}{MN}\bm{\Gamma}' \bm{\Gamma} )^{-1} \|_{F} = O_P(1)$, and the equality follows from Lemma \ref{lemma2}. 

Moreover, we have $\|\frac{1}{MN}\bm{\Gamma}' \bm{\Gamma}\|_F = O_P(1)$ by Assumption \ref{ass2}. We have shown that $\| (\textbf{V}_{g}^{\dag})^{-1} \|_F= O_P(1)$ since $\textbf{V}_{g}^{\dag}$ converges to a full rank matrix, and 

\begin{equation*}
\frac{1}{T}\| \textbf{G}' \widehat{\textbf{G}}^{\dag} \|_F \leq \frac{1}{\sqrt{T}}\| \textbf{G}\|_F \cdot \frac{1}{\sqrt{T}} \| \widehat{\textbf{G}}^{\dag} \|_F = O_P(1).
\end{equation*}
Thus, it follows that
\begin{eqnarray*}
\| \textbf{H} \|_F \le \frac{1}{MN}\| \bm{\Gamma}' \bm{\Gamma}\|_F  \cdot \frac{1}{T}\| \textbf{G}' \widehat{\textbf{G}}^{\dag} \|_F  \cdot \| (\textbf{V}_{g}^{\dag})^{-1} \|_F = O_P(1).
\end{eqnarray*}

(2). Let us now consider $\widehat{\rho}_{g,k} - \rho_{g,k}^{\triangledown}$. By construction of $\widehat{\rho}_{g,k}$ and $\rho_{g,k}^{\triangledown}$, write

\begin{eqnarray*}
\widehat{\rho}_{g,k} - \rho_{g,k}^{\triangledown}&=& \frac{1}{T}\widehat{\textbf{G}}_{k}^{\dag \prime} \widehat{\bm{\Sigma}} \widehat{\textbf{G}}_{k}^{\dag} - \frac{1}{T} \textbf{h}_k^{\prime} \textbf{G}^{\prime} \bm{\Sigma} \textbf{G}\textbf{h}_k \\
&=&  \frac{1}{T}(\widehat{\textbf{G}}_{k}^{\dag} - \textbf{G}\textbf{h}_k + \textbf{G}\textbf{h}_k )^{\prime} (\widehat{\bm{\Sigma}} - \bm{\Sigma} + \bm{\Sigma} ) ( \widehat{\textbf{G}}_{k}^{\dag} - \textbf{G}\textbf{h}_k + \textbf{G}\textbf{h}_k ) - \frac{1}{T} \textbf{h}_k^{\prime} \textbf{G}^{
\prime} \bm{\Sigma} \textbf{G}\textbf{h}_k \\
&=&  \frac{1}{T} (\widehat{\textbf{G}}_{k}^{\dag} - \textbf{G}\textbf{h}_k)^{\prime} (\widehat{\bm{\Sigma}} - \bm{\Sigma}) (\widehat{\textbf{G}}_{k}^{\dag} - \textbf{G}\textbf{h}_k) \\
&&+ \frac{2}{T}(\widehat{\textbf{G}}_{k}^{\dag} - \textbf{G}\textbf{h}_k)^{\prime} (\widehat{\bm{\Sigma}} - \bm{\Sigma}) \textbf{G}\textbf{h}_k \\
&& +  \frac{1}{T} (\widehat{\textbf{G}}_{k}^{\dag} - \textbf{G}\textbf{h}_k)^{\prime} \bm{\Sigma} (\widehat{\textbf{G}}_{k}^{\dag} - \textbf{G}\textbf{h}_k) \\
&&+  \frac{2}{T}(\widehat{\textbf{G}}_{k}^{\dag} - \textbf{G}\textbf{h}_k)^{\prime} \bm{\Sigma} \textbf{G}\textbf{h}_k \\
&& +  \frac{1}{T} \textbf{h}_k^{\prime} \textbf{G}^{\prime} (\widehat{\bm{\Sigma}} - \bm{\Sigma}) \textbf{G}\textbf{h}_k \\
&:=& J_1 + 2 J_2 + J_3 + 2 J_4 + J_5
\end{eqnarray*}

Start our analysis from $J_1$, and write

\begin{eqnarray*}
| J_1 | &=& \frac{1}{T} | (\widehat{\textbf{G}}_{k}^{\dag} - \textbf{G}\textbf{h}_k)^{\prime} (\widehat{\bm{\Sigma}} - \bm{\Sigma}) (\widehat{\textbf{G}}_{k}^{\dag} - \textbf{G}\textbf{h}_k) | \\
&\leq & \frac{1}{T} \| \widehat{\textbf{G}}_{k}^{\dag} - \textbf{G}\textbf{h}_k \|_F \cdot \| \widehat{\bm{\Sigma}} - \bm{\Sigma} \|_2 \cdot \| \widehat{\textbf{G}}_{k}^{\dag} - \textbf{G}\textbf{h}_k \|_F \\
&=& o_P(\| \widehat{\bm{\Sigma}} - \bm{\Sigma} \|_2) = o_P(|J_5|).
\end{eqnarray*}
Similarly, we have $|J_2|=o_P(|J_5|)$ and $|J_3|=o_P(|J_4|)$. It remains to consider the orders of $J_4$ and $J_5$.

For $|J_4|$, write

\begin{eqnarray*}
| J_4 |  &=& \frac{1}{T} | (\widehat{\textbf{G}}_{k}^{\dag} - \textbf{G}\textbf{h}_k)^{\prime} \bm{\Sigma} \textbf{G}\textbf{h}_k | \leq  \frac{1}{T} \| \widehat{\textbf{G}}_{k}^{\dag} - \textbf{G}\textbf{h}_k \|_F \cdot \| \bm{\Sigma} \textbf{G}\textbf{h}_k \|_F \\
&=& \frac{1}{T} \| \widehat{\textbf{G}}_{k}^{\dag} - \textbf{G}\textbf{h}_k \|_F \cdot \frac{1}{MNT}\| \textbf{G} \bm{\Gamma}' \bm{\Gamma} \textbf{G}' \textbf{G}\textbf{h}_k \|_F \\
&\leq &\frac{1}{T} \| \widehat{\textbf{G}}_{k}^{\dag} - \textbf{G}\textbf{h}_k \|_F \cdot \frac{1}{MNT} \| \textbf{G}\|_F \cdot \| \bm{\Gamma}' \bm{\Gamma} \|_F \cdot \| \textbf{G}' \textbf{G} \|_F \cdot \|\textbf{h}_k \|_F \\
&=& O_P \left( \frac{\sqrt{\ln(MN)}}{\min \{\sqrt{M}, \sqrt{N}, \sqrt{T}\}} \right),
\end{eqnarray*}
where the last equality follows from Lemma \ref{lemma_nof1}.

For $|J_5|$, write

\begin{eqnarray*}
\| \widehat{\bm{\Sigma}} - \bm{\Sigma} \|_2 &=& \left\| \frac{1}{MNT}\textbf{Y}'\textbf{Y} - \frac{1}{MNT} \textbf{G} \bm{\Gamma}' \bm{\Gamma} \textbf{G}' \right\|_2 \leq \| \textbf{A}_2 \|_2 + \cdots + \| \textbf{A}_{16} \|_2 \\
&=& O_P \bigg( \frac{\sqrt{\ln(MN)}}{\min \{\sqrt{M}, \sqrt{N}, \sqrt{T}\}} \bigg),
\end{eqnarray*}
where $\textbf{A}_2,\ldots, \textbf{A}_{16}$ have been defined in the proof of Lemma \ref{lemma_nof1}.

Thus, 

\begin{eqnarray*}
|J_5|  &= & \frac{1}{T} | \textbf{h}_k^{\prime} \textbf{G}^{\prime} (\widehat{\bm{\Sigma}} - \bm{\Sigma}) \textbf{G}\textbf{h}_k | \leq \frac{1}{T} \| \textbf{h}_k \|_F^2 \cdot \| \textbf{G} \|_F^{2} \cdot \| \widehat{\bm{\Sigma}} - \bm{\Sigma} \|_2 \\
&=& O_P \bigg( \frac{\sqrt{\ln(MN)}}{\min \{\sqrt{M}, \sqrt{N}, \sqrt{T}\}} \bigg).
\end{eqnarray*}
Hence, based on the above development,

\begin{equation*}
| \widehat{\rho}_{g,k} - \rho_{g,k}^{\triangledown} |  = O_P \left( \frac{\sqrt{\ln(MN)}}{\min \{\sqrt{M}, \sqrt{N}, \sqrt{T}\}} \right),
\end{equation*}
which completes the proof of the first result.

\medskip

(3).  This proof is based on Lemma \ref{lemma_nof0}. Let us denote $\textbf{G}^{\bot}$ as a $T \times (k_{\max} - r_g)$ matrix such that $\frac{1}{T}(\textbf{G}^{\bot}, \textbf{G}\textbf{R})'(\textbf{G}^{\bot}, \textbf{G}\textbf{R}) = \diag\{\textbf{I}_{k_{\max}-r_g}, \textbf{I}_{r_g}\}$, where $\textbf{R}$ is an $r_g \times r_g$ rotation matrix. The matrices $\frac{1}{\sqrt{T}}\textbf{G}^{\bot}$, $\frac{1}{\sqrt{T}}\textbf{G}\textbf{R}$, $\bm{\Sigma}$ and $\widehat{\bm{\Sigma}} - \bm{\Sigma}$ correspond to $\textbf{Q}_1$, $\textbf{Q}_2$, $\textbf{A}$ and $\textbf{E}$ of Lemma \ref{lemma_nof0}. The counterpart of the matrix $\textbf{Q}_1^{0}$ of Lemma \ref{lemma_nof0} is given by

\begin{equation*}
\widehat{\textbf{G}}^{\bot} = \frac{1}{\sqrt{T}}(\textbf{G}^{\bot} + \textbf{G}\textbf{R}\textbf{P})(\textbf{I}_{k_{\max} - r_g} + \textbf{P}'\textbf{P} )^{-1/2},
\end{equation*}
where

\begin{equation*}
\| \textbf{P} \|_2 
\leq \frac{4 \| \widehat{\bm{\Sigma}} - \bm{\Sigma} \|_2}{\mbox{sep}(0, \frac{1}{T}\textbf{G}'\bm{\Sigma}\textbf{G})}
\leq O_P(1) \cdot \| \widehat{\bm{\Sigma}} - \bm{\Sigma} \|_2 
= O_P \bigg( \frac{\sqrt{\ln(MN)}}{\min \{\sqrt{M}, \sqrt{N}, \sqrt{T}\}} \bigg).
\end{equation*}

Since $\widehat{\textbf{G}}^{\bot}$ is an orthonormal basis for a subspace that is invariant for $\widehat{\bm{\Sigma}}$, we have $\rho_{r_g+k} = \widehat{\textbf{G}}_k^{\bot \prime} \widehat{\bm{\Sigma}} \widehat{\textbf{G}}_k^{\bot}$, where $k=1,\ldots, k_{\max}-r$, and $\widehat{\textbf{G}}_k^{\bot}$ is the $k$-th column of $\widehat{\textbf{G}}^{\bot}$. Consider $\|\widehat{\textbf{G}}^{\bot} - \frac{1}{\sqrt{T}}\textbf{G}^{\bot} \|_2$, and write

\begin{eqnarray}\label{bp4}
\left\|\widehat{\textbf{G}}^{\bot} - \frac{1}{\sqrt{T}}\textbf{G}^{\bot}\right \|_2 
&=& \frac{1}{\sqrt{T}} \| [\textbf{G}^{\bot} + \textbf{G} \textbf{R} \textbf{P} - \textbf{G}^{\bot} (\textbf{I}_{k_{\max}-r_g} + \textbf{P}'\textbf{P})^{1/2}] (\textbf{I}_{k_{\max}-r_g} + \textbf{P}'\textbf{P})^{-1/2} \|_2 \nonumber \\
&\leq &\frac{1}{\sqrt{T}} \| \textbf{G}^{\bot} (\textbf{I}_{k_{\max}-r_g} - (\textbf{I}_{k_{\max}-r_g} + \textbf{P}'\textbf{P})^{1/2}) (\textbf{I}_{k_{\max}-r_g} + \textbf{P}'\textbf{P})^{-1/2} \|_2 \nonumber \\
    &&+ \frac{1}{\sqrt{T}} \| \textbf{G} \textbf{R} \textbf{P} (\textbf{I}_{k_{\max}-r_g} + \textbf{P}'\textbf{P})^{-1/2} \|_2\nonumber  \\
&\leq & \| (\textbf{I}_{k_{\max}-r_g} - (\textbf{I}_{k_{\max}-r_g} + \textbf{P}'\textbf{P})^{1/2}) (\textbf{I}_{k_{\max}-r_g} + \textbf{P}'\textbf{P})^{-1/2} \|_2 \nonumber \\
    && + \| \textbf{P} (\textbf{I}_{k_{\max}-r_g} + \textbf{P}'\textbf{P})^{-1/2} \|_2 \nonumber \\
&\leq & \| \textbf{I}_{k_{\max}-r_g} - (\textbf{I}_{k_{\max}-r_g} + \textbf{P}'\textbf{P})^{1/2} \|_2 + \|\textbf{P}\|_2\nonumber   \\
&\leq& 2\|\textbf{P}\|_2 = O_P \left( \frac{\sqrt{\ln(MN)}}{\min \{\sqrt{M}, \sqrt{N}, \sqrt{T}\}} \right).
\end{eqnarray}
where the second and third inequalities follow from Exercise 1 on page 231 and \cite{Magnus2019-ef}. Thus, for $k=1,\ldots,k_{\max}-r_g$,

\begin{eqnarray*}
| \widehat{\rho}_{r+k} | &=& |\widehat{\textbf{G}}_k^{\bot \prime} \widehat{\bm{\Sigma}} \widehat{\textbf{G}}_k^{\bot} | \\
&=& | (\widehat{\textbf{G}}_k^{\bot} - T^{-1/2}\textbf{G}_k^{\bot} + T^{-1/2}\textbf{G}_k^{\bot})' (\widehat{\bm{\Sigma}} - \bm{\Sigma} + \bm{\Sigma}) (\widehat{\textbf{G}}_k^{\bot} - T^{-1/2}\textbf{G}_k^{\bot} + T^{-1/2}\textbf{G}_k^{\bot} ) | \\
&\leq &\| \widehat{\textbf{G}}_k^{\bot} - T^{-1/2}\textbf{G}_k^{\bot} \|_F^2 \cdot \| \widehat{\bm{\Sigma}} - \bm{\Sigma} \|_2 \\
&&+ 2 \| \widehat{\textbf{G}}_k^{\bot} - T^{-1/2}\textbf{G}_k^{\bot} \|_F^2 \cdot \| \widehat{\bm{\Sigma}} - \bm{\Sigma} \|_2 \cdot \frac{1}{\sqrt{T}}\| \textbf{G}_k^{\bot} \|_F \\
&&+ \| \widehat{\textbf{G}}_k^{\bot} - T^{-1/2}\textbf{G}_k^{\bot} \|_F^2 \cdot \|\bm{\Sigma}\|_2 \\
&=& O_P \left( \frac{\ln(MN)}{\min \{ M, N, T \}} \right),
\end{eqnarray*}
where $\textbf{G}_{k}^{\bot}$ is the $k$-th column of $\textbf{G}^{\bot}$, and the last equality follows from \eqref{bp4} and the proof of the first result of this lemma. The proof is complete. \hspace*{\fill}{$\blacksquare$}

\bigskip

\noindent \textbf{Proof of Theorem \ref{theorem1}}:

(1). First, consider the case where $r_g = 0$. By Lemma \ref{lemma_nof1}.3, we have $\widehat{\rho}_{g,k} = O_P \left( \frac{\ln(MN)}{\min \{M,N,T\} }\right)$ for $k= 1,\ldots,k_{\max}$, which are less than $\omega_{MNT}$ w.p.a.1 for $k=1,\ldots, k_{\max}$. In connection with the fact that $\widehat{\rho}_{g,0}=1$ which is larger than $\omega_{MNT}$ and $\widehat{\rho}_{g,1} / \widehat{\rho}_{g,0} = O_P \left( \frac{\ln(MN)}{\min \{M,N,T\} }\right)$ which is less than 1 w.p.a.1, it follows that when $r_g=0$, we have $P(\widehat{r}_g = 0) \to 1$.

Next, we consider the case where $0<r_g<k_{\max}$. Note that for $k=1,\ldots,r_g$,

\begin{eqnarray*}
\rho_{g,k}^{\triangledown} &=& \frac{1}{T} \textbf{h}_k^{\prime} \textbf{G}^{\prime} \bm{\Sigma} \textbf{G}\textbf{h}_k
=  \frac{1}{T} \textbf{h}_k^{\prime} \textbf{G}^{\prime} \textbf{G} \cdot  \frac{1}{MN} \bm{\Gamma}' \bm{\Gamma} \cdot \frac{1}{T}  \textbf{G}' \textbf{G}\textbf{h}_k \\
&=  &\frac{1}{T} \widehat{\textbf{G}}_k^{\dag \prime} \textbf{G} \cdot \frac{1}{MN} \bm{\Gamma}' \bm{\Gamma} \cdot \frac{1}{T} \textbf{G}' \widehat{\textbf{G}}_k^{\dag} \cdot ( 1 + o_P(1) ) \asymp 1,
\end{eqnarray*}
where $\textbf{h}_k$, $\bm{\Sigma}$ and $\widehat{\textbf{G}}_k^{\dag} $ have been defined in the proof of Lemma \ref{lemma_nof0}, the third equality follows from Lemma \ref{lemma_nof1}.1, and the last step follows from (\ref{equivalence}).

Note that $\widehat{\rho}_{g,0}=1$ which is larger than $\omega_{MNT}$. Furthermore, by Lemma \ref{lemma_nof1}.2, $ \widehat{\rho}_{g,k} \asymp \rho_{g,k}^{\triangledown}$ for $k=1,\ldots,r_g$, which are larger than $\omega_{MNT}$ w.p.a.1. Moreover, for $k=0,\ldots,r_g-1$ we can conclude that

\begin{eqnarray*}
\frac{ \widehat{\rho}_{g, k+1}}{ \widehat{\rho}_{g, k}} \asymp 1
\end{eqnarray*}
since $\widehat{\rho}_{g,0}=1$, $ \widehat{\rho}_{g,k} \asymp \rho_{g,k}^{\triangledown}$ and $\rho_{g,k}^{\triangledown} \asymp 1$ for $k=1,\ldots,r_g$.

For $k=r_g+1,\ldots,k_{\max}$, by Lemma \ref{lemma_nof1}.3, $\widehat{\rho}_{g,k} = O_P \left( \frac{\ln(MN)}{\min \{M,N,T\} } \right)$, which are less than $\omega_{MNT}$ w.p.a.1. Thus, in connection with the fact that $ \widehat{\rho}_{g, r_g} \asymp 1$, we obtain that

\begin{eqnarray*}
\frac{ \widehat{\rho}_{g,r_g+1}}{ \widehat{\rho}_{g,r_g}} = O_P \left( \frac{\ln(MN)}{\min \{M,N,T\}} \right),
\end{eqnarray*}
which is less than 1 w.p.a.1.

Based on the above and (\ref{Eq2.8}), we are readily to conclude that $P(\widehat{r}_g = r_g) \to 1$. The first result then follows.

\medskip

(2). Having establishing the first result, the second step follows from almost an identical procedure of the proof for Lemma \ref{lemma_nof0}. Thus, omitted. The proof is complete. \hspace*{\fill}{$\blacksquare$}

\bigskip

\noindent \textbf{Proof of Lemma \ref{lemma_nof1_local}}: 

Before proceeding further, we denote a few notations for simplicity. Let $\bm{\Gamma}_{I,j}$ be denoted in the same way as $\widehat{\bm{\Gamma}}_{I,j}$, $\textbf{U}_{\bullet j \bullet} = \{u_{ijt} \}_{M\times T}$.

By expanding \eqref{Eq2.9}, we obtain that

\begin{eqnarray} \label{eq10}
\widehat{\textbf{F}}_{I,j} \textbf{V}_{I,j} 
&=& \frac{1}{MT} \left( \bm{\Gamma}_{I,j} \textbf{G}' - \widehat{\bm{\Gamma}}_{I,j} \widehat{\textbf{G}}' + \diag\{\bm{\Lambda}_{E,\bullet j}' \}' \textbf{F}_{E}' + \bm{\Lambda}_{I,\bullet j} \textbf{F}_{I,j}' + \textbf{U}_{\bullet j \bullet} \right)' \nonumber \\
&&\quad \cdot \left( \bm{\Gamma}_{I,j} \textbf{G}' - \widehat{\bm{\Gamma}}_{I,j} \widehat{\textbf{G}}' +\diag\{\bm{\Lambda}_{E,\bullet j}' \}' \textbf{F}_{E}' + \bm{\Lambda}_{I,\bullet j} \textbf{F}_{I,j}' + \textbf{U}_{\bullet j \bullet} \right) \widehat{\textbf{F}}_{I, j} \nonumber \\
&:=& (\textbf{B}_1 + \cdots + \textbf{B}_{16})  \widehat{\textbf{F}}_{I, j} 
\end{eqnarray}
where 

\begin{eqnarray*}
&&\textbf{B}_1 = \frac{1}{MT}\textbf{F}_{I,j}  \bm{\Lambda}_{I,\bullet j}'  \bm{\Lambda}_{I,\bullet j} \textbf{F}_{I,j}' , \quad
\textbf{B}_2 = \frac{1}{MT} (\bm{\Gamma}_{I,j} \textbf{G}' - \widehat{\bm{\Gamma}}_{I,j} \widehat{\textbf{G}}' )' \diag\{\bm{\Lambda}_{E,\bullet j}' \}'\textbf{F}_{E}', \nonumber \\
&&\textbf{B}_3 = \frac{1}{MT} (\bm{\Gamma}_{I,j} \textbf{G}' - \widehat{\bm{\Gamma}}_{I,j} \widehat{\textbf{G}}' )' \bm{\Lambda}_{I,\bullet j} \textbf{F}_{I,j}', \quad \textbf{B}_4 = \frac{1}{MT} (\bm{\Gamma}_{I,j} \textbf{G}' - \widehat{\bm{\Gamma}}_{I,j} \widehat{\textbf{G}}' )' \textbf{U}_{\bullet j\bullet}, \nonumber\\
&&\textbf{B}_5 =\textbf{B}_2', \quad \textbf{B}_6 = \frac{1}{MT} \textbf{F}_{E} \diag\{\bm{\Lambda}_{E,\bullet j}' \} \diag\{\bm{\Lambda}_{E,\bullet j}' \}'\textbf{F}_{E}', \quad \textbf{B}_7 = \frac{1}{MT}\textbf{F}_{E} \diag\{\bm{\Lambda}_{E,\bullet j}' \} \bm{\Lambda}_{I,\bullet j} \textbf{F}_{I,j}'\\
&&\textbf{B}_8 = \frac{1}{MT}\textbf{F}_{E} \diag\{\bm{\Lambda}_{E,\bullet j}' \} \textbf{U}_{\bullet j \bullet},\quad  \textbf{B}_9 =\textbf{B}_3', \quad \textbf{B}_{10} = \textbf{B}_7',\nonumber \\
&&\textbf{B}_{11} =  \frac{1}{MT}  (\bm{\Gamma}_{I,j} \textbf{G}' - \widehat{\bm{\Gamma}}_{I,j} \widehat{\textbf{G}}' )' (\bm{\Gamma}_{I,j} \textbf{G}' - \widehat{\bm{\Gamma}}_{I,j} \widehat{\textbf{G}}' ),\quad \textbf{B}_{12} = \frac{1}{MT} \textbf{F}_{I,j} \bm{\Lambda}_{I,\bullet j}'  \textbf{U}_{\bullet j \bullet},\nonumber \\
&& \textbf{B}_{13} = \textbf{B}_4', \quad
\textbf{B}_{14} =\textbf{B}_8',\quad \textbf{B}_{15} =\textbf{B}_{12}', \quad\textbf{B}_{16} = \frac{1}{MT} \textbf{U}_{\bullet j \bullet}^{\prime} \textbf{U}_{\bullet j \bullet}.
\end{eqnarray*}
We can then immediately obtain that
\begin{equation} \label{eq1}
\widehat{\textbf{F}}_{I,j} \textbf{V}_{I,j}  -  \textbf{F}_{I,j} \cdot \frac{1}{M}\bm{\Lambda}_{I,\bullet j}'  \bm{\Lambda}_{I,\bullet j} \cdot \frac{1}{T}\textbf{F}_{I,j}' \widehat{\textbf{F}}_{I, j}   = (\textbf{B}_2 + \cdots + \textbf{B}_{16}) \widehat{\textbf{F}}_{I, j}.
\end{equation}

By Lemma \ref{lemma8}, we obtain that

\begin{equation*}
\| \textbf{B}_2 + \cdots + \textbf{B}_{16} \|_{2} \leq \| \textbf{B}_2 \|_{2} + \cdots + \| \textbf{B}_{16} \|_{2} = O_P \left( \frac{\sqrt{\ln(MN)}}{\min \{\sqrt{M}, \sqrt{N}, \sqrt{T}\}} + T^{a_{I,j}}\right) .
\end{equation*}
Then left multiplying \eqref{eq1} by $\frac{1}{T}\textbf{F}_{I, j}'$ immediately yields that

\begin{eqnarray*}
&&\left\| \frac{1}{T}\textbf{F}_{I, j}' \widehat{\textbf{F}}_{I,j}\textbf{V}_{I,j} - \frac{1}{T}\textbf{F}_{I, j}' \textbf{F}_{I, j} \cdot \frac{1}{M}\bm{\Lambda}_{I,\bullet j}'  \bm{\Lambda}_{I,\bullet j} \cdot \frac{1}{T}\textbf{F}_{I, j}' \widehat{\textbf{F}}_{I,j}\right\|_F \\
&\leq& O(1) \frac{1}{\sqrt{T}}\|\textbf{F}_{I, j}'\|_F\cdot \|  \textbf{B}_2 + \cdots + \textbf{B}_{16} \|_2\cdot \frac{1}{\sqrt{T}}\| \widehat{\textbf{F}}_{I, j}\|_F =  O_P \left( \frac{\sqrt{\ln(MN)}}{\min \{\sqrt{M}, \sqrt{N}, \sqrt{T}\}} +T^{a_{I,j}}\right),
\end{eqnarray*}
where the first inequality follows from (\ref{eq1}) and $\|\bm{\Omega}\|_F \le \sqrt{\text{rank}(\bm{\Omega})} \cdot \|\bm{\Omega} \|_2$.

Furthermore, $ \frac{1}{T}\textbf{F}_{I, j}' \textbf{F}_{I, j} \cdot \frac{1}{M}\bm{\Lambda}_{I,\bullet j}'  \bm{\Lambda}_{I,\bullet j}  = \bm{\Sigma}_{\textbf{F}_{I,j}} \bm{\Sigma}_{\bm{\Lambda}_{I,\bullet j}} + o_P(1)$ by Assumptions \ref{ass4}. It follows that

\begin{equation*}  
 \bm{\Sigma}_{\textbf{F}_{I,j}} \bm{\Sigma}_{\bm{\Lambda}_{I,\bullet j}}  \cdot  \frac{1}{T}\textbf{F}_{I, j}' \widehat{\textbf{F}}_{I,j}=  \frac{1}{T}\textbf{F}_{I, j}' \widehat{\textbf{F}}_{I,j} \textbf{V}_{I,j} + o_P(1).
\end{equation*}
Then the first result follows in a manner similar to the first result of Lemma \ref{lemma_nof1}.

\medskip

(2)-(3). The second and third results follow the proofs similar to the second and third results of Lemma \ref{lemma_nof1}, so the details are omitted. The proof is now complete. \hspace*{\fill}{$\blacksquare$}

\bigskip

\noindent \textbf{Proof of Lemma \ref{lemma_nof2_local}:}

The proofs for the results of Lemma \ref{lemma_nof2_local} are identical to those for Lemma \ref{lemma_nof1_local}. Thus, omitted. \hspace*{\fill}{$\blacksquare$}

\noindent \textbf{Proof of Theorem \ref{theorem2}:}

(1). Having established Lemma \ref{lemma_nof1_local}, the proof for the first result of this theorem is identical to the proof of Theorem \ref{theorem1} 

(2). The proof for the second result of this theorem can be achieved in a manner similar to the first result, so omitted as well. \hspace*{\fill}{$\blacksquare$}

\subsection{On Asymptotic Distribution}

\noindent \textbf{Proof of Lemma \ref{lemma2_improved}}:

(1). The proof is largely the same as Lemma \ref{lemma_nof1}, but we improve the rates of convergence on the following terms. We first improve the convergence rates of the terms $ \frac{1}{\sqrt{T}}\| \textbf{A}_5 \widehat{\textbf{G}} \|_2 $, $ \frac{1}{\sqrt{T}}\| \textbf{A}_6 \widehat{\textbf{G}} \|_2 $, $ \frac{1}{\sqrt{T}}\| \textbf{A}_8 \widehat{\textbf{G}} \|_2 $ and $ \frac{1}{\sqrt{T}}\| \textbf{A}_{11} \widehat{\textbf{G}} \|_2 $ of Lemma \ref{lemma_nof1} using both Assumption \ref{ass4}.1.(a) and Assumption \ref{ass5}.1.

\begin{eqnarray*}
&& \frac{1}{\sqrt{T}} \| \textbf{A}_5 \widehat{\textbf{G}} \|_2 
\leq \frac{1}{MNT\sqrt{T}} \| \textbf{G} \|_F \cdot \| \bm{\Gamma}' \bm{\Lambda}_E \|_F \cdot \left( \| \textbf{F}_E' ( \widehat{\textbf{G}} - \textbf{G}\textbf{H} ) \|_2 + \| \textbf{F}_E'\textbf{G}\textbf{H} \|_F \right) \\
&\leq& O_P(1) \frac{1}{MNT\sqrt{T}} \cdot \sqrt{T} \cdot \sqrt{MN} \cdot \left( (\sqrt{M} \vee \sqrt{T}) \cdot \| \widehat{\textbf{G}} - \textbf{G}\textbf{H} \|_F + T\sqrt{M} \cdot T^{\max_{i}a_{E,i}} \right) \\
&=& o_P(1) \cdot \frac{1}{\sqrt{T}} \| \widehat{\textbf{G}} - \textbf{G}\textbf{H} \|_F + O_P\left( \frac{T^{\max_{i}a_{E,i}}}{\sqrt{N}} \right),
\end{eqnarray*}
where the second inequality follows from Assumption \ref{ass1}, Assumption \ref{ass4}.1.(a), Assumption \ref{ass5}.1. and $\| \textbf{H}\|_F=O_P(1)$.

\begin{eqnarray*}
&& \frac{1}{\sqrt{T}} \| \textbf{A}_6 \widehat{\textbf{G}} \|_2 
\leq \frac{1}{MNT\sqrt{T}} \| \textbf{G} \|_F \cdot \| \bm{\Gamma}' \bm{\Lambda}_I \|_F \cdot \left( \| \textbf{F}_I' ( \widehat{\textbf{G}} - \textbf{G}\textbf{H}) \|_2 + \| \textbf{F}_I' \textbf{G}\textbf{H} \|_F \right) \\
&\leq& O_P(1) \frac{1}{MNT\sqrt{T}} \cdot \sqrt{T} \cdot \sqrt{MN} \cdot \left( (\sqrt{N} \vee \sqrt{T}) \cdot \| \widehat{\textbf{G}} - \textbf{G}\textbf{H} \|_F + T\sqrt{N} \cdot T^{\max_{j}a_{I,j}} \right) \\
&=& o_P(1) \cdot \frac{1}{\sqrt{T}} \| \widehat{\textbf{G}} - \textbf{G}\textbf{H} \|_F + O_P\left( \frac{T^{\max_{j}a_{I,j}}}{\sqrt{M}} \right),
\end{eqnarray*}
where the second inequality follows from Assumption \ref{ass1}, Assumption \ref{ass4}.1.(a), Assumption \ref{ass5}.1 and $\| \textbf{H}\|_F=O_P(1)$.

\begin{eqnarray*}
&& \frac{1}{\sqrt{T}} \| \textbf{A}_8 \widehat{\textbf{G}} \|_2 
\leq \frac{1}{MNT\sqrt{T}} \| \textbf{F}_E \|_2 \cdot \| \bm{\Lambda}_E' \bm{\Gamma} \|_F \cdot \| \textbf{G}' \widehat{\textbf{G}} \|_F  \\
&\leq& O_P(1) \frac{1}{MNT\sqrt{T}} (\sqrt{M} \vee \sqrt{T}) \cdot \sqrt{MN} \cdot T \\
&=& O_P(1) \left( \frac{1}{\sqrt{NT}} + \frac{1}{\sqrt{MN}} \right),
\end{eqnarray*}
where the second inequality follows from Assumption \ref{ass1} and Assumption \ref{ass5}.1.

\begin{eqnarray*}
&& \frac{1}{\sqrt{T}} \| \textbf{A}_{11} \widehat{\textbf{G}} \|_2 
\leq \frac{1}{MNT\sqrt{T}} \| \textbf{F}_I \|_2 \cdot \| \bm{\Lambda}_I' \bm{\Gamma} \|_F \cdot \| \textbf{G}' \widehat{\textbf{G}} \|_F  \\
&\leq& O_P(1) \frac{1}{MNT\sqrt{T}} (\sqrt{N} \vee \sqrt{T}) \cdot \sqrt{MN} \cdot T \\
&=& O_P(1) \left( \frac{1}{\sqrt{MT}} + \frac{1}{\sqrt{MN}} \right),
\end{eqnarray*}
where the second inequality follows from Assumption \ref{ass1} and Assumption \ref{ass5}.1.

Next, we improve the convergence rates of the terms $ \frac{1}{\sqrt{T}}\| \textbf{A}_2 \widehat{\textbf{G}} \|_2 $, $ \frac{1}{\sqrt{T}}\| \textbf{A}_3 \widehat{\textbf{G}} \|_2 $, $ \frac{1}{\sqrt{T}}\| \textbf{A}_9 \widehat{\textbf{G}} \|_2 $, $ \frac{1}{\sqrt{T}}\| \textbf{A}_{12} \widehat{\textbf{G}} \|_2 $, $ \frac{1}{\sqrt{T}}\| \textbf{A}_{15} \widehat{\textbf{G}} \|_2 $ and $ \frac{1}{\sqrt{T}}\| \textbf{A}_{16} \widehat{\textbf{G}} \|_2 $ of Lemma \ref{lemma_nof1} by making use of Assumption \ref{ass4}.1.(a).

\begin{eqnarray*}
&& \frac{1}{\sqrt{T}} \| \textbf{A}_2 \widehat{\textbf{G}} \|_2 \leq \frac{1}{MNT \sqrt{T}} \| \textbf{F}_E \|_2 \cdot \| \bm{\Lambda}_E \|_2^2 \cdot \left( \| \textbf{F}_E' ( \widehat{\textbf{G}} - \textbf{G}\textbf{H} ) \|_2 + \| \textbf{F}_E'\textbf{G}\textbf{H} \|_F \right) \\
&\leq&  \frac{1}{MNT\sqrt{T}} O_P(1) (\sqrt{M} \vee \sqrt{T}) \cdot N \ln(MN) \\
&&\cdot \left( (\sqrt{M} \vee \sqrt{T}) \cdot \| \widehat{\textbf{G}} - \textbf{G}\textbf{H} \|_F + T\sqrt{M} \cdot T^{\max_{i}a_{E,i}} \right) \\
&=& o_P(1) \cdot \frac{1}{\sqrt{T}} \| \widehat{\textbf{G}} - \textbf{G}\textbf{H} \|_F + \ln(MN) \cdot O_P \left( \frac{T^{\max_{i}a_{E,i}}}{\sqrt{M} \wedge \sqrt{T}} \right),
\end{eqnarray*}
where the second inequality follows from Assumption \ref{ass1}, Assumption \ref{ass4}.1.(a) and the facts that $\|\textbf{H}\|_F=O_P(1)$ and $\|\bm{\Lambda}_E\|_2^2 = O_P(N\ln(MN))$.

\begin{eqnarray*}
&& \frac{1}{\sqrt{T}} \| \textbf{A}_3 \widehat{\textbf{G}} \|_2 \leq \frac{1}{MNT \sqrt{T}} \| \textbf{F}_I \|_2 \cdot \| \bm{\Lambda}_I \|_2^2 \cdot \left( \| \textbf{F}_I' ( \widehat{\textbf{G}} - \textbf{G}\textbf{H}) \|_2 + \| \textbf{F}_I' \textbf{G}\textbf{H} \|_F \right) \\
&\leq&  \frac{1}{MNT\sqrt{T}} O_P(1) (\sqrt{N} \vee \sqrt{T}) \cdot M \ln(MN) \\
&&\cdot \left( (\sqrt{N} \vee \sqrt{T}) \cdot \| \widehat{\textbf{G}} - \textbf{G}\textbf{H} \|_F + T\sqrt{N} \cdot T^{\max_{j}a_{I,j}} \right) \\
&=& o_P(1) \cdot \frac{1}{\sqrt{T}} \| \widehat{\textbf{G}} - \textbf{G}\textbf{H} \|_F + \ln(MN) \cdot O_P \left( \frac{T^{\max_{j}a_{I,j}}}{\sqrt{N} \wedge \sqrt{T}} \right),
\end{eqnarray*}
where the second inequality follows from Assumption \ref{ass1}, Assumption \ref{ass4}.1.(a) and the facts that $\|\textbf{H}\|_F = O_P(1)$ and $\|\bm{\Lambda}_I\|_2^2 = O_P(M\ln(MN))$.

\begin{eqnarray*}
&& \frac{1}{\sqrt{T}} \| \textbf{A}_9 \widehat{\textbf{G}} \|_2 \leq \frac{1}{MNT \sqrt{T}} \| \textbf{F}_{E} \bm{\Lambda}_E' \bm{\Lambda}_I \|_2 \cdot \left( \| \textbf{F}_I' ( \widehat{\textbf{G}} - \textbf{G}\textbf{H}) \|_2 + \| \textbf{F}_I' \textbf{G}\textbf{H} \|_F \right) \\
&\leq&  \frac{1}{MNT\sqrt{T}} O_P(1) (\sqrt{M} \vee \sqrt{T}) \cdot \sqrt{M \ln(MN)} \cdot \sqrt{N \ln(MN)} \\
&&\cdot \left( (\sqrt{N} \vee \sqrt{T}) \cdot \| \widehat{\textbf{G}} - \textbf{G}\textbf{H} \|_F + T\sqrt{N} \cdot T^{\max_{j}a_{I,j}} \right) \\
&=& o_P(1) \cdot \frac{1}{\sqrt{T}} \| \widehat{\textbf{G}} - \textbf{G}\textbf{H} \|_F + \ln(MN) \cdot O_P \left( \frac{T^{\max_{j}a_{I,j}}}{\sqrt{M} \wedge \sqrt{T}} \right),
\end{eqnarray*}
where the second inequality follows from Assumption \ref{ass1}, Assumption \ref{ass4}.1.(a) and the facts that $\| \textbf{H}\|_F=O_P(1)$, $\|\bm{\Lambda}_E\|_2 = O_P(\sqrt{N\ln(MN)})$ and $\|\bm{\Lambda}_I\|_2 = O_P(\sqrt{M\ln(MN)})$.

\begin{eqnarray*}
&& \frac{1}{\sqrt{T}} \| \textbf{A}_{12} \widehat{\textbf{G}} \|_2 \leq \frac{1}{MNT \sqrt{T}} \| \textbf{F}_{I} \bm{\Lambda}_I' \bm{\Lambda}_E \|_2 \cdot \left( \| \textbf{F}_E' ( \widehat{\textbf{G}} - \textbf{G}\textbf{H} ) \|_2 + \| \textbf{F}_E'\textbf{G}\textbf{H} \|_F \right) \\
&\leq&  \frac{1}{MNT\sqrt{T}} O_P(1) (\sqrt{N} \vee \sqrt{T}) \cdot \sqrt{N \ln(MN)} \cdot \sqrt{M \ln(MN)} \\
&&\cdot \left( (\sqrt{M} \vee \sqrt{T}) \cdot \| \widehat{\textbf{G}} - \textbf{G}\textbf{H} \|_F + T\sqrt{M} \cdot T^{\max_{i}a_{E,i}} \right) \\
&=& o_P(1) \cdot \frac{1}{\sqrt{T}} \| \widehat{\textbf{G}} - \textbf{G}\textbf{H} \|_F + \ln(MN) \cdot O_P \left( \frac{T^{\max_{i}a_{E,i}}}{\sqrt{N} \wedge \sqrt{T}} \right),
\end{eqnarray*}
where the second inequality follows from Assumption \ref{ass1}, Assumption \ref{ass4}.1.(a) and the facts that $\|\bm{\Lambda}_E\|_2 = O_P(\sqrt{N\ln(MN)})$, $\|\bm{\Lambda}_I\|_2 = O_P(\sqrt{M\ln(MN)})$ and $\| \textbf{H}\|_F=O_P(1)$.

\begin{eqnarray*}
&& \frac{1}{\sqrt{T}} \| \textbf{A}_{15} \widehat{\textbf{G}} \|_2
\leq \frac{1}{MNT\sqrt{T}} \| \textbf{U}' \bm{\Lambda}_E \|_2 \cdot \left( \| \textbf{F}_E' ( \widehat{\textbf{G}} - \textbf{G}\textbf{H} ) \|_2 + \| \textbf{F}_E'\textbf{G}\textbf{H} \|_F \right) \\
&\leq& \frac{1}{MNT\sqrt{T}} O_P(1) (\sqrt{MN} \vee \sqrt{T}) \cdot \sqrt{N\ln(MN)}  \\
&&\cdot \left( (\sqrt{M} \vee \sqrt{T}) \cdot \| \widehat{\textbf{G}} - \textbf{G}\textbf{H} \|_F + T\sqrt{M} \cdot T^{\max_{i}a_{E,i}} \right) \\
&=& o_P(1) \cdot \frac{1}{\sqrt{T}} \| \widehat{\textbf{G}} - \textbf{G}\textbf{H} \|_F + \sqrt{\ln(MN)} \cdot O_P \left( \frac{T^{\max_{i}a_{E,i}}}{\sqrt{MN} \wedge \sqrt{T}} \right),
\end{eqnarray*}
where the second inequality follows from Assumption \ref{ass1}, Assumption \ref{ass4}.1.(a) and the facts that $\|\textbf{U}\|_2 = O_P(\sqrt{MN} \vee \sqrt{T})$, $\|\bm{\Lambda}_E\|_2 = O_P(\sqrt{N\ln(MN)})$ and $\| \textbf{H}\|_F=O_P(1)$.

\begin{eqnarray*}
&& \frac{1}{\sqrt{T}} \| \textbf{A}_{16} \widehat{\textbf{G}} \|_2
\leq \frac{1}{MNT\sqrt{T}} \| \textbf{U}' \bm{\Lambda}_I \|_2 \cdot \left( \| \textbf{F}_I' ( \widehat{\textbf{G}} - \textbf{G}\textbf{H}) \|_2 + \| \textbf{F}_I' \textbf{G}\textbf{H} \|_F \right) \\
&\leq& \frac{1}{MNT\sqrt{T}} O_P(1) (\sqrt{MN} \vee \sqrt{T}) \cdot \sqrt{M\ln(MN)} \\
&&\cdot \left( (\sqrt{N} \vee \sqrt{T}) \cdot \| \widehat{\textbf{G}} - \textbf{G}\textbf{H} \|_F + T\sqrt{N} \cdot T^{\max_{j}a_{I,j}} \right) \\
&=& o_P(1) \cdot \frac{1}{\sqrt{T}} \| \widehat{\textbf{G}} - \textbf{G}\textbf{H} \|_F + \sqrt{\ln(MN)} \cdot O_P \left( \frac{T^{\max_{j}a_{I,j}}}{\sqrt{MN} \wedge \sqrt{T}} \right),
\end{eqnarray*}
where the second inequality follows from Assumption \ref{ass4}.1.(a) and the facts that $\|\bm{\Lambda}_I\|_2 = O_P(\sqrt{M\ln(MN)})$, $\|\textbf{U}\|_2 = O_P(\sqrt{MN} \vee \sqrt{T})$ and $\| \textbf{H}\|_F=O_P(1)$.

Then we improve the convergence rates of the terms $ \frac{1}{\sqrt{T}}\| \textbf{A}_4 \widehat{\textbf{G}} \|_2 $, $ \frac{1}{\sqrt{T}}\| \textbf{A}_7 \widehat{\textbf{G}} \|_2 $, $ \frac{1}{\sqrt{T}}\| \textbf{A}_{10} \widehat{\textbf{G}} \|_2 $ and $ \frac{1}{\sqrt{T}}\| \textbf{A}_{13} \widehat{\textbf{G}} \|_2 $ of Lemma \ref{lemma_nof1} by bounding $ \| \textbf{A}_{k} \widehat{\textbf{G}} \|_2 $ directly for $k=4,7,10$ and $13$.

\begin{eqnarray*}
&& \frac{1}{\sqrt{T}} \| \textbf{A}_4 \widehat{\textbf{G}} \|_2 
\le  \frac{1}{MNT \sqrt{T}} \|\textbf{U}\|_2 \cdot \left( \|\textbf{U} \|_2 \cdot \| \widehat{\textbf{G}} - \textbf{G}\textbf{H} \|_F + \|\textbf{U}\textbf{G}\textbf{H} \|_F \right) \\
&\leq& O_P(1) \frac{1}{MNT \sqrt{T}} (\sqrt{MN} \vee \sqrt{T}) \left(  (\sqrt{MN} \vee \sqrt{T}) \cdot \| \widehat{\textbf{G}} - \textbf{G}\textbf{H} \|_F + \sqrt{MNT} \right) \\
&=& o_P(1) \cdot \frac{1}{\sqrt{T}} \| \widehat{\textbf{G}} - \textbf{G}\textbf{H} \|_F + O_P\left( \frac{1}{T} + \frac{1}{\sqrt{MNT}} \right),
\end{eqnarray*}
where the second inequality follows from $\|\textbf{U}\|_2 =O_P(\sqrt{MN}\vee \sqrt{T})$ and the fact that

\begin{eqnarray*}
&&\frac{1}{M^2 N^2 T^2} E \| \textbf{G}'\textbf{U}' \|_{F}^2= \frac{1}{M^2N^2T^2} \sum_{i=1}^M \sum_{j=1}^N E\left\| \sum_{t=1}^T \bm{g}_{t} u_{ijt}\right\|_F^2 \\
&=& \frac{1}{M^2 N^2 T^2}   \sum_{i=1}^{M} \sum_{j=1}^{N} \sum_{t,s=1}^{T} E[u_{ijt}u_{ijs}  \bm{g}_{t}' \bm{g}_{s}  ]   \\
&\le & O(1)  \frac{1}{MNT^2}  \sum_{t,s=1}^{T} |E[u_{ijt}u_{ijs}]|=  O\left(\frac{1}{MNT} \right),
\end{eqnarray*}
in which the last equality follows from the mixing condition of Assumption \ref{ass3}.1.

\begin{eqnarray*}
&& \frac{1}{\sqrt{T}} \| \textbf{A}_7 \widehat{\textbf{G}} \|_2
\leq \frac{1}{MNT\sqrt{T}} \| \textbf{G}\|_F \| \bm{\Gamma}'\textbf{U}  \widehat{\textbf{G}} \|_F \\
&\leq& \frac{1}{MNT\sqrt{T}} \|\textbf{G}\|_F \cdot \left( \| \bm{\Gamma}'\textbf{U} \|_F \cdot \| \widehat{\textbf{G}} - \textbf{G}\textbf{H} \|_F + \|\bm{\Gamma}'\textbf{U} \textbf{G}\|_F \cdot \|\textbf{H} \|_F \right) \\
&=& O_P(1) \frac{1}{MNT} \left( \sqrt{MNT} \cdot \| \widehat{\textbf{G}} - \textbf{G}\textbf{H} \|_F + \sqrt{MNT} \right) \\
&=& o_P(1) \cdot \frac{1}{\sqrt{T}} \| \widehat{\textbf{G}} - \textbf{G}\textbf{H} \|_F + O_P(1)\left( \frac{1}{\sqrt{MNT}} \right),
\end{eqnarray*}
where the first equality follows from $\| \bm{\Gamma}'\textbf{U} \|_F=O_P(\sqrt{MNT})$, $\|\textbf{H}\|_F=O_P(1)$ and the fact that
\begin{eqnarray*}
&& \frac{1}{M^2N^2T^2} E \| \textbf{G}' \textbf{U}' \bm{\Gamma} \|_F^2 
= \frac{1}{M^2N^2T^2} E \left\| \sum_{i=1}^{M} \sum_{j=1}^{N} \sum_{t=1}^{T} \bm{g}_t u_{ijt} \bm{\gamma}_{ij}' \right \|_F^2  \\
&\le &O(1) \frac{1}{M^2N^2T^2} \sum_{i,m=1}^{M} \sum_{j,n=1}^{N} \sum_{t,s=1}^{T} |E[ u_{ijt}u_{mns}] |= O \left(\frac{1}{MNT} \right),
\end{eqnarray*}
where the last equality follows from the mixing condition of Assumption \ref{ass3}.1.

\begin{eqnarray*}
&& \frac{1}{\sqrt{T}} \| \textbf{A}_{10} \widehat{\textbf{G}} \|_2
\leq \frac{1}{MNT\sqrt{T}} \| \textbf{F}_{E} \bm{\Lambda}_E' \|_2 \cdot \left( \|\textbf{U} \|_2 \cdot \| \widehat{\textbf{G}} - \textbf{G}\textbf{H} \|_F + \|\textbf{U}\textbf{G}\textbf{H} \|_F \right) \\
&\leq& \frac{1}{MNT\sqrt{T}} O_P(1) (\sqrt{M} \vee \sqrt{T}) \cdot \sqrt{N\ln(MN)} \\
&&\cdot \left( (\sqrt{MN} \vee \sqrt{T}) \cdot \| \widehat{\textbf{G}} - \textbf{G}\textbf{H} \|_F + \sqrt{MNT} \right) \\
&=& o_P(1) \cdot \frac{1}{\sqrt{T}} \| \widehat{\textbf{G}} - \textbf{G}\textbf{H} \|_F + \sqrt{\ln(MN)} \cdot O_P \left( \frac{1}{T} + \frac{1}{\sqrt{MT}} \right),
\end{eqnarray*}
where the second inequality follows from Assumption \ref{ass1} and the facts that $\| \bm{\Lambda}_E\|_2 =\sqrt{N\ln(MN)}$, $\|\textbf{U}\textbf{G}\|_F=O_P(\sqrt{MNT})$ and $\|\textbf{H}\|_F=O_P(1)$.

\begin{eqnarray*}
&& \frac{1}{\sqrt{T}} \| \textbf{A}_{13} \widehat{\textbf{G}} \|_2
\leq \frac{1}{MNT\sqrt{T}} \| \textbf{F}_{I} \bm{\Lambda}_I' \|_2 \cdot \left( \|\textbf{U} \|_2 \cdot \| \widehat{\textbf{G}} - \textbf{G}\textbf{H} \|_F + \|\textbf{U}\textbf{G}\textbf{H} \|_F \right) \\
&\leq& \frac{1}{MNT\sqrt{T}} O_P(1) (\sqrt{N} \vee \sqrt{T}) \cdot \sqrt{M\ln(MN)} \\
&&\cdot \left( (\sqrt{MN} \vee \sqrt{T}) \cdot \| \widehat{\textbf{G}} - \textbf{G}\textbf{H} \|_F + \sqrt{MNT} \right) \\
&=& o_P(1) \cdot \frac{1}{\sqrt{T}} \| \widehat{\textbf{G}} - \textbf{G}\textbf{H} \|_F + \sqrt{\ln(MN)} \cdot O_P \left( \frac{1}{T} + \frac{1}{\sqrt{NT}} \right),
\end{eqnarray*}
where the second inequality follows from Assumption \ref{ass1} and the facts that $\| \bm{\Lambda}_I\|_2 =\sqrt{M\ln(MN)}$, $\|\textbf{U}\textbf{G}\|_F=O_P(\sqrt{MNT})$ and $\|\textbf{H}\|_F=O_P(1)$.

Finally, given $\frac{1}{\sqrt{T}} \| \textbf{A}_{14} \widehat{\textbf{G}} \|_2 = O_P\left( \frac{1}{\sqrt{MN}} \right)$ from Lemma \ref{lemma_nof1}, we have

\begin{eqnarray*}
&& \frac{1}{\sqrt{T}} \| \widehat{\textbf{G}} - \textbf{G}\textbf{H} \|_F 
\leq O_P(1) \cdot \frac{1}{\sqrt{T}} \left( \| \textbf{A}_2 \widehat{\textbf{G}} \|_{2} + \cdots + \| \textbf{A}_{16} \widehat{\textbf{G}} \|_{2} \right) \\
&=& o_P(1) \cdot \frac{1}{\sqrt{T}} \| \widehat{\textbf{G}} - \textbf{G}\textbf{H} \|_F 
+ O_P\left( \frac{\sqrt{\ln(MN)}}{\min \{T, \sqrt{MT}, \sqrt{NT} \}} + \frac{1}{\sqrt{MN}} + \Delta_{g, MNT} \right),
\end{eqnarray*}
where $\Delta_{g, MNT}$ is defined in the body of this lemma. The first result then follows.

\medskip

(2). Denote $\delta_{g,MNT} = O_P\left( \frac{\sqrt{\ln(MN)}}{\min \{T, \sqrt{MT}, \sqrt{NT} \}} + \frac{1}{\sqrt{MN}} + \Delta_{g, MNT} \right)$ for notational simplicity. Note that

\begin{eqnarray*} 
&& \frac{1}{T} \| \textbf{G}' (\widehat{\textbf{G}}  - \textbf{G} \textbf{H}) \|_{F}  
\leq \frac{1}{T} ( \| \textbf{G}' \textbf{A}_2 \widehat{\textbf{G}} \|_2 + \cdots + \| \textbf{G}' \textbf{A}_{16} \widehat{\textbf{G}} \|_2 ) \cdot \| \textbf{V}_{g}^{-1} \|_{F},
\end{eqnarray*}
where $\textbf{H} = ( \frac{\bm{\Gamma}' \bm{\Gamma}}{MN} ) ( \frac{\textbf{G}' \widehat{\textbf{G}}}{T} ) \textbf{V}_{g}^{-1}$. Below, we reconsider the terms on the right hand side, and provide an improved rate under extra conditions of Assumption \ref{ass5}.1. Note that $\| \textbf{V}_{g}^{-1} \|_{F}=O_P(1)$, so we focus on $\frac{1}{T}\| \textbf{G}' \textbf{A}_2 \widehat{\textbf{G}} \|_2, \ldots,  \frac{1}{T}\| \textbf{G}' \textbf{A}_{16} \widehat{\textbf{G}} \|_2$.

\begin{eqnarray*}
&&\frac{1}{T} \| \textbf{G}' \textbf{A}_2 \widehat{\textbf{G}} \|_2
\leq \frac{1}{T} \frac{1}{MNT} \|  \textbf{G}' \textbf{F}_E \|_F \cdot \| \bm{\Lambda}_E\|_2^2 \cdot \left( \| \textbf{F}_E ( \widehat{\textbf{G}} - \textbf{G} \textbf{H}) \|_2 + \| \textbf{F}_E \textbf{G} \textbf{H} \|_F \right) \\
&\leq& O_P(1) \frac{1}{T} \frac{1}{MNT} \cdot T\sqrt{M} \cdot T^{\max_{i}a_{E,i}} \cdot N \ln(MN) \\
&&\cdot \left( (\sqrt{M} \vee \sqrt{T}) \sqrt{T} \cdot \delta_{g,MNT}
+ T\sqrt{M} \cdot T^{\max_{i}a_{E,i}} \right) \\
&=& \ln(MN) \cdot O_P \left( \frac{\delta_{g,MNT} \cdot T^{\max_{i}a_{E,i}} }{\sqrt{M} \wedge \sqrt{T}} + T^{ 2\max_{i}a_{E,i}} \right),
\end{eqnarray*}
where the second inequality follows from Assumption \ref{ass4}.1.(a) and the facts that $\| \bm{\Lambda}_E\|_2^2=N\ln(MN)$, $\frac{1}{\sqrt{T}}\| \widehat{\textbf{G}} - \textbf{G} \textbf{H}\|_F = \delta_{g,MNT}$ and $\|\textbf{H}\|_F=O_P(1)$.

\begin{eqnarray*}
&&\frac{1}{T} \| \textbf{G}' \textbf{A}_3 \widehat{\textbf{G}} \|_2
\leq \frac{1}{T} \frac{1}{MNT} \|  \textbf{G}' \textbf{F}_I \|_F \cdot \| \bm{\Lambda}_I \|_2^2 \cdot \left( \| \textbf{F}_I' ( \widehat{\textbf{G}} - \textbf{G}\textbf{H}) \|_2 + \| \textbf{F}_I' \textbf{G}\textbf{H} \|_F \right) \\
&\leq& O_P(1) \frac{1}{T} \frac{1}{MNT} \cdot T\sqrt{N} \cdot T^{\max_{j}a_{I,j}} \cdot M \ln(MN) \\
&&\cdot \left( (\sqrt{N} \vee \sqrt{T}) \sqrt{T} \cdot \delta_{g,MNT} + T\sqrt{N} \cdot T^{\max_{j}a_{I,j}} \right) \\
&=& \ln(MN) \cdot O_P \left( \frac{\delta_{g,MNT} \cdot T^{\max_{j}a_{I,j}} }{\sqrt{N} \wedge \sqrt{T}} + T^{ 2\max_{j}a_{I,j}} \right),
\end{eqnarray*}
where the second inequality follows from Assumption \ref{ass4}.1.(a) and the facts that $\| \bm{\Lambda}_I\|_2^2=M\ln(MN)$, $\frac{1}{\sqrt{T}}\| \widehat{\textbf{G}} - \textbf{G} \textbf{H}\|_F = \delta_{g,MNT}$ and $\|\textbf{H}\|_F=O_P(1)$.

\begin{eqnarray*}
&&\frac{1}{T} \| \textbf{G}' \textbf{A}_9 \widehat{\textbf{G}} \|_2
\leq \frac{1}{T} \frac{1}{MNT} \|  \textbf{G}' \textbf{F}_E \|_F \cdot \| \bm{\Lambda}_E' \bm{\Lambda}_I \|_2 \cdot \left( \| \textbf{F}_I' ( \widehat{\textbf{G}} - \textbf{G}\textbf{H}) \|_2 + \| \textbf{F}_I' \textbf{G}\textbf{H} \|_F \right) \\
&\leq& O_P(1) \frac{1}{T} \frac{1}{MNT} \cdot T\sqrt{M} \cdot T^{\max_{i}a_{E,i}} \cdot \ln(MN) \sqrt{MN} \\
&&\cdot \left( (\sqrt{N} \vee \sqrt{T}) \sqrt{T} \cdot \delta_{g,MNT}
+ T\sqrt{N} \cdot T^{\max_{j}a_{I,j}} \right) \\
&=& \sqrt{\ln(MN)} \cdot O_P \left( \frac{\delta_{g,MNT} \cdot T^{\max_{i}a_{E,i}} }{\sqrt{N} \wedge \sqrt{T}} + T^{\max_{i}a_{E,i} + \max_{j}a_{I,j}} \right),
\end{eqnarray*}
where the second inequality follows from Assumption \ref{ass4}.1.(a) and the facts that $\| \bm{\Lambda}_I\|_2=\sqrt{M\ln(MN)}$, $\| \bm{\Lambda}_E\|_2=\sqrt{N\ln(MN)}$, $\frac{1}{\sqrt{T}}\| \widehat{\textbf{G}} - \textbf{G} \textbf{H}\|_F = \delta_{g,MNT}$ and $\|\textbf{H}\|_F=O_P(1)$.

\begin{eqnarray*}
&&\frac{1}{T} \| \textbf{G}' \textbf{A}_{12} \widehat{\textbf{G}} \|_2
\leq \frac{1}{T} \frac{1}{MNT} \|  \textbf{G}' \textbf{F}_I \|_F \cdot \| \bm{\Lambda}_I' \bm{\Lambda}_E \|_2 \cdot \left( \| \textbf{F}_E' ( \widehat{\textbf{G}} - \textbf{G}\textbf{H} ) \|_2 + \| \textbf{F}_E'\textbf{G}\textbf{H} \|_F \right) \\
&\leq& O_P(1) \frac{1}{T} \frac{1}{MNT} \cdot T\sqrt{N} \cdot T^{\max_{j}a_{I,j}} \cdot \ln(MN)\sqrt{MN} \\
&&\cdot \left( (\sqrt{M} \vee \sqrt{T}) \sqrt{T} \cdot \delta_{g,MNT}
+ T\sqrt{M} \cdot T^{\max_{i}a_{E,i}} \right) \\
&=& \sqrt{\ln(MN)} \cdot O_P \left( \frac{\delta_{g,MNT} \cdot T^{\max_{j}a_{I,j}} }{\sqrt{M} \wedge \sqrt{T}} + T^{\max_{i}a_{E,i} + \max_{j}a_{I,j}} \right),
\end{eqnarray*}
where the second inequality follows from Assumption \ref{ass4}.1.(a) and the facts that $\| \bm{\Lambda}_I\|_2=\sqrt{M\ln(MN)}$, $\| \bm{\Lambda}_E\|_2=\sqrt{N\ln(MN)}$, $\frac{1}{\sqrt{T}}\| \widehat{\textbf{G}} - \textbf{G} \textbf{H}\|_F = \delta_{g,MNT}$ and $\|\textbf{H}\|_F=O_P(1)$.

\begin{eqnarray*}
&&\frac{1}{T} \| \textbf{G}' \textbf{A}_{10} \widehat{\textbf{G}} \|_2
\leq \frac{1}{T} \frac{1}{MNT} \|  \textbf{G}' \textbf{F}_{E} \|_F \cdot \| \bm{\Lambda}_E \|_2 \cdot \left( \|\textbf{U} \|_2 \cdot \| \widehat{\textbf{G}} - \textbf{G}\textbf{H} \|_F + \|\textbf{U}\textbf{G}\textbf{H} \|_F \right) \\
&\leq& O_P(1) \frac{1}{T} \frac{1}{MNT} \cdot T\sqrt{M} \cdot T^{\max_{i}a_{E,i}} \cdot \sqrt{N \ln(MN)} \\
&&\cdot \left(  (\sqrt{MN} \vee \sqrt{T}) \sqrt{T} \cdot \delta_{g,MNT} + \sqrt{MNT} \right) \\
&=& \sqrt{\ln(MN)} \cdot O_P \left( \frac{\delta_{g,MNT} \cdot T^{\max_{i}a_{E,i}} }{\sqrt{MN} \wedge \sqrt{T}} + \frac{ T^{\max_{i}a_{E,i}}}{\sqrt{T}} \right),
\end{eqnarray*}
where the second inequality follows from Assumption \ref{ass4}.1.(a) and the facts that $\| \bm{\Lambda}_E\|_2=\sqrt{N\ln(MN)}$, $\frac{1}{\sqrt{T}}\| \widehat{\textbf{G}} - \textbf{G} \textbf{H}\|_F = \delta_{g,MNT}$, $\|\textbf{U}\|_2=O_P(\sqrt{MN}\vee \sqrt{T})$, $\|\textbf{U}\textbf{G}\|_F=O_P(\sqrt{MNT})$ and $\|\textbf{H}\|_F=O_P(1)$.

\begin{eqnarray*}
&&\frac{1}{T} \| \textbf{G}' \textbf{A}_{13} \widehat{\textbf{G}} \|_2
\leq \frac{1}{T} \frac{1}{MNT} \|  \textbf{G}' \textbf{F}_{I} \|_F \cdot \| \bm{\Lambda}_I \|_2 \cdot \left( \|\textbf{U} \|_2 \cdot \| \widehat{\textbf{G}} - \textbf{G}\textbf{H} \|_F + \|\textbf{U}\textbf{G}\textbf{H} \|_F \right) \\
&\leq& O_P(1) \frac{1}{T} \frac{1}{MNT} \cdot T\sqrt{N} \cdot T^{\max_{j}a_{I,j}} \cdot \sqrt{M\ln(MN)} \\
&&\cdot \left(  (\sqrt{MN} \vee \sqrt{T}) \sqrt{T} \cdot \delta_{g,MNT} + \sqrt{MNT} \right) \\
&=& \sqrt{\ln(MN)} \cdot O_P \left( \frac{\delta_{g,MNT} \cdot T^{\max_{j}a_{I,j}} }{\sqrt{MN} \wedge \sqrt{T}} + \frac{ T^{\max_{j}a_{I,j}}}{\sqrt{T}} \right),
\end{eqnarray*}
where the second inequality follows from Assumption \ref{ass4}.1.(a) and the facts that $\| \bm{\Lambda}_I\|_2=\sqrt{M\ln(MN)}$, $\frac{1}{\sqrt{T}}\| \widehat{\textbf{G}} - \textbf{G} \textbf{H}\|_F = \delta_{g,MNT}$, $\|\textbf{U}\|_2=O_P(\sqrt{MN}\vee \sqrt{T})$, $\|\textbf{U}\textbf{G}\|_F=O_P(\sqrt{MNT})$ and $\|\textbf{H}\|_F=O_P(1)$.

For $k=5$ and $6$, we apply the inequality $\frac{1}{T} \| \textbf{G}' \textbf{A}_{k} \widehat{\textbf{G}} \|_2 \leq \frac{1}{T} \| \textbf{G} \|_F \cdot \| \textbf{A}_{k} \widehat{\textbf{G}} \|_2$ and then use the results in Lemma \ref{lemma2_improved}.1. For $k=8$ and $11$, we bound $\frac{1}{T} \| \textbf{G}' \textbf{A}_{k} \widehat{\textbf{G}} \|_2$ directly and use Assumption \ref{ass4}.1.(a) and Assumption \ref{ass5}.1.

\begin{eqnarray*}
&& \frac{1}{T} \| \textbf{G}' \textbf{A}_8 \widehat{\textbf{G}} \|_2 
\leq \frac{1}{T}\frac{1}{MNT} \| \textbf{G}'\textbf{F}_E \|_F \cdot \| \bm{\Lambda}_E' \bm{\Gamma} \|_F \cdot \| \textbf{G}' \widehat{\textbf{G}} \|_F  \\
&\leq& O_P(1) \frac{1}{T} \frac{1}{MNT} T\sqrt{M} \cdot T^{\max_{i}a_{E,i}} \cdot \sqrt{MN} \cdot T \\
&=& O_P \left( \frac{T^{\max_{i}a_{E,i}}}{\sqrt{N}} \right),
\end{eqnarray*}
where the second inequality follows from Assumption \ref{ass4}.1.(a) and Assumption \ref{ass5}.1.

\begin{eqnarray*}
&& \frac{1}{T} \| \textbf{G}' \textbf{A}_{11} \widehat{\textbf{G}} \|_2 
\leq \frac{1}{T}\frac{1}{MNT} \| \textbf{G}'\textbf{F}_I \|_F \cdot \| \bm{\Lambda}_I' \bm{\Gamma} \|_F \cdot \| \textbf{G}' \widehat{\textbf{G}} \|_F  \\
&=& O_P(1) \frac{1}{T} \frac{1}{MNT} T\sqrt{N} \cdot T^{\max_{j}a_{I,j}} \cdot \sqrt{MN} \cdot T \\
&=& O_P \left( \frac{T^{\max_{j}a_{I,j}}}{\sqrt{M}} \right),
\end{eqnarray*}
where the second inequality follows from Assumption \ref{ass4}.1.(a) and Assumption \ref{ass5}.1.

Next, for $k=4,7,14,15$ and $16$, we bound $\frac{1}{T} \| \textbf{G}' \textbf{A}_{k} \widehat{\textbf{G}} \|_2$ directly and use the convergence rate of $\frac{1}{\sqrt{T}}\|\widehat{\textbf{G}} - \textbf{G}\textbf{H}\|$ achieved in the first result of this lemma.

\begin{eqnarray*}
&& \frac{1}{T} \| \textbf{G}' \textbf{A}_4 \widehat{\textbf{G}} \|_2 
\leq \frac{1}{T} \frac{1}{MNT}\| \textbf{G}' \textbf{U}'\|_F \cdot \left( \|\textbf{U} \|_2 \cdot \| \widehat{\textbf{G}} - \textbf{G}\textbf{H} \|_F + \|\textbf{U}\textbf{G}\textbf{H} \|_F \right) \\
&\leq& O_P(1) \frac{\sqrt{MNT}}{MNT^2} \left(  (\sqrt{MN} \vee \sqrt{T}) \cdot \sqrt{T} \cdot \delta_{g,MNT} + \sqrt{MNT} \right) \\
&=& O_P \left( \frac{\delta_{g,MNT}}{T \wedge \sqrt{MNT}} + \frac{1}{T} \right),
\end{eqnarray*}
where the second inequality follows from the facts that $\|\textbf{U}\textbf{G}\|_F = O_P(\sqrt{MNT})$, $\|\textbf{U}\|_2=O_P(\sqrt{MN}\vee \sqrt{T})$, $\frac{1}{\sqrt{T}}\| \widehat{\textbf{G}} - \textbf{G} \textbf{H}\|_F = \delta_{g,MNT}$ and $\|\textbf{H}\|_F=O_P(1)$.

Consider the term involving $\textbf{A}_7$, write

\begin{eqnarray*}
&& \frac{1}{T} \| \textbf{G}' \textbf{A}_7 \widehat{\textbf{G}} \|_2
\leq \frac{1}{T} \frac{1}{MNT} \| \textbf{G}' \textbf{G}\|_F \| \bm{\Gamma}'\textbf{U}  \widehat{\textbf{G}} \|_F \\
&\leq& \frac{1}{T} \frac{1}{MNT} \|\textbf{G}\|_F^2 \cdot \left( \| \bm{\Gamma}'\textbf{U} \|_F \cdot \| \widehat{\textbf{G}} - \textbf{G}\textbf{H} \|_F + \|\bm{\Gamma}'\textbf{U} \textbf{G}\|_F \cdot \|\textbf{H} \|_F \right) \\
&\leq& O_P(1) \frac{1}{MNT} \left( \sqrt{MNT}\sqrt{T} \cdot \delta_{g,MNT} + \sqrt{MNT} \right) \\
&=& O_P \left( \frac{\delta_{g,MNT}}{\sqrt{MN}} + \frac{1}{\sqrt{MNT}} \right),
\end{eqnarray*}
where the third inequality follows from the facts that $\|\bm{\Gamma}'\textbf{U}\|_F=O_P(\sqrt{MNT})$, $\| \textbf{G}' \textbf{U}' \bm{\Gamma} \|_F = O_P (\sqrt{MNT})$, $\frac{1}{\sqrt{T}}\| \widehat{\textbf{G}} - \textbf{G} \textbf{H}\|_F = \delta_{g,MNT}$ and $\|\textbf{H}\|_F=O_P(1)$. Similarly,

\begin{eqnarray*} 
\frac{1}{T} \| \textbf{G}' \textbf{A}_{14} \widehat{\textbf{G}} \|_2 = O_P\left(\frac{1}{\sqrt{MNT}}\right). 
\end{eqnarray*}
For the terms involving $\textbf{A}_{15}$ and $\textbf{A}_{16}$, write

\begin{eqnarray*}
&& \frac{1}{T} \| \textbf{G}' \textbf{A}_{15} \widehat{\textbf{G}} \|_2
\leq \frac{1}{T} \frac{1}{MNT} \| \textbf{G}' \textbf{U}' \|_F \cdot \| \bm{\Lambda}_E \|_2 \cdot \left( \| \textbf{F}_E' ( \widehat{\textbf{G}} - \textbf{G}\textbf{H} ) \|_2 + \| \textbf{F}_E'\textbf{G}\textbf{H} \|_F \right) \\
&\leq& \frac{1}{T} \frac{1}{MNT} O_P(1) \sqrt{MNT} \cdot \sqrt{N\ln(MN)}  \\
&&\cdot \left( (\sqrt{M} \vee \sqrt{T}) \cdot \sqrt{T} \cdot  \delta_{g,MNT} + T\sqrt{M} \cdot T^{\max_{i}a_{E,i}} \right) \\
&=& \sqrt{\ln(MN)} \cdot O_P\left(\frac{\delta_{g,MNT}}{ T \wedge \sqrt{MT}} + \frac{T^{\max_{i}a_{E,i}}}{\sqrt{T}}\right),
\end{eqnarray*}
where the second inequality follows from Assumption \ref{ass1}, Assumption \ref{ass4}.1.(a), and the facts that $\|\textbf{U}\textbf{G}\|_F = O_P(\sqrt{MNT})$, $\|\bm{\Lambda}_E\|_2 = O_P(\sqrt{N\ln(MN)})$, $\frac{1}{\sqrt{T}}\| \widehat{\textbf{G}} - \textbf{G} \textbf{H}\|_F = \delta_{g,MNT}$ and $\| \textbf{H}\|_F=O_P(1)$.

\begin{eqnarray*}
&& \frac{1}{T} \| \textbf{G}' \textbf{A}_{16} \widehat{\textbf{G}} \|_2
\leq \frac{1}{T} \frac{1}{MNT} \| \textbf{G}' \textbf{U}'\|_F \cdot \| \bm{\Lambda}_I \|_2 \cdot \left( \| \textbf{F}_I' ( \widehat{\textbf{G}} - \textbf{G}\textbf{H}) \|_2 + \| \textbf{F}_I' \textbf{G}\textbf{H} \|_F \right) \\
&=& \frac{1}{T} \frac{1}{MNT} O_P(1) \sqrt{MNT} \cdot \sqrt{M\ln(MN)} \\
&&\cdot \left( (\sqrt{N} \vee \sqrt{T}) \cdot \sqrt{T} \cdot \delta_{g,MNT}
+ T\sqrt{N} \cdot T^{\max_{j}a_{I,j}} \right) \\
&=& \sqrt{\ln(MN)} \cdot O_P\left(\frac{\delta_{g,MNT}}{ T \wedge \sqrt{NT}} + \frac{T^{\max_{j}a_{I,j}}}{\sqrt{T}}\right),
\end{eqnarray*}
where the second inequality follows from Assumption \ref{ass1}, Assumption \ref{ass4}.1.(a), and the facts that $\|\textbf{U}\textbf{G}\|_F = O_P(\sqrt{MNT})$, $\|\bm{\Lambda}_ I\|_2 = O_P(\sqrt{M\ln(MN)})$, $\frac{1}{\sqrt{T}}\| \widehat{\textbf{G}} - \textbf{G} \textbf{H}\|_F=\delta_{g,MNT}$ and $\| \textbf{H}\|_F=O_P(1)$.

Based on the above development, we obtain that
\begin{eqnarray*}
\frac{1}{T} \| \textbf{G}' (\widehat{\textbf{G}}  - \textbf{G} \textbf{H}) \|_F
&=& O_P \left( \frac{1}{MN} + \frac{1}{T} + \Delta_{g, MNT}^{*} \right),
\end{eqnarray*}
where $\Delta_{g, MNT}^{*}$ is defined in the body of this lemma. The second result then follows.

\medskip

(3). 
By the second result of this lemma and the condition $\frac{1}{T} \textbf{G}'\textbf{G} = \textbf{I}_{r_g}$,  we immediately obtain that

\begin{eqnarray} \label{clt_e0}
\frac{1}{T}\textbf{G}'\widehat{\textbf{G}} - \textbf{H} =  O_P \left( \frac{1}{MN} + \frac{1}{T} + \Delta_{g, MNT}^{*} \right).
\end{eqnarray}
Left multiplying both sides of \eqref{clt_e0} by $\textbf{H}$ and using $\| \textbf{H} \|_F=O_P(1)$,  we have

\begin{eqnarray} \label{clt_e1}
\frac{1}{T}\textbf{H}' \textbf{G}'\widehat{\textbf{G}}  - \textbf{H}'\textbf{H} = O_P \left(\frac{1}{MN} + \frac{1}{T} + \Delta_{g, MNT}^{*} \right).
\end{eqnarray}
Furthermore, \eqref{e2} of Lemma \ref{lemma3} becomes

\begin{eqnarray} \label{clt_e2}
\frac{1}{T} \widehat{\textbf{G}}'\widehat{\textbf{G}}  - \frac{1}{T} \widehat{\textbf{G}}'\textbf{G} \textbf{H} 
= \textbf{I}_{r_g} - \frac{1}{T} \widehat{\textbf{G}}'\textbf{G} \textbf{H} = O_P \left(\frac{1}{MN} + \frac{1}{T} + \Delta_{g, MNT}^{*} \right).
\end{eqnarray}

Summing up (\ref{clt_e1}) and (\ref{clt_e2}),

\begin{eqnarray} \label{clt_e3}
\textbf{I}_{r_g} - \textbf{H}'\textbf{H}=O_P \left( \frac{1}{MN} + \frac{1}{T} + \Delta_{g, MNT}^{*} \right).
\end{eqnarray}
(\ref{clt_e3}) shows that $\textbf{H}$ is an orthogonal matrix w.p.a.1 and hence its eigenvalues are either $1$ or $-1$.  By the definition of $\textbf{H}$,  we have

\begin{eqnarray} \label{clt_e4}
\textbf{H} \cdot \textbf{V}_{g} =  \left( \frac{\bm{\Gamma}' \bm{\Gamma}}{MN} \right) \textbf{H} + O_P \left( \frac{1}{MN} + \frac{1}{T} + \Delta_{g, MNT}^{*} \right),
\end{eqnarray}
which implies that $\textbf{H}$ is a matrix consisting of eigenvectors of $\frac{1}{MN}\bm{\Gamma}' \bm{\Gamma}$ in the limit. By Assumption \ref{ass5}.2, $\bm{\Gamma}' \bm{\Gamma}$ is diagonal with distinct eigenvalues. It follows that each eigenvalue is associated with a unique unitary eigenvector (up to a sign change) and each eigenvector has a single non-zero element. This implies that $\textbf{H}$ is a diagonal matrix up to the order $O_P \left(\frac{1}{MN} + \frac{1}{T} + \Delta_{g, MNT}^{*} \right)$.  In connection with the fact that the eigenvalues of $\textbf{H}$ are $1$ or $-1$, $\textbf{H}$ is a diagonal matrix with $1$ or $-1$ as its elements. Without loss of generality, we assume all elements are $1$.  This implies that

\begin{eqnarray*}
\textbf{H} = \textbf{I}_{r_g} + O_P \left( \frac{1}{MN} + \frac{1}{T} + \Delta_{g, MNT}^{*} \right). 
\end{eqnarray*}
Moreover,  in connection with \eqref{clt_e4},  we have $\textbf{V}_{g} = \nonumber \frac{\bm{\Gamma}' \bm{\Gamma}}{MN} + o_P(1)$.  The proof is now complete.  \hspace*{\fill}{$\blacksquare$}

\bigskip

\noindent \textbf{Proof of Lemma \ref{lemma6_improved}}:

(1).  Based on Assumptions \ref{ass1}-\ref{ass5} and Lemma \ref{lemma2_improved},  we provide faster rates for $\textbf{M}_{\widehat{\textbf{G}}} (\textbf{G}-\widehat{\textbf{G}} \textbf{H}^{-1}) \bm{\Gamma}_{I, j}'$, $\textbf{D}_1$, $\textbf{D}_2$, and $\textbf{D}_3$ below.

First consider $\textbf{M}_{\widehat{\textbf{G}}} (\textbf{G}-\widehat{\textbf{G}} \textbf{H}^{-1}) \bm{\Gamma}_{I, j}'$ and write
\begin{eqnarray*}
&& \frac{1}{\sqrt{MT}}\|\textbf{M}_{\widehat{\textbf{G}}} (\textbf{G}-\widehat{\textbf{G}} \textbf{H}^{-1}) \bm{\Gamma}_{I, j}' \|_F 
\le O(1) \cdot \frac{1}{\sqrt{MT}} \|\textbf{M}_{\widehat{\textbf{G}}} (\textbf{A}_2 + \cdots + \textbf{A}_{16}) \widehat{\textbf{G}}\|_2 \cdot \|  \textbf{K} \|_F \cdot \|\bm{\Gamma}_{I, j} \|_F, \\
&\leq& O_P(1)\frac{1}{\sqrt{MT}} \cdot  \sqrt{T} \left( \frac{\sqrt{\ln(MN)}}{\min \{T, \sqrt{MT}, \sqrt{NT} \}} + \frac{1}{\sqrt{MN}} + \Delta_{g, MNT} \right) \cdot \sqrt{M} \\
&=& O_P \left( \frac{\sqrt{\ln(MN)}}{\min \{T, \sqrt{MT}, \sqrt{NT} \}} + \frac{1}{\sqrt{MN}} + \Delta_{g, MNT} \right),
\end{eqnarray*}
where the first inequality follows from $\|\bm{\Omega}\|_F \le \sqrt{\text{rank}(\bm{\Omega})} \cdot \|\bm{\Omega} \|_2$, the second inequality follows from Lemma \ref{lemma2_improved}, $\|\widehat{\textbf{G}} \|_F =O(\sqrt{T})$ and $\|\bm{\Gamma}_{I, j} \|_F =O_P(\sqrt{M})$. $\Delta_{g, MNT}$ is defined in the body of Lemma \ref{lemma2_improved}.

Consider $\textbf{D}_1$, and write

\begin{eqnarray*}
&&\frac{1}{\sqrt{MT}}\| \textbf{D}_1 \|_F = \frac{1}{\sqrt{MT}}\| \diag\{\bm{\Lambda}_{E,\bullet j}' \}' \textbf{F}_{E}' \textbf{P}_{\widehat{\textbf{G}}} \|_F
\leq O(1) \frac{1}{\sqrt{MT}}\| \diag\{\bm{\Lambda}_{E,\bullet j}' \}' \textbf{F}_{E}' \textbf{P}_{\widehat{\textbf{G}}} \|_2 \\
&\leq& O(1)\frac{1}{\sqrt{MT}T}\| \diag\{\bm{\Lambda}_{E,\bullet j}' \} \|_2 \cdot \|\textbf{F}_{E}' \widehat{\textbf{G}}\|_F \cdot \|\widehat{\textbf{G}} \|_F \\
&\leq& O(1) \frac{1}{\sqrt{MT}T}\| \diag\{\bm{\Lambda}_{E,\bullet j}' \} \|_2 \cdot \left( \|\textbf{F}_{E}' ( \widehat{\textbf{G}} - \textbf{G}\textbf{H} ) \|_F + \| \textbf{F}_{E}' \textbf{G}\textbf{H} \|_F \right) \cdot \|\widehat{\textbf{G}} \|_F \\
&=& O_P\left( \frac{\sqrt{\ln(MN)}}{\sqrt{T} \wedge \sqrt{M}} \right) \left( \frac{\sqrt{\ln(MN)}}{\min \{T, \sqrt{MT}, \sqrt{NT} \}} + \frac{1}{\sqrt{MN}} + \Delta_{g, MNT} \right) + O_P\left(\sqrt{\ln(MN)} \cdot T^{\max_i a_{E,i}}\right),
\end{eqnarray*}
where the first inequality follows from $\|\bm{\Omega}\|_F \le \sqrt{\text{rank}(\bm{\Omega})} \cdot \|\bm{\Omega} \|_2$
, the second inequality follows from $\textbf{P}_{\widehat{\textbf{G}}} = \frac{1}{T}\widehat{\textbf{G}} \widehat{\textbf{G}}^{\prime}$, and the last equality follows from Lemma \ref{lemma2_improved} and Assumption \ref{ass4}.1.(a).

Next, consider $\textbf{D}_2$, and write

\begin{eqnarray*}
&&\frac{1}{\sqrt{MT}}\| \textbf{D}_2 \|_F 
= \frac{1}{\sqrt{MT}}\| \bm{\Lambda}_{I,\bullet j} \textbf{F}_{I,j}' \textbf{P}_{\widehat{\textbf{G}}} \|_F 
\leq \frac{1}{\sqrt{MT}T}\| \bm{\Lambda}_{I,\bullet j} \textbf{F}_{I,j}'  \widehat{\textbf{G}} \|_F \cdot \| \widehat{\textbf{G}}\|_F \\ 
&\le & \frac{1}{\sqrt{MT}T}\| \bm{\Lambda}_{I,\bullet j} \textbf{F}_{I,j}'  (\widehat{\textbf{G}} - \textbf{G} \textbf{H})\|_F \cdot \| \widehat{\textbf{G}} \|_F + \frac{1}{\sqrt{MT}T}\| \bm{\Lambda}_{I,\bullet j} \textbf{F}_{I,j}' \textbf{G} \textbf{H}\|_F \cdot \| \widehat{\textbf{G}} \|_F.
\end{eqnarray*}
Note that 

\begin{eqnarray*}
&&\frac{1}{\sqrt{MT}T}\| \bm{\Lambda}_{I,\bullet j} \textbf{F}_{I,j}'  (\widehat{\textbf{G}} - \textbf{G} \textbf{H})\|_F \cdot \| \widehat{\textbf{G}} \|_F  = O_P \left( \frac{\sqrt{\ln(MN)}}{\min \{T, \sqrt{MT}, \sqrt{NT} \}} + \frac{1}{\sqrt{MN}} + \Delta_{g, MNT} \right), \\
&&\frac{1}{\sqrt{MT}T}\| \bm{\Lambda}_{I,\bullet j} \textbf{F}_{I,j}' \textbf{G} \textbf{H}\|_F \cdot \| \widehat{\textbf{G}} \|_F = O_P( T^{a_{I,j}} ),
\end{eqnarray*}
where we have used Assumption \ref{ass2}, Assumption \ref{ass4}.1.(a), Lemma \ref{lemma2_improved}, and $ \|  \textbf{F}_{I, j}\|_F=O_P(\sqrt{T})$.

Then we consider $\textbf{D}_3$ and write

\begin{eqnarray*}
&& \frac{1}{\sqrt{MT}}\| \textbf{D}_3 \|_F 
= \frac{1}{\sqrt{MT}}\| \textbf{U}_{\bullet j \bullet} \textbf{P}_{\widehat{\textbf{G}}} \|_F
\leq O(1) \frac{1}{\sqrt{MT}}\| \textbf{U}_{\bullet j \bullet} \textbf{P}_{\widehat{\textbf{G}}} \|_2 \\
&\leq& O(1) \frac{1}{\sqrt{MT}T}\| \textbf{U}_{\bullet j \bullet} \widehat{\textbf{G}} \| \cdot \| \widehat{\textbf{G}}\|_F
\leq O(1) \frac{1}{\sqrt{MT}T} \left( \| \textbf{U}_{\bullet j \bullet} ( \widehat{\textbf{G}} - \textbf{G}\textbf{H}) \|_2 + \| \textbf{U}_{\bullet j \bullet} \textbf{G}\textbf{H} \|_F \right) \cdot \| \widehat{\textbf{G}} \|_F \\
&=& O_P(1) \left( \frac{1}{\sqrt{M}} \vee \frac{1}{\sqrt{T}} \right) \left( \frac{\sqrt{\ln(MN)}}{\min \{T, \sqrt{MT}, \sqrt{NT} \}} + \frac{1}{\sqrt{MN}} + \Delta_{g, MNT} \right) + O_P\left( \frac{1}{\sqrt{T}} \right),
\end{eqnarray*}
where the first inequality follows from $\|\bm{\Omega}\|_F \le \sqrt{\text{rank}(\bm{\Omega})} \cdot \|\bm{\Omega} \|_2$
, the second inequality follows from $\textbf{P}_{\widehat{\textbf{G}}} = \frac{1}{T}\widehat{\textbf{G}} \widehat{\textbf{G}}^{\prime}$, and the last equality follows from $\| \textbf{U}_{\bullet j \bullet} \textbf{G} \|_F = O_P(\sqrt{MT})$, which is similar to the result of $\| \textbf{G}'\textbf{U}' \|_F$ in the proof of Lemma \ref{lemma2_improved}(i).

\medskip

Based on the above development, we obtain that

\begin{eqnarray*}
\frac{1}{\sqrt{MT}}\left\|\bm{\Gamma}_{I, j}\textbf{G}' - \widehat{\bm{\Gamma}}_{I, j} \widehat{\textbf{G}}'\right\|_F = O_P \left( \frac{1}{\sqrt{MN}} + \frac{1}{\sqrt{T}} + \Delta_{Ij, MNT} \right),
\end{eqnarray*}
where $\Delta_{Ij, MNT}$ is defined in the body of this lemma. This completes the proof of the first result.

\medskip

(2). The second result can be proved in exactly the same way as the first result. Thus, omitted.

\medskip

(3). Consider (1) in Lemma \ref{lemma8}. Write

\begin{eqnarray*}
&&\frac{1}{MT} \|(\bm{\Gamma}_{I, j} \textbf{G}' - \widehat{\bm{\Gamma}}_{I, j} \widehat{\textbf{G}}' )' \diag\{\bm{\Lambda}_{E,\bullet j}' \}'\textbf{F}_{E}' \|_2\\
&\leq& \frac{1}{MT} \|\bm{\Gamma}_{I, j} \textbf{G}' - \widehat{\bm{\Gamma}}_{I, j} \widehat{\textbf{G}}' \|_2 \cdot \|  \diag\{\bm{\Lambda}_{E,\bullet j}' \} \|_2 \cdot \| \textbf{F}_E \|_2 \\
&=& \frac{1}{MT} \|\bm{\Gamma}_{I, j} \textbf{G}' - \widehat{\bm{\Gamma}}_{I, j} \widehat{\textbf{G}}'  \|_2 \cdot \sqrt{ \max_{i\ge 1, j\ge 1}\| \bm{\lambda}_{E,ij}\|_{F}^2 } \cdot \| \textbf{F}_{E} \|_2 \\
&=& O_P(1)\frac{1}{\sqrt{MT}} \cdot \left( \frac{1}{\sqrt{MN}} + \frac{1}{\sqrt{T}} + \Delta_{Ij, MNT} \right) \cdot \sqrt{\ln (MN)} \cdot (\sqrt{T}\vee \sqrt{M})\\
&=& O_P(1)\frac{\sqrt{\ln(MN)}}{\sqrt{M} \wedge \sqrt{T}} \cdot \left( \frac{1}{\sqrt{MN}} + \frac{1}{\sqrt{T}} + \Delta_{Ij, MNT} \right),
\end{eqnarray*}
where the second equality follows from the first result of this lemma, Assumption \ref{ass1} and Assumption \ref{ass2}.

\medskip

Consider (2) in Lemma \ref{lemma8}. Write

\begin{eqnarray*}
&& \frac{1}{MT}\| (\bm{\Gamma}_{I, j} \textbf{G}' - \widehat{\bm{\Gamma}}_{I, j} \widehat{\textbf{G}}' )' \bm{\Lambda}_{I,\bullet j} \textbf{F}_{I,j}' \|_2 \\
&\le &\frac{1}{MT} \| \bm{\Gamma}_{I, j} \textbf{G}' - \widehat{\bm{\Gamma}}_{I, j} \widehat{\textbf{G}}'\|_2 \cdot \|  \bm{\Lambda}_{I,\bullet j}  \|_2 \cdot \| \textbf{F}_{I,j} \|_F \\
&=& O_P(1)\frac{1}{\sqrt{MT}} \cdot \left( \frac{1}{\sqrt{MN}} + \frac{1}{\sqrt{T}} + \Delta_{Ij, MNT} \right) \cdot \sqrt{M}\cdot \sqrt{T} \\
&=& O_P\left( \frac{1}{\sqrt{MN}} + \frac{1}{\sqrt{T}} + \Delta_{Ij, MNT} \right),
\end{eqnarray*}
where the first equality follows from the first result of this lemma.

\medskip

Consider (3) in Lemma \ref{lemma8}. Write

\begin{eqnarray*}
&& \frac{1}{MT}\| (\bm{\Gamma}_{I, j} \textbf{G}' - \widehat{\bm{\Gamma}}_{I, j} \widehat{\textbf{G}}' )' \textbf{U}_{\bullet j \bullet} \|_2\\
&\le &\frac{1}{MT} \| \bm{\Gamma}_{I, j} \textbf{G}' - \widehat{\bm{\Gamma}}_{I, j} \widehat{\textbf{G}}'\|_2 \cdot \| \textbf{U}_{\bullet j \bullet}  \|_2\\
&=& O_P(1)\frac{1}{\sqrt{MT}} \cdot \left( \frac{1}{\sqrt{MN}} + \frac{1}{\sqrt{T}} + \Delta_{Ij, MNT} \right) \cdot (\sqrt{M} \vee \sqrt{T}) \\
&=& O_P(1) \frac{1}{ \sqrt{M} \wedge \sqrt{T} } \left( \frac{1}{\sqrt{MN}} + \frac{1}{\sqrt{T}} + \Delta_{Ij, MNT} \right),
\end{eqnarray*}
where the first equality follows from the first result of this lemma.

\medskip

Consider (7) in Lemma \ref{lemma8}. Write

\begin{eqnarray*}
&&\frac{1}{MT}\| (\bm{\Gamma}_{I, j} \textbf{G}' - \widehat{\bm{\Gamma}}_{I, j} \widehat{\textbf{G}}' )' (\bm{\Gamma}_{I, j} \textbf{G}' - \widehat{\bm{\Gamma}}_{I, j} \widehat{\textbf{G}}' )\|_2\le \frac{1}{MT} \| \bm{\Gamma}_{I, j} \textbf{G}' - \widehat{\bm{\Gamma}}_{I, j} \widehat{\textbf{G}}' \|_2^2\\
&=& O_P \left( \frac{1}{MN} + \frac{1}{T}
+ \Delta_{Ij, MNT}^2 \right),
\end{eqnarray*}
where the first equality follows from the first result of this lemma.

\medskip

The results for other terms are the same as those in Lemma \ref{lemma8}. The rest of the proofs are in the same spirit as Lemma \ref{lemma_nof1_local}. Thus, omitted.

\medskip

(4). The result can be proved in exactly the same way as in the third result. The proof is now complete. \hspace*{\fill}{$\blacksquare$}

\bigskip

\noindent \textbf{Proof of Lemma \ref{lemma_clt0_local}}:

(1). We left and right multiply (\ref{eq1}) by $\frac{1}{T} \textbf{F}_{I, j}'$ and $\textbf{V}_{I, j}^{-1}$ respectively to obtain that

\begin{equation} \label{eq.a6.1}
\frac{1}{T} \textbf{F}_{I, j}' (\widehat{\textbf{F}}_{I,j} -  \textbf{F}_{I,j} \textbf{H}_{I,j} ) = \frac{1}{T} \textbf{F}_{I, j}' (\textbf{B}_2 + \cdots + \textbf{B}_{16}) \widehat{\textbf{F}}_{I, j} \textbf{V}_{I, j}^{-1},
\end{equation}
where $\textbf{H}_{I,j} = \frac{1}{M}\bm{\Lambda}_{I, \bullet j}' \bm{\Lambda}_{I, \bullet j}\cdot \frac{1}{T}\textbf{F}_{I,j}' \widehat{\textbf{F}}_{I,j} \cdot \textbf{V}_{I,j}^{-1}$. Therefore, 

\begin{eqnarray*} 
&& \frac{1}{T} \| \textbf{F}_{I, j}' (\widehat{\textbf{F}}_{I,j} -  \textbf{F}_{I,j} \textbf{H}_{I,j} ) \|_{F}  
\leq O(1) \cdot \frac{1}{T} \left( \| \textbf{F}_{I, j}' \textbf{B}_2 \widehat{\textbf{F}}_{I,j} \|_2 + \cdots + \| \textbf{F}_{I, j}' \textbf{B}_{16} \widehat{\textbf{F}}_{I,j} \|_2 \right) \cdot \| \textbf{V}_{I, j}^{-1} \|_{F},
\end{eqnarray*}
where the first inequality follows from (\ref{eq.a6.1}) and $\|\bm{\Omega}\|_F \le \sqrt{\text{rank}(\bm{\Omega})} \cdot \|\bm{\Omega} \|_2$.

Note that $\| \textbf{V}_{I,j}^{-1} \|_{F}=O_P(1)$, so we focus on $\frac{1}{T}\| \textbf{F}_{I, j}' \textbf{B}_2 \widehat{\textbf{F}}_{I,j} \|_2, \ldots,  \frac{1}{T}\| \textbf{F}_{I, j}' \textbf{B}_{16} \widehat{\textbf{F}}_{I,j} \|_2$. For $k=6-8, 10, 12, 14-16$, we carefully investigate $\| \textbf{F}_{I, j}' \textbf{B}_{k} \widehat{\textbf{F}}_{I,j} \|_2$ in order to achieve sharper bounds. For other terms, we apply the inequality $\frac{1}{T}\| \textbf{F}_{I, j}' \textbf{B}_k \widehat{\textbf{F}}_{I,j} \|_2 \leq \frac{1}{T}\| \textbf{F}_{I, j}\|_F \cdot \| \textbf{B}_k \|_2 \cdot \| \widehat{\textbf{F}}_{I,j} \|_F$, and use the updated results for $\textbf{B}_k$ in the proof of Lemma \ref{lemma6_improved}.

\medskip

That said, we now start the investigation. For the term involving $\textbf{B}_6$, write

\begin{eqnarray*} 
&& \frac{1}{T} \| \textbf{F}_{I, j}' \textbf{B}_{6} \widehat{\textbf{F}}_{I,j} \|_2
= \frac{1}{MT} \frac{1}{T} \| \textbf{F}_{I, j}' \textbf{F}_{E} \diag\{\bm{\Lambda}_{E,\bullet j}' \} \diag\{\bm{\Lambda}_{E,\bullet j}' \}' \textbf{F}_{E}' \widehat{\textbf{F}}_{I,j} \|_2 \\
&\leq& \frac{1}{MT} \frac{1}{T} \| \textbf{F}_{I, j}' \textbf{F}_{E} \|_F \cdot \| \diag\{\bm{\Lambda}_{E,\bullet j}' \} \|_2^2 \cdot \left( \| \textbf{F}_{E}' ( \widehat{\textbf{F}}_{I,j} - \textbf{F}_{I,j} \textbf{H}_{I,j} )\|_2 + \| \textbf{F}_{E}' \textbf{F}_{I,j} \textbf{H}_{I,j} \|_F \right) \\
&\le & O_P(1) \frac{1}{MT} \sqrt{M} \cdot T^{\max_i b_{EI, ij}} \cdot \ln (MN) \\
&&\cdot \left( (\sqrt{M} \vee \sqrt{T}) \sqrt{T} \left( \frac{\sqrt{\ln (MN)}}{ \sqrt{M} \wedge \sqrt{T}} + \Delta_{Ij, MNT} \right) + T\sqrt{M} \cdot T^{\max_i b_{EI, ij}} \right),
\end{eqnarray*}
where the second inequality follows from Assumption \ref{ass6}.1, Lemma \ref{lemma6_improved}.3, $\|\textbf{F}_E\|_2=O_P(\sqrt{M} \vee \sqrt{T})$, $\|\textbf{H}_{I,j}\|_F=O_P(1)$ and $\| \diag\{\bm{\Lambda}_{E,\bullet j}' \} \|_2 = O_P(\sqrt{\ln(MN)})$.

For the term involving $\textbf{B}_7$, write

\begin{eqnarray*} 
&& \frac{1}{T} \| \textbf{F}_{I, j}' \textbf{B}_{7} \widehat{\textbf{F}}_{I,j} \|_2
= \frac{1}{MT} \frac{1}{T} \| \textbf{F}_{I, j}' \textbf{F}_{E} \diag\{\bm{\Lambda}_{E,\bullet j}' \} \bm{\Lambda}_{I,\bullet j} \textbf{F}_{I,j}' \widehat{\textbf{F}}_{I,j} \|_2 \\
&\leq& \frac{1}{MT} \frac{1}{T} \| \textbf{F}_{I, j}' \textbf{F}_{E} \|_F \cdot \| \diag\{\bm{\Lambda}_{E,\bullet j}' \} \|_2 \cdot \| \bm{\Lambda}_{I,\bullet j} \|_F \cdot \| \textbf{F}_{I,j} \|_F \cdot \| \widehat{\textbf{F}}_{I,j} \|_F \\
&\le & O_P(1) \frac{1}{MT} \sqrt{M} \cdot T^{\max_i b_{EI, ij}} \cdot \sqrt{\ln (MN)} \cdot \sqrt{M} \cdot T \\
&=& O_P(1) \sqrt{\ln (MN)} \cdot T^{\max_i b_{EI, ij}},
\end{eqnarray*}
where the second inequality follows from Assumption \ref{ass6}.1, $\| \diag\{\bm{\Lambda}_{E,\bullet j}' \} \|_2 = O_P(\sqrt{\ln(MN)})$, $\| \bm{\Lambda}_{I,\bullet j} \|_F=O_P(\sqrt{M})$, $\| \textbf{F}_{I,j} \|_F=O_P(\sqrt{T})$ and $\| \widehat{\textbf{F}}_{I,j} \|_F=O_P(\sqrt{T})$.

Consider the term involving $\textbf{B}_8$, first write

\begin{eqnarray*} 
&& \frac{1}{M^2T^2} E \| \textbf{U}_{\bullet j \bullet}\textbf{F}_{I,j} \|_F^2
= \frac{1}{M^2T^2} \sum_{i=1}^{M} E \left\| \sum_{t=1}^{T} \bm{f}_{I,jt} u_{ijt} \right \|_F^2  \\
&=& \frac{1}{M^2T^2} \sum_{i=1}^{M} \sum_{t,s=1}^{T} E [ u_{ijt}u_{ijs} \bm{f}_{I,jt}'\bm{f}_{I,js} ] 
\leq O(1) \frac{1}{MT^2} \sum_{t,s=1}^{T} | E [u_{ijt}u_{ijs}] | = O\left(\frac{1}{MT}\right).
\end{eqnarray*}
Thus, $\frac{1}{MT}\| \textbf{U}_{\bullet j \bullet}\textbf{F}_{I,j} \|_F = O_P\left(\frac{1}{\sqrt{MT}}\right)$. It follows that

\begin{eqnarray*} 
&& \frac{1}{T} \| \textbf{F}_{I, j}' \textbf{B}_{8} \widehat{\textbf{F}}_{I,j} \|_2
= \frac{1}{MT} \frac{1}{T} \| \textbf{F}_{I, j}' \textbf{F}_{E} \diag\{\bm{\Lambda}_{E,\bullet j}' \} \textbf{U}_{\bullet j \bullet} \widehat{\textbf{F}}_{I,j} \|_2 \\
&\leq& \frac{1}{MT} \frac{1}{T} \| \textbf{F}_{I, j}' \textbf{F}_{E} \|_F \cdot \| \diag\{\bm{\Lambda}_{E,\bullet j}' \} \|_2 \cdot \left( \| \textbf{U}_{\bullet j \bullet} ( \widehat{\textbf{F}}_{I,j} - \textbf{F}_{I,j} \textbf{H}_{I,j} )\|_2 + \| \textbf{U}_{\bullet j \bullet} \textbf{F}_{I,j} \textbf{H}_{I,j} \|_F \right) \\
&\le & O_P(1) \frac{1}{MT} \sqrt{M} \cdot T^{\max_i b_{EI, ij}} \cdot \sqrt{\ln(MN)} \\
&&\cdot \left( (\sqrt{M} \vee \sqrt{T}) \sqrt{T} \left( \frac{\sqrt{\ln (MN)}}{ \sqrt{M} \wedge \sqrt{T}} + \Delta_{Ij, MNT} \right) + \sqrt{MT} \right),
\end{eqnarray*}
where the second inequality follows from Assumption \ref{ass6}.1, Lemma \ref{lemma6_improved}.3, $\|\textbf{U}_{\bullet j \bullet}\|_2=O_P(\sqrt{M} \vee \sqrt{T})$, $\|\textbf{H}_{I,j}\|_F=O_P(1)$ and $\| \diag\{\bm{\Lambda}_{E,\bullet j}' \} \|_2 = O_P(\sqrt{\ln(MN)})$.

For the term involving $\textbf{B}_{10}$, write

\begin{eqnarray*} 
&& \frac{1}{T} \| \textbf{F}_{I, j}' \textbf{B}_{10} \widehat{\textbf{F}}_{I,j} \|_2
= \frac{1}{MT} \frac{1}{T} \| \textbf{F}_{I,j}' \textbf{F}_{I,j} \bm{\Lambda}_{I,\bullet j}' \diag\{\bm{\Lambda}_{E,\bullet j}' \}' \textbf{F}_{E}' \widehat{\textbf{F}}_{I,j} \|_2 \\
&\leq& \frac{1}{MT} \frac{1}{T} \| \textbf{F}_{I,j} \|_F^2 \cdot \| \bm{\Lambda}_{I,\bullet j} \|_F \cdot \| \diag\{\bm{\Lambda}_{E,\bullet j}' \} \|_2 \cdot \left( \| \textbf{F}_{E}' ( \widehat{\textbf{F}}_{I,j} - \textbf{F}_{I,j} \textbf{H}_{I,j} )\|_2 + \| \textbf{F}_{E}' \textbf{F}_{I,j} \textbf{H}_{I,j} \|_F \right) \\
&\le & O_P(1) \frac{1}{MT} \cdot \sqrt{M} \cdot \sqrt{\ln(MN)} \\
&&\cdot \left( (\sqrt{M} \vee \sqrt{T}) \sqrt{T} \left( \frac{\sqrt{\ln (MN)}}{ \sqrt{M} \wedge \sqrt{T}} + \Delta_{Ij, MNT} \right) + T\sqrt{M} \cdot T^{\max_i b_{EI, ij}} \right) \\
&=& O_P(1) \frac{{\sqrt{\ln(MN)}}}{\sqrt{M} \wedge \sqrt{T}} \cdot \left( \frac{\sqrt{\ln (MN)}}{ \sqrt{M} \wedge \sqrt{T}} + \Delta_{Ij, MNT} \right) + O_P(1) \sqrt{\ln(MN)} \cdot T^{\max_i b_{EI, ij}}
\end{eqnarray*}
where the second inequality follows from Lemma \ref{lemma6_improved}.3 and Assumption \ref{ass6}.1.

For the terms involving $\textbf{B}_{12}$ and $\textbf{B}_{15}$, write

\begin{eqnarray*} 
&& \frac{1}{M^2T^2} E \| \bm{\Lambda}_{I,\bullet j}'  \textbf{U}_{\bullet j \bullet}\textbf{F}_{I,j} \|_F^2
= \frac{1}{M^2T^2} E \left\| \sum_{i=1}^{M} \sum_{t=1}^{T} \bm{f}_{I,jt} u_{ijt} \bm{\lambda}_{I,ij}' \right \|_F^2  \\
&=& \frac{1}{M^2T^2} \sum_{i,m=1}^{M} \sum_{t,s=1}^{T} E \| \bm{\lambda}_{I, ij} \bm{f}_{I,jt}'\bm{f}_{I,js} \bm{\lambda}_{I, mj}'  u_{ijt}u_{mjs} \|_F = O\left(\frac{1}{MT}\right),
\end{eqnarray*}
where the last equality follows from the mixing condition of Assumption \ref{ass3}.1. Thus, \\
$\frac{1}{MT}\| \bm{\Lambda}_{I,\bullet j}'  \textbf{U}_{\bullet j \bullet}\textbf{F}_{I,j} \|_F = O_P\left(\frac{1}{\sqrt{MT}}\right)$. It follows that

\begin{eqnarray*} 
&& \frac{1}{T} \| \textbf{F}_{I, j}' \textbf{B}_{12} \widehat{\textbf{F}}_{I,j} \|_2
= \frac{1}{MT} \frac{1}{T}\| \textbf{F}_{I, j}' \textbf{F}_{I,j} \bm{\Lambda}_{I,\bullet j}'  \textbf{U}_{\bullet j \bullet} \widehat{\textbf{F}}_{I,j} \|_2 \\
&\leq& \frac{1}{MT} \frac{1}{T}\| \textbf{F}_{I, j} \|_F^2 \cdot \|\bm{\Lambda}_{I,\bullet j}'  \textbf{U}_{\bullet j \bullet} \widehat{\textbf{F}}_{I,j} \|_2 \\
&\leq& \frac{1}{MT} \frac{1}{T}\| \textbf{F}_{I, j} \|_F^2 \cdot \left( \|\bm{\Lambda}_{I,\bullet j}'  \textbf{U}_{\bullet j \bullet} \|_F \cdot \| \widehat{\textbf{F}}_{I,j} - \textbf{F}_{I,j}\textbf{H}_{I,j} \|_2 + \|\bm{\Lambda}_{I,\bullet j}'  \textbf{U}_{\bullet j \bullet}\textbf{F}_{I,j} \textbf{H}_{I,j} \|_F \right) \\
&\leq& O_P(1) \frac{1}{MT} \left( \sqrt{MT} \sqrt{T} \left( \frac{\sqrt{\ln (MN)}}{ \sqrt{M} \wedge \sqrt{T}} + \Delta_{Ij, MNT} \right) + \sqrt{MT} \right) \\
&=& O_P(1) \left( \frac{\sqrt{\ln (MN)}}{M} + \frac{\sqrt{\ln (MN)}}{\sqrt{MT}} \right) + O_P \left( \frac{\Delta_{Ij, MNT}}{\sqrt{M}} \right),
\end{eqnarray*}
where the last inequality follows from $\|\bm{\Lambda}_{I,\bullet j}'  \textbf{U}_{\bullet j \bullet} \|_F=O_P(\sqrt{MT})$ and $\| \bm{\Lambda}_{I,\bullet j}'  \textbf{U}_{\bullet j \bullet}\textbf{F}_{I,j} \|_F = O_P\left(\sqrt{MT}\right)$. Similarly,

\begin{eqnarray*} 
&& \frac{1}{T} \| \textbf{F}_{I, j}' \textbf{B}_{15} \widehat{\textbf{F}}_{I,j} \|_2
= \frac{1}{MT} \frac{1}{T}\| \textbf{F}_{I, j}' \textbf{U}_{\bullet j \bullet}' \bm{\Lambda}_{I,\bullet j} \textbf{F}_{I,j}' \widehat{\textbf{F}}_{I,j} \|_2 \\
&\leq& \frac{1}{MT} \frac{1}{T}\| \textbf{F}_{I, j}' \textbf{U}_{\bullet j \bullet}' \bm{\Lambda}_{I,\bullet j} \|_F \cdot \| \textbf{F}_{I,j} \|_F \cdot \| \widehat{\textbf{F}}_{I,j} \|_F =  O_P\left( \frac{1}{\sqrt{MT}} \right),
\end{eqnarray*}
where the first inequality follows from $\| \bm{\Lambda}_{I,\bullet j}'  \textbf{U}_{\bullet j \bullet}\textbf{F}_{I,j} \|_F = O_P\left(\sqrt{MT}\right)$.

For the term involving $\textbf{B}_{14}$, write

\begin{eqnarray*} 
&& \frac{1}{T} \| \textbf{F}_{I, j}' \textbf{B}_{14} \widehat{\textbf{F}}_{I,j} \|_2
= \frac{1}{MT} \frac{1}{T} \| \textbf{F}_{I, j}' \textbf{U}_{\bullet j \bullet}' \diag\{\bm{\Lambda}_{E,\bullet j}' \}' \textbf{F}_{E}' \widehat{\textbf{F}}_{I,j} \|_2 \\
&\leq& \frac{1}{MT} \frac{1}{T} \| \textbf{F}_{I, j}' \textbf{U}_{\bullet j \bullet}' \|_F \cdot \| \diag\{\bm{\Lambda}_{E,\bullet j}' \} \|_2 \cdot \left( \| \textbf{F}_{E}' ( \widehat{\textbf{F}}_{I,j} - \textbf{F}_{I,j} \textbf{H}_{I,j} )\|_2 + \| \textbf{F}_{E}' \textbf{F}_{I,j} \textbf{H}_{I,j} \|_F \right) \\
&\le & O_P(1) \frac{1}{MT}  \frac{1}{T} \sqrt{MT} \cdot \sqrt{\ln(MN)} \\
&&\cdot \left( (\sqrt{M} \vee \sqrt{T}) \sqrt{T} \left( \frac{\sqrt{\ln (MN)}}{ \sqrt{M} \wedge \sqrt{T}} + \Delta_{Ij, MNT} \right) + T\sqrt{M} \cdot T^{\max_i b_{EI, ij}} \right),
\end{eqnarray*}
where the second inequality follows from Assumption \ref{ass6}.1, Lemma \ref{lemma6_improved}.3, $\|\textbf{U}_{\bullet j \bullet}\|_2=O_P(\sqrt{M} \vee \sqrt{T})$, $\|\textbf{H}_{I,j}\|_F=O_P(1)$, $\| \diag\{\bm{\Lambda}_{E,\bullet j}' \} \|_2 = O_P(\sqrt{\ln(MN)})$ and $\| \textbf{U}_{\bullet j \bullet} \textbf{F}_{I,j}\|_F=O_P(\sqrt{MT})$.

For the term involving $\textbf{B}_{16}$, write
\begin{eqnarray*} 
&& \frac{1}{T} \| \textbf{F}_{I, j}' \textbf{B}_{16} \widehat{\textbf{F}}_{I,j} \|_2 
= \frac{1}{MT} \frac{1}{T} \| \textbf{F}_{I, j}' \textbf{U}_{\bullet j \bullet}^{\prime} \textbf{U}_{\bullet j \bullet} \widehat{\textbf{F}}_{I,j} \|_2 \\
&\leq& \frac{1}{MT} \frac{1}{T} \| \textbf{F}_{I, j}' \textbf{U}_{\bullet j \bullet}^{\prime} \|_F \cdot \left( \| \textbf{U}_{\bullet j \bullet} ( \widehat{\textbf{F}}_{I,j} - \textbf{F}_{I,j}\textbf{H}_{I,j} )\|_2 + \| \textbf{U}_{\bullet j \bullet} \textbf{F}_{I,j}\textbf{H}_{I,j} \|_F \right) \\
&\leq& O_P(1) \frac{1}{MT} \frac{1}{T} \sqrt{MT} \cdot \left( (\sqrt{M} \vee \sqrt{T}) \sqrt{T} \left( \frac{\sqrt{\ln (MN)}}{ \sqrt{M} \wedge \sqrt{T}} + \Delta_{Ij, MNT} \right) + \sqrt{MT}  \right) \\
&=& O_P \left( \frac{\sqrt{\ln(MN)}}{M\sqrt{T}} + \frac{1}{T} + \frac{ \Delta_{Ij, MNT} }{T \wedge \sqrt{MT}} \right),
\end{eqnarray*}
where the second inequality follows from Lemma \ref{lemma6_improved}.3, $\|\textbf{U}_{\bullet j \bullet}\|_2 = O_P(\sqrt{M} \vee \sqrt{T})$, $\textbf{H}_{I,j} = O_P(1)$ and $\| \textbf{U}_{\bullet j \bullet} \textbf{F}_{I,j}\|_F=O_P(\sqrt{MT})$.

Based on the above development, we obtain that
\begin{eqnarray*} 
\frac{1}{T} \| \textbf{F}_{I, j}' ( \widehat{\textbf{F}}_{I,j} - \textbf{F}_{I, j} \textbf{H}_{I,j} ) \|_{F} = O_P\left( \frac{\ln(MN)}{M} + \frac{1}{\sqrt{MN}} + \frac{1}{\sqrt{T}} + \Delta_{Ij, MNT} + \sqrt{\ln(MN)} \cdot T^{\max_i b_{EI, ij}} \right). 
\end{eqnarray*}

\medskip

(2). The second result can be proved in exactly the same way as the first result. Thus, omitted.

\medskip

(3)-(4). Given the first result of this lemma, the proof can be done in exactly the same way as in Lemma \ref{lemma2_improved}, thus omitted. \hspace*{\fill}{$\blacksquare$}

\bigskip

\noindent \textbf{Proof of Theorem \ref{theorem_global_clt}}:

(1). Let $\upsilon_{ijt} = \bm{\lambda}_{E, ij}' \bm{f}_{E, it} + \bm{\lambda}_{I,ij}' \bm{f}_{I,j t} + u_{ijt}$. In matrix notation, we have $\bm{\Upsilon} = \bm{\Lambda}_E\textbf{F}_E'+ \bm{\Lambda}_I\textbf{F}_I' + \textbf{U}$. Then $\textbf{Y} = \bm{\Gamma}\textbf{G}' + \bm{\Upsilon}$. If $\frac{\sqrt{MN}}{T} \to 0$,  we have

\begin{eqnarray*}
&&\sqrt{MN} (\widehat{\bm{g}}_t - \textbf{H}' \bm{g}_t)  
= \textbf{V}_g^{-1} \cdot \frac{\widehat{\textbf{G}}'\textbf{G}}{T}\cdot \frac{1}{\sqrt{MN}} \sum_{i=1}^{M} \sum_{j=1}^{N} \bm{\gamma}_{ij} \upsilon_{ijt} + o_P(1) \\
&=& \textbf{H}' \left( \frac{\bm{\Gamma}'\bm{\Gamma}}{MN} \right)^{-1} \frac{1}{\sqrt{MN}} \sum_{i=1}^{M} \sum_{j=1}^{N} \bm{\gamma}_{ij} \upsilon_{ijt} + o_P(1) \\
&=& \left( \frac{\bm{\Gamma}'\bm{\Gamma}}{MN} \right)^{-1} \frac{1}{\sqrt{MN}} \sum_{i=1}^{M} \sum_{j=1}^{N} \bm{\gamma}_{ij} \upsilon_{ijt} + o_P(1)
\end{eqnarray*}
where the second equality follows from the fact that
\begin{eqnarray*}
\textbf{V}_g^{-1}\left(\frac{\widehat{\textbf{G}}'\textbf{G}}{T}\right) = \textbf{V}_g^{-1}\left(\frac{\widehat{\textbf{G}}'\textbf{G}}{T}\right) \left( \frac{\bm{\Gamma}'\bm{\Gamma}}{MN} \right) \left( \frac{\bm{\Gamma}'\bm{\Gamma}}{MN} \right)^{-1} = \textbf{H}' \left( \frac{\bm{\Gamma}'\bm{\Gamma}}{MN} \right)^{-1},
\end{eqnarray*}
and the third equality follows from Lemma \ref{lemma2_improved}. 

Note that

\begin{eqnarray*}
&& \sqrt{MN} (\widehat{\bm{g}}_t - \bm{g}_t) = \sqrt{MN} (\widehat{\bm{g}}_t - \textbf{H}' \bm{g}_t) + \sqrt{MN} (\textbf{H}' - \textbf{I}_{r_g}) \bm{g}_t \\
&=& \sqrt{MN} (\widehat{\bm{g}}_t - \textbf{H}' \bm{g}_t) + \sqrt{MN} \cdot O_P \left( \frac{1}{MN} + \frac{1}{T} + \Delta_{g, MNT}^{*} \right)\\
&=& \left( \frac{\bm{\Gamma}'\bm{\Gamma}}{MN} \right)^{-1} \frac{1}{\sqrt{MN}} \sum_{i=1}^{M} \sum_{j=1}^{N} \bm{\gamma}_{ij} \upsilon_{ijt} + o_P(1),
\end{eqnarray*}
if $\frac{\sqrt{MN}}{T} \to 0$ and $\sqrt{MN} \cdot \Delta_{g, MNT}^{*} \to 0$. Then by Assumption \ref{ass2}.1 and Assumption \ref{ass5}.3,  the result follows immediately. 

\medskip

(2). Let $\upsilon_{I,ijt}=\bm{\gamma}_{ij}' \bm{g}_t - \widehat{\bm{\gamma}}_{ij}' \widehat{\bm{g}}_t + \bm{\lambda}_{E, ij}' \bm{f}_{E, it} + u_{ijt}$ for $i=1,\ldots,M$. In matrix notation, we have $\bm{\Upsilon}_{I,j} = \bm{\Gamma}_{I, j} \textbf{G}' - \widehat{\bm{\Gamma}}_{I, j} \widehat{\textbf{G}}' + \diag\{\bm{\Lambda}_{E,\bullet j}' \}' \textbf{F}_{E}' + \textbf{U}_{\bullet j \bullet}$. Then $\textbf{Y}_{I,j} = \bm{\Lambda}_{I,\bullet j} \textbf{F}_{I,j}' + \bm{\Upsilon}_{I,j}$. If $\frac{\sqrt{M}}{T} \to 0$, we have

\begin{eqnarray*}
&&\sqrt{M} (\widehat{\bm{f}}_{I,jt} - \textbf{H}_{I,j}' \bm{f}_{I,jt}) 
= \textbf{V}_{I,j}^{-1} \left(\frac{\widehat{\textbf{F}}_{I,j}'\textbf{F}_{I,j}}{T}\right) \frac{1}{\sqrt{M}} \sum_{i=1}^{M} \bm{\lambda}_{I,ij} \upsilon_{I,ijt} + o_P(1) \\
&=& \textbf{H}_{I,j}' \left( \frac{\bm{\Lambda}_{I, \bullet j}'\bm{\Lambda}_{I, \bullet j}}{M} \right)^{-1} \frac{1}{\sqrt{M}} \sum_{i=1}^{M} \bm{\lambda}_{I,ij} \upsilon_{I,ijt} + o_P(1) \\
&=& \left( \frac{\bm{\Lambda}_{I, \bullet j}'\bm{\Lambda}_{I, \bullet j}}{M} \right)^{-1} \frac{1}{\sqrt{M}} \sum_{i=1}^{M} \bm{\lambda}_{I,ij} \upsilon_{I,ijt} + o_P(1) \\
&=& \left( \frac{\bm{\Lambda}_{I, \bullet j}'\bm{\Lambda}_{I, \bullet j}}{M} \right)^{-1} \frac{1}{\sqrt{M}} \sum_{i=1}^{M} \bm{\lambda}_{I,ij} (\bm{\gamma}_{ij}' \bm{g}_t - \widehat{\bm{\gamma}}_{ij}' \widehat{\bm{g}}_t + \bm{\lambda}_{E, ij}' \bm{f}_{E, it} + u_{ijt} ) + o_P(1) \\
&=& \left( \frac{\bm{\Lambda}_{I, \bullet j}'\bm{\Lambda}_{I, \bullet j}}{M} \right)^{-1} \frac{1}{\sqrt{M}} \left( \sum_{i=1}^{M} \bm{\lambda}_{I,ij} ( \bm{\lambda}_{E, ij}' \bm{f}_{E, it} + u_{ijt} ) + \sum_{i=1}^{M} \bm{\lambda}_{I,ij}(\bm{\gamma}_{ij}' \bm{g}_t - \widehat{\bm{\gamma}}_{ij}' \widehat{\bm{g}}_t ) \right) + o_P(1) \\
&=& \left( \frac{\bm{\Lambda}_{I, \bullet j}'\bm{\Lambda}_{I, \bullet j}}{M} \right)^{-1} \frac{1}{\sqrt{M}} \sum_{i=1}^{M} \bm{\lambda}_{I,ij} (\bm{\lambda}_{E, ij}' \bm{f}_{E, it} + u_{ijt} ) \\
&&+ \left( \frac{\bm{\Lambda}_{I, \bullet j}'\bm{\Lambda}_{I, \bullet j}}{M} \right)^{-1} \frac{1}{\sqrt{M}} \left( M \cdot O_P \left(\frac{1}{ \sqrt{MN} } + \frac{1}{\sqrt{T}} + \Delta_{Ij, MNT} \right) \right) + o_P(1),
\end{eqnarray*}
where the second equality follows from the fact that
\begin{eqnarray*}
&&\textbf{V}_{I,j}^{-1}\left(\frac{\widehat{\textbf{F}}_{I,j}'\textbf{F}_{I,j}}{T}\right) = \textbf{V}_{I,j}^{-1}\left(\frac{\widehat{\textbf{F}}_{I,j}'\textbf{F}_{I,j}}{T}\right) \left( \frac{\bm{\Lambda}_{I, \bullet j}'\bm{\Lambda}_{I, \bullet j}}{M} \right) \left( \frac{\bm{\Lambda}_{I, \bullet j}'\bm{\Lambda}_{I, \bullet j}}{M} \right)^{-1} 
= \textbf{H}_{I,j}' \left( \frac{\bm{\Lambda}_{I, \bullet j}'\bm{\Lambda}_{I, \bullet j}}{M} \right)^{-1},
\end{eqnarray*}
the third equality follows from $\textbf{H}_{I,j} = \textbf{I}_{r_{I,j}} + o_P(1)$, and the last equality follows from Lemma  \ref{lemma6_improved}.1. Thus,
\begin{eqnarray*}
&& \sqrt{M} (\widehat{\bm{f}}_{I,jt} - \bm{f}_{I,jt}) = \sqrt{M} (\widehat{\bm{f}}_{I,jt} - \textbf{H}_{I,j}' \bm{f}_{I,jt}) + \sqrt{M} (\textbf{H}_{I,j}' - \textbf{I}_{r_{I,j}}) \bm{f}_{I,jt} \\
&=& \sqrt{M} (\widehat{\bm{f}}_{I,jt} - \textbf{H}_{I,j}' \bm{f}_{I,jt}) \\
&& + \sqrt{M} \cdot O_P \left( \frac{\ln(MN)}{M} + \frac{1}{\sqrt{MN}} + \frac{1}{\sqrt{T}} + \Delta_{Ij, MNT} + \sqrt{\ln(MN)} \cdot T^{\max_i c_{EI, ij}} \right) \\
&=& \left( \frac{\bm{\Lambda}_{I, \bullet j}'\bm{\Lambda}_{I, \bullet j}}{M} \right)^{-1} \frac{1}{\sqrt{M}} \sum_{i=1}^{M} \bm{\lambda}_{I,ij} ( \bm{\lambda}_{E, ij}' \bm{f}_{E, it} + u_{ijt} ) + o_P(1),
\end{eqnarray*}
if $\frac{\ln(MN)}{\sqrt{M}} \to 0$, $\frac{\sqrt{M}}{\sqrt{T}} \to 0$ and $\sqrt{M} \left(\Delta_{Ij, MNT} + \sqrt{\ln(MN)} \cdot T^{\max_i c_{EI, ij}} \right) \to 0$.
Then by Assumption \ref{ass4}.1(c) and Assumption \ref{ass6}.4(a), the result follows immediately.

\medskip

(3). The second result can be proved in exactly the same way as the first result. Thus, omitted. \hspace*{\fill}{$\blacksquare$}


\section*{Appendix B}

\renewcommand{\theequation}{B.\arabic{equation}}
\renewcommand{\thesection}{B.\arabic{section}}
\renewcommand{\thefigure}{B.\arabic{figure}}
\renewcommand{\thetable}{B.\arabic{table}}
\renewcommand{\thelemma}{B.\arabic{lemma}}
\renewcommand{\theassumption}{B.\arabic{assumption}}
\renewcommand{\thetheorem}{B.\arabic{theorem}}

\setcounter{equation}{0}
\setcounter{lemma}{0}
\setcounter{section}{0}
\setcounter{table}{0}
\setcounter{figure}{0}
\setcounter{assumption}{0}

\small

In this Appendix, we provide the secondary lemmas and the corresponding proofs.

\section{Preliminary Lemmas}\label{AppendixB.2}

\begin{lemma} \label{lemma_nof0}
Suppose that $\normalfont\textbf{A}$ and $\normalfont\textbf{A}+\textbf{E}$ are $n \times n$ symmetric matrices and that $\normalfont\textbf{Q}=(\textbf{Q}_1, \textbf{Q}_2)$, where $\normalfont\textbf{Q}_1$ is $n \times r$ and $\normalfont\textbf{Q}_2$ is $n\times(n-r)$, is an orthogonal matrix such that $\normalfont\mbox{span}(\textbf{Q}_1)$ is an invariant subspace for $\normalfont\textbf{A}$; that is, $\normalfont\textbf{A} \times \mbox{span}(\textbf{Q}_1) \subset \mbox{span}(\textbf{Q}_1)$. Decompose $\normalfont\textbf{Q}'\textbf{A}\textbf{Q}$ and $\normalfont\textbf{Q}'\textbf{E}\textbf{Q}$ as $\normalfont\textbf{Q}'\textbf{A}\textbf{Q}=\mbox{diag}(\textbf{D}_1, \textbf{D}_2)$ and

\begin{equation*}
\normalfont\textbf{Q}'\textbf{E}\textbf{Q} = 
\begin{pmatrix}
\textbf{E}_{11} & \textbf{E}_{21}^{\prime} \\
\textbf{E}_{21} & \textbf{E}_{22}  \\
\end{pmatrix}.
\end{equation*}
Let 
\begin{equation*}
\normalfont\mbox{sep}(\textbf{D}_1, \textbf{D}_2) = \min_{ \lambda_1 \in \lambda(\textbf{D}_1), \mbox{ } \lambda_2 \in \lambda(\textbf{D}_2) } |\lambda_1 - \lambda_2|,
\end{equation*}
where $\lambda(\normalfont\textbf{B})$ denotes the set of eigenvalues of the matrix $\normalfont\textbf{B}$. If $\normalfont\mbox{sep}(\textbf{D}_1, \textbf{D}_2) > 0$ and $\normalfont\|\textbf{E}\|_2 \leq \mbox{sep}(\textbf{D}_1, \textbf{D}_2)/5$, then there exists a $(n-r)\times r$ matrix $\normalfont\textbf{P}$ with $\normalfont\|\textbf{P}\|_2 \leq 4\|\textbf{E}_{21}\|_2/\mbox{sep}(\textbf{D}_1, \textbf{D}_2)$, such that the columns of $\normalfont\textbf{Q}_1^{0} = (\textbf{Q}_1 + \textbf{Q}_2\textbf{P})(\textbf{I}_r + \textbf{P}'\textbf{P})^{-1/2}$ define an orthonormal basis for a subspace that is invariant for $\normalfont\textbf{A}+\textbf{E}$.
\end{lemma}

\begin{lemma} \label{lemma2}
Under Assumptions \ref{ass1}-\ref{ass3}, as $(M,N,T) \to (\infty,\infty,\infty)$,

\begin{enumerate}
\item $\normalfont  \frac{1}{MNT} \|\textbf{F}_E \bm{\Lambda}_E' \bm{\Lambda}_E \textbf{F}_E'\|_2 =O_P \left(  \ln(MN) \cdot (M^{-1} \vee T^{-1}) \right)$, 
\item $\normalfont \frac{1}{MNT} \| \textbf{F}_I\bm{\Lambda}_I^{\prime} \bm{\Lambda}_I \textbf{F}_I^{\prime}  \|_2 = O_P \left(\ln(MN) \cdot( N^{-1} \vee T^{-1} )\right)$, 
\item $\normalfont \frac{1}{MNT}\|\textbf{U}'\textbf{U} \|_2 = O_P \left( T^{-1/2} \vee (MN)^{-1/2} \right)$, 
\item $ \normalfont \frac{1}{M N T} \| \textbf{G} \bm{\Gamma}' \bm{\Lambda}_E \textbf{F}_E' \|_2 = O_P \left(  \sqrt{\ln(MN)}\cdot(M^{-1/2} \vee T^{-1/2} )\right)$, 
\item $\normalfont \frac{1}{M N T} \| \textbf{G} \bm{\Gamma}' \bm{\Lambda}_I \textbf{F}_I'  \|_2 =O_P \left( \sqrt{\ln(MN)} \cdot (N^{-1/2} \vee T^{-1/2}) \right) $, 
\item $\normalfont \frac{1}{MNT} \| \textbf{G} \bm{\Gamma}' \textbf{U} \|_2 = O_P \left(\frac{1}{\sqrt{MN}}\right) $, 
\item $\normalfont \frac{1}{MNT} \| \textbf{F}_{E} \bm{\Lambda}_E' \bm{\Lambda}_I \textbf{F}_{I}' \|_2 = O_P \left( \frac{ (\sqrt{T} \vee \sqrt{M}) \cdot (\sqrt{T} \vee \sqrt{N}) \cdot \ln  (MN)}{\sqrt{MN}T} \right) $, 
\item $\normalfont \frac{1}{M N T} \| \textbf{F}_{E} \bm{\Lambda}_E' \textbf{U} \|_2 = O_P \left( \frac{ (\sqrt{T} \vee \sqrt{M}) \cdot (\sqrt{T} \vee \sqrt{MN}) \cdot \sqrt{\ln(MN)} }{M\sqrt{N}T} \right)$, 
\item $\normalfont \frac{1}{M N T} \| \textbf{F}_{I} \bm{\Lambda}_I' \textbf{U} \|_2 =O_P \left( \frac{ (\sqrt{T} \vee \sqrt{N}) \cdot (\sqrt{T} \vee \sqrt{MN}) \cdot \sqrt{\ln(MN)} }{\sqrt{M}NT} \right)$.
\end{enumerate}
\end{lemma}

Having established Theorem \ref{theorem1}, we suppose that the number of global factors has been successfully identified. In order to keep the notation simple, for the following lemmas we suppress the dagger superscript in $\widehat{\textbf{G}}^{\dag}$ and $\textbf{V}_{g}^{\dag}$.

\begin{lemma} \label{lemma3}
Under Assumptions \ref{ass1}-\ref{ass3}, as $(M,N,T) \to (\infty,\infty,\infty)$,

\begin{enumerate}
\item $\normalfont\frac{1}{T} \| \textbf{G}' (\widehat{\textbf{G}} - \textbf{G} \textbf{H}) \|_{F} = O_P \left( \frac{\sqrt{\ln(MN)}}{\min \{\sqrt{M}, \sqrt{N}, \sqrt{T}\}} \right)$, 

\item $\normalfont\frac{1}{T} \| \widehat{\textbf{G}}' (\widehat{\textbf{G}} - \textbf{G} \textbf{H}) \|_{F} =  O_P \left( \frac{\sqrt{\ln(MN)}}{\min \{\sqrt{M}, \sqrt{N}, \sqrt{T}\}} \right)$, 

\item $\normalfont\| \textbf{P}_{\widehat{\textbf{G}}} - \textbf{P}_{\textbf{G}} \|_F^{2} = O_P \left( \frac{\sqrt{\ln(MN)}}{\min \{\sqrt{M}, \sqrt{N}, \sqrt{T}\}} \right)$, 
where $\normalfont\textbf{H} $ has been defined in Lemma \ref{lemma_nof1}.
\end{enumerate}

\end{lemma}

\begin{lemma}\label{lemma6}
Under Assumptions \ref{ass1}-\ref{ass4}, as $(M,N,T) \to (\infty,\infty,\infty)$, 

\begin{enumerate}
\item for $j=1,\ldots, N$, $\normalfont\frac{1}{\sqrt{MT}}\| \bm{\Gamma}_{I,j} \textbf{G}' - \widehat{\bm{\Gamma}}_{I,j}\widehat{\textbf{G}}' \|_F = O_P\left( \frac{\sqrt{\ln(MN)}}{ \min \{\sqrt{M}, \sqrt{N}, \sqrt{T}\}}+T^{a_{I,j}}   \right)$, where $\bm{\Gamma}_{I,j}$ is denoted in the same way as $\widehat{\bm{\Gamma}}_{I,j}$;

\item for $i=1,\ldots, M$, $\normalfont\frac{1}{\sqrt{NT}}\| \bm{\Gamma}_{E,i} \textbf{G}' - \widehat{\bm{\Gamma}}_{E,i}\widehat{\textbf{G}}' \|_F = O_P\left( \frac{\sqrt{\ln(MN)}}{ \min \{\sqrt{M}, \sqrt{N}, \sqrt{T}\}}+T^{a_{E,i}}   \right)$, where $\bm{\Gamma}_{E,i}$ be denoted in the same way as $\widehat{\bm{\Gamma}}_{E,i}$.
\end{enumerate}
 
\end{lemma}

\begin{lemma} \label{lemma8}
Under Assumptions \ref{ass1}-\ref{ass4}, as $(M,N,T) \to (\infty,\infty,\infty)$, 

\begin{enumerate}
\item $\normalfont \frac{1}{MT} \|(\bm{\Gamma}_{I,j} \textbf{G}' - \widehat{\bm{\Gamma}}_{I,j} \widehat{\textbf{G}}' )' \diag\{\bm{\Lambda}_{E,\bullet j}' \}'\textbf{F}_{E}' \|_2=O_P(1)\frac{\sqrt{\ln(MN)}}{\sqrt{M} \wedge \sqrt{T}}\cdot\left( \frac{\sqrt{\ln(MN)}}{ \min \{\sqrt{M}, \sqrt{N}, \sqrt{T}\}}+T^{a_{I,j}}   \right)  $,

\item $\normalfont \frac{1}{MT}\| (\bm{\Gamma}_{I,j} \textbf{G}' - \widehat{\bm{\Gamma}}_{I,j} \widehat{\textbf{G}}' )' \bm{\Lambda}_{I,\bullet j} \textbf{F}_{I,j}' \|_2 =O_P\left( \frac{\sqrt{\ln(MN)}}{ \min \{\sqrt{M}, \sqrt{N}, \sqrt{T}\}}+T^{a_{I,j}}   \right)$, 

\item $\normalfont \frac{1}{MT} \|(\bm{\Gamma}_{I,j} \textbf{G}' - \widehat{\bm{\Gamma}}_{I,j} \widehat{\textbf{G}}' )' \textbf{U}_{\bullet j\bullet} \|_2=O_P(1) \left( \frac{1}{\sqrt{M} \wedge \sqrt{T}} \right) \left( \frac{\sqrt{\ln(MN)}}{ \min \{\sqrt{M}, \sqrt{N}, \sqrt{T}\}}+T^{a_{I,j}}   \right)$,
 
\item $\normalfont \frac{1}{MT}\|  \textbf{F}_{E} \diag\{\bm{\Lambda}_{E,\bullet j}' \} \diag\{\bm{\Lambda}_{E,\bullet j}' \}'\textbf{F}_{E}' \|_2 = O_P \left(\frac{\ln (MN)}{M\wedge T} \right)$, 

\item $\normalfont \frac{1}{MT} \|\textbf{F}_{E} \diag\{\bm{\Lambda}_{E,\bullet j}' \} \bm{\Lambda}_{I,\bullet j} \textbf{F}_{I,j}'\|_2 = O_P\left(\frac{\sqrt{\ln (MN)}}{ \sqrt{M} \wedge \sqrt{T}} \right)$, 

\item  $\normalfont \frac{1}{MT}\| \textbf{F}_{E} \diag\{\bm{\Lambda}_{E,\bullet j}' \} \textbf{U}_{\bullet j \bullet}\|_2= O_P\left(\frac{\sqrt{\ln (MN)}}{ M \wedge T} \right) $, 

\item  $\normalfont \frac{1}{MT}\| (\bm{\Gamma}_{I,j} \textbf{G}' - \widehat{\bm{\Gamma}}_{I,j} \widehat{\textbf{G}}' )' (\bm{\Gamma}_{I,j} \textbf{G}' - \widehat{\bm{\Gamma}}_{I,j} \widehat{\textbf{G}}' )\|_2 =O_P\left( \frac{\ln(MN)}{ \min \{M, N, T\}}+T^{2a_{I,j}}   \right)$, 

\item  $\normalfont \frac{1}{MT} \|\textbf{F}_{I,j} \bm{\Lambda}_{I,\bullet j}'  \textbf{U}_{\bullet j \bullet}\|_2=O_P\left(\frac{1}{\sqrt{M}} \right)$, 

\item  $\normalfont\frac{1}{MT} \| \textbf{U}_{\bullet j \bullet}^{\prime} \textbf{U}_{\bullet j \bullet} \|_2 =O_P\left(\frac{1}{\sqrt{M}} \vee \frac{1}{\sqrt{T}} \right)$.
\end{enumerate}

\end{lemma}

\section{Proofs of Lemma \ref{lemma2} to Lemma \ref{lemma8}}\label{ProofofLB}

\noindent \textbf{Proof of Lemma \ref{lemma_nof0}}: 

This is Lemma 3 of \cite{LYB2011}. The proof is therefore omitted. \hspace*{\fill}{$\blacksquare$}

\bigskip

\noindent \textbf{Proof of Lemma \ref{lemma2}}:

(1). Write

\begin{eqnarray*}
&&\frac{1}{MNT}\| \textbf{F}_E \bm{\Lambda}_E^{\prime} \bm{\Lambda}_E \textbf{F}_E^{\prime} \|_2 
\le \frac{1}{MNT}\| \textbf{F}_{E}\|_2^2 \cdot \| \bm{\Lambda}_E \|_2^2 \\
&=& \frac{1}{MNT}\| \textbf{F}_{E}\|_2^2 \cdot \lambda_{\max} \left\{ \bm{\Lambda}_E'\bm{\Lambda}_E \right\}
= \frac{1}{MNT}\| \textbf{F}_E\|_2^2 \cdot \lambda_{\max} \left\{ \sum_{j=1}^N \diag\{\bm{\Lambda}_{E,\bullet j}' \} \diag\{\bm{\Lambda}_{E,\bullet j}' \}' \right\}   \\
&\le& \frac{1}{MNT} \|\textbf{F}_{E}\|_2^2 \cdot  \sum_{j=1}^N\lambda_{\max}\left\{ \diag\{\bm{\Lambda}_{E,\bullet j}' \} \diag\{\bm{\Lambda}_{E,\bullet j}' \}' \right\} \\
&=& \frac{1}{MNT}\| \textbf{F}_{E}\|_2^2 \cdot \sum_{j=1}^N \max_{1\le i\le M} \lambda_{\max}\left\{ \bm{\lambda}_{E,ij} \bm{\lambda}_{E,ij}^{\prime} \right\}
= \frac{1}{MNT}\| \textbf{F}_{E}\|_2^2 \cdot \sum_{j=1}^N \max_{1\le i\le M} \lambda_{\max}\left\{ \bm{\lambda}_{E,ij}^{\prime} \bm{\lambda}_{E,ij} \right\} \\
&=& \frac{1}{MNT}\| \textbf{F}_{E}\|_2^2 \cdot \sum_{j=1}^N \max_{1\le i\le M} \| \bm{\lambda}_{E,ij} \|_F^2
\le \frac{1}{M T}\| \textbf{F}_{E}\|_2^2 \cdot \max_{1\le i\le M, 1\le j\le N}\| \bm{\lambda}_{E,ij}\|_F^2  \\
&=& \frac{1}{MT} \cdot O_P(T\vee M) \cdot O_P(\ln(MN)) = \ln(MN) \cdot O_P \left( M^{-1} \vee T^{-1} \right),
\end{eqnarray*}
where the second equality follows from the definitions under \eqref{Eq2.3}, and the last equality follows from Assumption \ref{ass1}.2 and Assumption \ref{ass2}.2.

\medskip

(2). Write

\begin{eqnarray*}
&&\frac{1}{MNT}\|\textbf{F}_I\bm{\Lambda}_I^{\prime} \bm{\Lambda}_I \textbf{F}_I^{\prime} \|_2 
\leq \frac{1}{MNT}\|\textbf{F}_I\|_2^2 \cdot \| \bm{\Lambda}_I \|_2^2 \\
&=& \frac{1}{MNT}\|\textbf{F}_I\|_2^2 \cdot \lambda_{\max} \{ \bm{\Lambda}_I^{\prime} \bm{\Lambda}_I \} 
= \frac{1}{MNT}\|\textbf{F}_I\|_2^2 \cdot \lambda_{\max} \left \{ \diag\{\bm{\Lambda}_{I, \bullet 1}^{\prime} \bm{\Lambda}_{I, \bullet 1}, \ldots, \bm{\Lambda}_{I, \bullet N}^{\prime} \bm{\Lambda}_{I, \bullet N} \} \right \} \\
&=& \frac{1}{MNT} \|\textbf{F}_I\|_2^2 \cdot\max_{1\leq j\leq N} \lambda_{\max} \left \{ \bm{\Lambda}_{I,\bullet j}^{\prime} \bm{\Lambda}_{I,\bullet j} \right \}
= \frac{1}{MNT} \|\textbf{F}_I\|_2^2 \cdot\max_{1\leq j\leq N} \lambda_{\max} \left \{ \sum_{i=1}^{M} \bm{\lambda}_{I,ij}\bm{\lambda}_{I,ij}^{\prime} \right\}  \\
&\leq& \frac{1}{MNT}\| \mathbf{F}_I\|_2^2 \cdot  \max_{1\leq j\leq N}\sum_{i=1}^M \lambda_{\max} \left\{ \bm{\lambda}_{I,ij} \bm{\lambda}_{I,ij}^{\prime} \right\} 
= \frac{1}{MNT}\| \mathbf{F}_I\|_2^2 \cdot  \max_{1\leq j\leq N}\sum_{i=1}^M \lambda_{\max} \left\{ \bm{\lambda}_{I,ij}^{\prime} \bm{\lambda}_{I,ij} \right\} \\
&=& \frac{1}{MNT}\| \mathbf{F}_I\|_2^2 \cdot  \max_{1\leq j\leq N}\sum_{i=1}^M \| \bm{\lambda}_{I,ij}\|_F^2
\le \frac{1}{N T} \| \textbf{F}_{I}\|_2^2 \cdot \max_{1\le i\le M, 1\le j\le N}\| \bm{\lambda}_{I,ij}\|_F^2    \\
&=& \ln(MN) \cdot O_P \left( N^{-1} \vee T^{-1} \right),
\end{eqnarray*} 
where the second equality follows from the definitions under \eqref{Eq2.3}, and the last equality follows from  Assumption \ref{ass1}.2 and Assumption \ref{ass2}.2.

\medskip

(3) Write

\begin{eqnarray*}
&&\frac{1}{M^2 N^2 T^2} E \| \textbf{U}'\textbf{U} \|_{F}^2 = \frac{1}{M^2 N^2 T^2} \sum_{t,s=1}^{T} \sum_{i,m=1}^{M} \sum_{j,n=1}^{N} E [u_{ijt} u_{mnt} u_{ijs} u_{mns}] \\
&=& \frac{1}{M^2 N^2 T^2} \sum_{t,s=1}^{T} \bigg(\sum_{i=1}^{M}\sum_{j=1}^{N} E [u_{ijt}^2 u_{ijs}^2] + \sum_{(i,j)\ne (m,n)} E[(u_{ijt} u_{mnt}-\sigma_{ijmn}) (u_{ijs} u_{mns}-\sigma_{ijmn})] \bigg) \\
&&+\frac{1}{M^2N^2} \sum_{(i,j)\ne (m,n)} \sigma_{ijmn}^2 \\
&=& \frac{1}{M^2 N^2 T^2} \sum_{t=1}^{T} \bigg(\sum_{i=1}^{M}\sum_{j=1}^{N} E [u_{ijt}^4] + \sum_{(i,j)\ne (m,n)} E[(u_{ijt} u_{mnt}-\sigma_{ijmn})^2] \bigg) \\
& &+ \frac{1}{M^2 N^2 T^2} \sum_{t\ne s} \bigg(\sum_{i=1}^{M}\sum_{j=1}^{N} E [u_{ijt}^2 u_{ijs}^2] + \sum_{(i,j)\ne (m,n)} E[(u_{ijt} u_{mnt}-\sigma_{ijmn}) (u_{ijs} u_{mns}-\sigma_{ijmn})] \bigg) \\
&&+\frac{1}{M^2N^2} \sum_{(i,j)\ne (m,n)} \sigma_{ijmn}^2 \\
&=&O\left(\frac{1}{T} +\frac{1}{MN}\right),
\end{eqnarray*}
where the last equality follows from Assumption \ref{ass3}.

\medskip

(4). Recall that in the first result of this lemma we have shown that $\| \bm{\Lambda}_E \|_2^2 = O_P(N\ln(MN))$. Then write

\begin{eqnarray*}
&& \frac{1}{M N T} \| \textbf{G} \bm{\Gamma}' \bm{\Lambda}_E \textbf{F}_E' \|_2  \leq \frac{1}{MNT} \| \textbf{G} \|_F  \cdot \| \bm{\Gamma} \|_F \cdot \| \bm{\Lambda}_E \|_2 \cdot \| \textbf{F}_E \|_2  \\
&=& \frac{1}{MNT} \cdot O_P(\sqrt{T}) \cdot O_P(\sqrt{MN}) \cdot O_P(\sqrt{N\ln(MN)}) \cdot O_P(\sqrt{T} \vee \sqrt{M}) \\
&=& O_P \left( \sqrt{\ln(MN)} \cdot ( M^{-1/2} \vee T^{-1/2} ) \right),
\end{eqnarray*}
where the first equality uses the following facts that $\| \textbf{G} \|_F = O_P(\sqrt{T})$, $\| \bm{\Gamma} \|_F = O_P(\sqrt{MN})$, and $\| \textbf{F}_{E} \|_2 = O_P(\sqrt{T} \vee \sqrt{M})$.  

\medskip

(5).  Similar to the fourth result of this lemma, we write

\begin{eqnarray*}
&&  \frac{1}{MNT} \| \textbf{G} \bm{\Gamma}' \bm{\Lambda}_I \textbf{F}_I' \|_2 
\leq \frac{1}{MNT} \| \textbf{G} \|_F \cdot \| \bm{\Gamma} \|_F \cdot \| \bm{\Lambda}_{I} \|_2 \cdot \| \textbf{F}_{I} \|_2  \\
&=& \frac{1}{MNT} \cdot O_P(\sqrt{T}) \cdot O_P(\sqrt{MN}) \cdot O_P(\sqrt{M\ln(MN)}) \cdot O_P(\sqrt{T} \vee \sqrt{N})  \\
&=& O_P \left( \sqrt{\ln(MN)} \cdot ( N^{-1/2} \vee T^{-1/2} ) \right),
\end{eqnarray*}
where the first equality uses the following facts that $\| \textbf{G} \|_F = O_P(\sqrt{T})$, $\| \bm{\Gamma} \|_F = O_P(\sqrt{MN})$, $\| \bm{\Lambda}_{I} \|_2 = O_P(\sqrt{M\ln(MN)})$ and $\| \textbf{F}_{I} \|_2 = O_P(\sqrt{T} \vee \sqrt{N})$.

\medskip

(6). Write

\begin{eqnarray*}
&&\frac{1}{M^2 N^2 T^2} E\| \bm{\Gamma}' \textbf{U} \|_{F}^2= \frac{1}{M^2N^2T^2}\sum_{t=1}^TE\left\| \sum_{i=1}^M \sum_{j=1}^N \bm{\gamma}_{ij}u_{ijt}\right\|_F^2 \\
&=& \frac{1}{M^2 N^2 T^2}   \sum_{i,m=1}^{M} \sum_{j,n=1}^{N} \sum_{t=1}^{T} E[u_{ijt}u_{mnt}  \bm{\gamma}_{ij}' \bm{\gamma}_{mn}  ] \le O(1) \frac{1}{M^2 N^2 T}  \sum_{i,m=1}^{M} \sum_{j,n=1}^{N}|\sigma_{ijmn} |\\
&=&  O\left(\frac{1}{MNT} \right),
\end{eqnarray*}
where the last line follows from Assumption \ref{ass3}. Thus, we further write

\begin{eqnarray*}
\frac{1}{MNT} \| \textbf{G} \bm{\Gamma}' \textbf{U} \|_{F} \leq \frac{1}{M N T} \| \textbf{G} \|_{F} \cdot \| \bm{\Gamma}' \textbf{U} \|_{F} = O_P \left(\frac{1}{\sqrt{MN}} \right).
\end{eqnarray*}

\medskip

(7). Recall that in the first two results of this lemma we have shown that $\| \bm{\Lambda}_E \|_2^2 = O_P(N\ln(MN))$ and $\| \bm{\Lambda}_I \|_2^2 = O_P(M\ln(MN))$. Write

\begin{eqnarray*}
&& \frac{1}{MNT} \| \textbf{F}_{E} \bm{\Lambda}_E' \bm{\Lambda}_I \textbf{F}_{I}' \|_2 \leq \frac{1}{MNT} \| \textbf{F}_{E} \|_2  \cdot \| \bm{\Lambda}_E \|_2 \cdot \| \bm{\Lambda}_I \|_2 \cdot \| \textbf{F}_{I} \|_2  \\
&=&O_P(1) \frac{1}{MNT} \cdot (\sqrt{T} \vee \sqrt{M}) \cdot \sqrt{N \ln(MN)} \cdot \sqrt{M \ln(MN)} \cdot (\sqrt{T} \vee \sqrt{N})  \\
&=& O_P \left( \frac{ (\sqrt{T} \vee \sqrt{M}) \cdot (\sqrt{T} \vee \sqrt{N}) \cdot \ln(MN)}{\sqrt{MN}T} \right) ,
\end{eqnarray*}
where the first equality follows from Assumption \ref{ass1} and Assumption \ref{ass2}.

\medskip

(8). Similar to the seventh result, we have

\begin{eqnarray*}
&& \frac{1}{M N T} \| \textbf{F}_{E} \bm{\Lambda}_E' \textbf{U} \|_2 \leq \frac{1}{M N T} \| \textbf{F}_E \|_2 \cdot \| \bm{\Lambda}_E \|_2 \cdot \| \textbf{U} \|_2 \\
&=& O_P(1)\frac{1}{M N T} \cdot (\sqrt{T} \vee \sqrt{M}) \cdot \sqrt{N \ln(MN)} \cdot (\sqrt{T} \vee \sqrt{MN}) \\
&=& O_P \left( \frac{ (\sqrt{T} \vee \sqrt{M}) \cdot (\sqrt{T} \vee \sqrt{MN}) \cdot \sqrt{\ln(MN)} }{M\sqrt{N}T} \right),
\end{eqnarray*}
where the first equality follows from Assumption \ref{ass1}, Assumption \ref{ass2} and the development in the third result of this lemma.

\medskip

(9) The proof is similar to the eighth result of this lemma. Thus, omitted. 
\hspace*{\fill}{$\blacksquare$}

\bigskip

\noindent \textbf{Proof of Lemma \ref{lemma3}}: 

(1). We have

\begin{eqnarray*}
\frac{1}{T} \| \textbf{G}' (\widehat{\textbf{G}}  - \textbf{G} \textbf{H}) \|_{F} \leq \frac{1}{\sqrt{T}} \| \textbf{G} \|_F \cdot \frac{1}{\sqrt{T}} \| \widehat{\textbf{G}}  - \textbf{G} \textbf{H} \|_{F} = O_P \left( \frac{\sqrt{\ln(MN)}}{\min \{\sqrt{M}, \sqrt{N}, \sqrt{T}\}} \right),
\end{eqnarray*}
where the equality follows from Assumption \ref{ass1} and Lemma \ref{lemma_nof1}. 

\medskip

(2). We have

\begin{eqnarray*}
&&\frac{1}{T} \| \widehat{\textbf{G}}' (\widehat{\textbf{G}} - \textbf{G} \textbf{H}) \|_{F} = \frac{1}{T} \| (\widehat{\textbf{G}} - \textbf{G} \textbf{H} + \textbf{G} \textbf{H})' (\widehat{\textbf{G}} - \textbf{G} \textbf{H}) \|_{F} \\
&=& \frac{1}{T} \| (\widehat{\textbf{G}} - \textbf{G} \textbf{H})' (\widehat{\textbf{G}} - \textbf{G} \textbf{H}) + (\textbf{G} \textbf{H})' (\widehat{\textbf{G}} - \textbf{G} \textbf{H}) \|_{F} \\
&\leq &\frac{1}{T} \| \widehat{\textbf{G}} - \textbf{G} \textbf{H} \|_F^2  + \| \textbf{H} \|_F \cdot \frac{1}{T} \| \textbf{G}' (\widehat{\textbf{G}} - \textbf{G} \textbf{H}) \|_{F} \\
&=&  O_P \left( \frac{\sqrt{\ln(MN)}}{\min \{\sqrt{M}, \sqrt{N}, \sqrt{T}\}} \right),
\end{eqnarray*}
where the third equality follows from the first result of this lemma and the fact that $\|\textbf{H} \|_F=O_P(1)$.

\medskip

(3). By the first result of this lemma, we obtain that

\begin{equation} \label{e0}
\frac{1}{T}\textbf{G}'\widehat{\textbf{G}}  - \frac{1}{T} \textbf{G}'\textbf{G} \textbf{H} =  O_P \left( \frac{\sqrt{\ln(MN)}}{\min \{\sqrt{M}, \sqrt{N}, \sqrt{T}\}} \right),
\end{equation}
which in connection with $\|\textbf{H} \|_F=O_P(1)$ implies

\begin{equation} \label{e1}
\frac{1}{T}\textbf{H}' \textbf{G}'\widehat{\textbf{G}}  - \frac{1}{T}\textbf{H}' \textbf{G}'\textbf{G} \textbf{H} = O_P \left( \frac{\sqrt{\ln(MN)}}{\min \{\sqrt{M}, \sqrt{N}, \sqrt{T}\}} \right).
\end{equation}
By the second result of this lemma, we obtain that

\begin{equation} \label{e2}
\frac{1}{T} \widehat{\textbf{G}}'\widehat{\textbf{G}}  - \frac{1}{T} \widehat{\textbf{G}}'\textbf{G} \textbf{H} = \textbf{I}_{r_g} - \frac{1}{T} \widehat{\textbf{G}}'\textbf{G} \textbf{H} = O_P \left( \frac{\sqrt{\ln(MN)}}{\min \{\sqrt{M}, \sqrt{N}, \sqrt{T}\}} \right),
\end{equation}
Summing up (\ref{e1}) and (\ref{e2}) yields

\begin{equation} \label{e3}
\textbf{I}_{r_g} - \frac{1}{T}\textbf{H}' \textbf{G}'\textbf{G} \textbf{H} = O_P \left( \frac{\sqrt{\ln(MN)}}{\min \{\sqrt{M}, \sqrt{N}, \sqrt{T}\}} \right).
\end{equation}

Furthermore, write

\begin{eqnarray*}
&&\| \textbf{P}_{\widehat{\textbf{G}}} - \textbf{P}_{\textbf{G}} \|_F^{2}
= \mbox{tr} \left\{ (\textbf{P}_{\widehat{\textbf{G}}} - \textbf{P}_{\textbf{G}})' (\textbf{P}_{\widehat{\textbf{G}}} - \textbf{P}_{\textbf{G}}) \right\} \\
&=& \mbox{tr} \left\{ \textbf{P}_{\widehat{\textbf{G}}} - \textbf{P}_{\widehat{\textbf{G}}} \textbf{P}_{\textbf{G}} - \textbf{P}_{\textbf{G}} \textbf{P}_{\widehat{\textbf{G}}} + \textbf{P}_{\textbf{G}} \right\} \\
&=& \mbox{tr} \left\{ \frac{1}{T} \widehat{\textbf{G}} \widehat{\textbf{G}}' - \textbf{P}_{\widehat{\textbf{G}}} \textbf{P}_{\textbf{G}} - \textbf{P}_{\textbf{G}} \textbf{P}_{\widehat{\textbf{G}}} + \textbf{G}(\textbf{G}' \textbf{G})^{-1} \textbf{G}' \right\} \\
&=& \mbox{tr} \left\{ \frac{1}{T} \widehat{\textbf{G}} \widehat{\textbf{G}}' \right\} - 2 \cdot \mbox{tr} \left\{ \textbf{P}_{\widehat{\textbf{G}}} \textbf{P}_{\textbf{G}} \right\} + \mbox{tr} \left\{ \textbf{G}(\textbf{G}' \textbf{G})^{-1} \textbf{G}' \right\} \\
&=& \mbox{tr} \left\{ \frac{1}{T} \widehat{\textbf{G}}'\widehat{\textbf{G}} \right\} - 2 \cdot \mbox{tr} \left\{ \frac{1}{T} \widehat{\textbf{G}} \widehat{\textbf{G}}' \textbf{P}_{\textbf{G}} \right\} + \mbox{tr} \left\{ \textbf{G}'\textbf{G}(\textbf{G}' \textbf{G})^{-1} \right\} \\
&=& \mbox{tr} \left\{ \textbf{I}_{r_g} \right\} - 2 \cdot \mbox{tr} \left\{ \frac{1}{T} \widehat{\textbf{G}}' \textbf{P}_{\textbf{G}}\widehat{\textbf{G}} \right\} + \mbox{tr} \left\{ \textbf{I}_{r_g} \right\} 
= 2 \cdot \mbox{tr} \left\{ \textbf{I}_{r_g} - \frac{1}{T} \widehat{\textbf{G}}' \textbf{P}_{\textbf{G}}\widehat{\textbf{G}} \right\}.
\end{eqnarray*}
Using the results in (\ref{e0}), it can be shown that

\begin{eqnarray*}
&& \frac{1}{T} \widehat{\textbf{G}}' \textbf{P}_{\textbf{G}}\widehat{\textbf{G}}
= \frac{1}{T} \widehat{\textbf{G}}' \textbf{G}(\textbf{G}'\textbf{G})^{-1}\textbf{G}' \widehat{\textbf{G}} \\
&=& \left( \frac{1}{T} \widehat{\textbf{G}}' \textbf{G} \right) \left( \frac{1}{T} \textbf{G}'\textbf{G} \right)^{-1} \left( \frac{1}{T} \textbf{G}' \widehat{\textbf{G}} \right) \\
&=& \frac{1}{T} \textbf{H}' \textbf{G}'\textbf{G} \textbf{H} + O_P \left( \frac{\sqrt{\ln(MN)}}{\min \{\sqrt{M}, \sqrt{N}, \sqrt{T}\}} \right),
\end{eqnarray*}
which in connection with (\ref{e3}) yields
\begin{equation*}
\frac{1}{T} \widehat{\textbf{G}}' \textbf{P}_{\textbf{G}}\widehat{\textbf{G}} = \textbf{I}_{r_g} + O_P \left( \frac{\sqrt{\ln(MN)}}{\min \{\sqrt{M}, \sqrt{N}, \sqrt{T}\}} \right).
\end{equation*}
It follows that

\begin{equation*}
\| \textbf{P}_{\widehat{\textbf{G}}} - \textbf{P}_{\textbf{G}} \|_F^{2} =  O_P \left( \frac{\sqrt{\ln(MN)}}{\min \{\sqrt{M}, \sqrt{N}, \sqrt{T}\}} \right).
\end{equation*}
The proof is now complete. \hspace*{\fill}{$\blacksquare$}

It is worth noting that we do not impose any orthogonality conditions between global and local factors (loadings) in proving the above results. 

\bigskip

\noindent \textbf{Proof of Lemma \ref{lemma6}}:

(1). First, using $\widehat{\bm{\Gamma}}_{I,j} = \frac{1}{T}\textbf{Y}_{I, j} \widehat{\textbf{G}} $, we note that

\begin{eqnarray*}
&&\frac{1}{\sqrt{MT}}\left\|\bm{\Gamma}_{I,j}\textbf{G}' - \widehat{\bm{\Gamma}}_{I,j} \widehat{\textbf{G}}'\right\|_F =\frac{1}{\sqrt{MT}}\left\| \bm{\Gamma}_{I,j} \textbf{G}' - \textbf{Y}_{I, j} \textbf{P}_{\widehat{\textbf{G}}} \right\|_F \\
&=& \frac{1}{\sqrt{MT}}\left\|\bm{\Gamma}_{I,j}\textbf{G}' - \left( \bm{\Gamma}_{I,j} \textbf{G}' + \diag\{\bm{\Lambda}_{E,\bullet j}' \}' \textbf{F}_{E}' + \bm{\Lambda}_{I,\bullet j} \textbf{F}_{I,j}' + \textbf{U}_{\bullet j \bullet} \right) \textbf{P}_{\widehat{\textbf{G}}} \right\|_F  \\
&=& \frac{1}{\sqrt{MT}}\left\|\bm{\Gamma}_{I,j} \textbf{G}' \textbf{M}_{\widehat{\textbf{G}}} - \left( \diag\{\bm{\Lambda}_{E,\bullet j}' \}' \textbf{F}_{E}' + \bm{\Lambda}_{I,\bullet j} \textbf{F}_{I,j}' + \textbf{U}_{\bullet j \bullet} \right) \textbf{P}_{\widehat{\textbf{G}}} \right\|_F \\
&=&\frac{1}{\sqrt{MT}}\left\| \bm{\Gamma}_{I,j} (\textbf{G} - \widehat{\textbf{G}} \textbf{H}^{-1})' \textbf{M}_{\widehat{\textbf{G}}} - \left( \diag\{\bm{\Lambda}_{E,\bullet j}' \}' \textbf{F}_{E}' + \bm{\Lambda}_{I,\bullet j} \textbf{F}_{I,j}' + \textbf{U}_{\bullet j \bullet} \right) \textbf{P}_{\widehat{\textbf{G}}}\right\|_F.
\end{eqnarray*}
For the first term on the right hand side above, we have

\begin{eqnarray} \label{eq.b4.1}
\textbf{M}_{\widehat{\textbf{G}}} (\textbf{G}-\widehat{\textbf{G}} \textbf{H}^{-1}) \bm{\Gamma}_{I,j}'  & =& -\textbf{M}_{\widehat{\textbf{G}}} (\textbf{A}_2 + \cdots + \textbf{A}_{16}) \widehat{\textbf{G}} \left( \frac{1}{T} \textbf{G}' \widehat{\textbf{G}} \right)^{-1} \left(\frac{1}{MN}\bm{\Gamma}' \bm{\Gamma} \right)^{-1}\bm{\Gamma}_{I,j}',
\end{eqnarray}
where $\textbf{A}_2,\ldots, \textbf{A}_{16}$ are defined in Lemma \ref{lemma_nof1} already. In addition, for notational simplicity, let $\textbf{K} = \left( \frac{1}{T} \textbf{G}' \widehat{\textbf{G}} \right)^{-1} \left(\frac{1}{MN}\bm{\Gamma}' \bm{\Gamma} \right)^{-1}$ in what follows. It is easy to know that $\| \textbf{K} \|_F = O_P(1)$.

\medskip

Write

\begin{eqnarray*}
&& \frac{1}{\sqrt{MT}}\|\textbf{M}_{\widehat{\textbf{G}}} (\textbf{G}-\widehat{\textbf{G}} \textbf{H}^{-1}) \bm{\Gamma}_{I,j}' \|_F 
\leq O(1) \cdot \frac{1}{\sqrt{MT}} \|\textbf{M}_{\widehat{\textbf{G}}} (\textbf{A}_2 + \cdots + \textbf{A}_{16}) \widehat{\textbf{G}} \textbf{K} \bm{\Gamma}_{I,j}' \|_2 \\
&\le&  O(1) \cdot \frac{1}{\sqrt{MT}} \|\textbf{M}_{\widehat{\textbf{G}}} (\textbf{A}_2 + \cdots + \textbf{A}_{16}) \widehat{\textbf{G}}\|_2 \cdot \|  \textbf{K} \|_F \cdot \| \bm{\Gamma}_{I,j} \|_F,\\
&\leq& O_P(1)\frac{1}{\sqrt{MT}}\cdot  \frac{\sqrt{\ln(MN)}}{\min \{\sqrt{M}, \sqrt{N}, \sqrt{T}\}}  \cdot \sqrt{T} \cdot \sqrt{M}\\
&=& O_P\left( \frac{\sqrt{\ln(MN)}}{ \min \{\sqrt{M}, \sqrt{N}, \sqrt{T}\}}  \right),
\end{eqnarray*}
where the first inequality follows from (\ref{eq.b4.1}) and $\|\bm{\Omega}\|_F \le \sqrt{\text{rank}(\bm{\Omega})} \cdot \|\bm{\Omega} \|_2$, the third inequality follows from Lemma \ref{lemma2} and the facts that $\|\widehat{\textbf{G}} \|_F =O(\sqrt{T})$ and $\|\bm{\Gamma}_{I,j} \|_F =O_P(\sqrt{M})$.

Next, we consider the second expansion.

\begin{eqnarray*}
( \diag\{\bm{\Lambda}_{E,\bullet j}' \}' \textbf{F}_{E}' + \bm{\Lambda}_{I,\bullet j} \textbf{F}_{I,j}' + \textbf{U}_{\bullet j \bullet} ) \textbf{P}_{\widehat{\textbf{G}}} &=&\diag\{\bm{\Lambda}_{E,\bullet j}' \}' \textbf{F}_{E}' \textbf{P}_{\widehat{\textbf{G}}} + \bm{\Lambda}_{I,\bullet j} \textbf{F}_{I,j}' \textbf{P}_{\widehat{\textbf{G}}} + \textbf{U}_{\bullet j \bullet}  \textbf{P}_{\widehat{\textbf{G}}} \\
&:=& \textbf{D}_1 + \textbf{D}_2 + \textbf{D}_3. 
\end{eqnarray*}

For $\textbf{D}_1$, write

\begin{eqnarray*}
&&\frac{1}{\sqrt{MT}}\| \textbf{D}_1 \|_F 
= \frac{1}{\sqrt{MT}}\| \diag\{\bm{\Lambda}_{E,\bullet j}' \}' \textbf{F}_{E}' \textbf{P}_{\widehat{\textbf{G}}} \|_F \\
&\leq& O(1) \frac{1}{\sqrt{MT}}\| \diag\{\bm{\Lambda}_{E,\bullet j}' \}' \textbf{F}_{E}' \textbf{P}_{\widehat{\textbf{G}}} \|_2
\leq O(1) \frac{1}{\sqrt{MT}}\| \diag\{\bm{\Lambda}_{E,\bullet j}' \}\|_2\cdot \|\textbf{F}_{E}\|_2 \\
&\leq& O_P(1) \frac{1}{\sqrt{MT}}\cdot\sqrt{\ln(MN)} \cdot(\sqrt{M} \vee \sqrt{T})
= O_P\left(\frac{\sqrt{\ln(MN)}}{\sqrt{M} \wedge \sqrt{T}} \right),
\end{eqnarray*}
where the first inequality follows from $\|\bm{\Omega}\|_F \le \sqrt{\text{rank}(\bm{\Omega})} \cdot \|\bm{\Omega} \|_2$
, the second inequality follows from $\|\textbf{P}_{\widehat{\textbf{G}}} \|_2 = 1$
, and the last inequality follows from Assumption \ref{ass1} and Assumption \ref{ass2}. 

Consider $\textbf{D}_2$, and write

\begin{eqnarray*}
&&\frac{1}{\sqrt{MT}}\| \textbf{D}_2 \|_F = \frac{1}{\sqrt{MT}}\| \bm{\Lambda}_{I,\bullet j} \textbf{F}_{I,j}' \textbf{P}_{\widehat{\textbf{G}}} \|_F \leq  \frac{1}{\sqrt{MT}T}\| \bm{\Lambda}_{I,\bullet j} \textbf{F}_{I,j}'  \widehat{\textbf{G}} \|_F \cdot \| \widehat{\textbf{G}}\|_F \\ 
&\leq & \frac{1}{\sqrt{MT}T}\| \bm{\Lambda}_{I,\bullet j} \textbf{F}_{I,j}'  (\widehat{\textbf{G}} - \textbf{G} \textbf{H} + \textbf{G} \textbf{H})\|_F \cdot \| \widehat{\textbf{G}} \|_F \\ 
&\le & \frac{1}{\sqrt{MT}T}\| \bm{\Lambda}_{I,\bullet j} \textbf{F}_{I,j}'  (\widehat{\textbf{G}} - \textbf{G} \textbf{H})\|_F \cdot \| \widehat{\textbf{G}} \|_F + \frac{1}{\sqrt{MT}T}\| \bm{\Lambda}_{I,\bullet j} \textbf{F}_{I,j}' \textbf{G} \textbf{H}\|_F \cdot \| \widehat{\textbf{G}} \|_F.
\end{eqnarray*}
Note that 

\begin{eqnarray*}
&& \frac{1}{\sqrt{MT}T}\| \bm{\Lambda}_{I,\bullet j} \textbf{F}_{I,j}'  (\widehat{\textbf{G}} - \textbf{G} \textbf{H})\|_F \cdot \| \widehat{\textbf{G}} \|_F \\
&\le &O_P(1)\frac{1}{\sqrt{M}}\cdot \sqrt{M}\cdot \frac{\sqrt{\ln(MN)}}{\min \{\sqrt{M}, \sqrt{N}, \sqrt{T}\}} =O_P\left( \frac{ \sqrt{\ln(MN)}}{ \min \{\sqrt{M}, \sqrt{N}, \sqrt{T}\}}  \right),
\end{eqnarray*}
where the inequality follows from Assumption \ref{ass2}, Lemma \ref{lemma_nof1}, and the fact that $ \|  \textbf{F}_{I, j}\|_F=O_P(\sqrt{T})$. Note further that

\begin{eqnarray*}
&&\frac{1}{\sqrt{MT}T}\| \bm{\Lambda}_{I,\bullet j} \textbf{F}_{I,j}' \textbf{G} \textbf{H}\|_F \cdot \| \widehat{\textbf{G}} \|_F \\
&\le &\frac{1}{\sqrt{MT}T}\| \bm{\Lambda}_{I,\bullet j}\|_F \cdot \| \textbf{F}_{I,j}' \textbf{G}\|_F \cdot \| \textbf{H}\|_F \cdot \| \widehat{\textbf{G}} \|_F\\
&\le&O_P(1) \frac{1}{\sqrt{M}}\cdot \sqrt{M} \cdot T^{a_{I,j}} =O_P( T^{a_{I,j}} ),
\end{eqnarray*}
where the second inequality follows from Assumption \ref{ass4}.1.(a).

Consider $\textbf{D}_3$ and write

\begin{eqnarray*}
&& \frac{1}{\sqrt{MT}}\| \textbf{D}_3 \|_F 
= \frac{1}{\sqrt{MT}}\| \textbf{U}_{\bullet j \bullet} \textbf{P}_{\widehat{\textbf{G}}} \|_F
\leq O(1) \frac{1}{\sqrt{MT}}\| \textbf{U}_{\bullet j \bullet} \textbf{P}_{\widehat{\textbf{G}}} \|_2 \\
&\leq& O(1) \frac{1}{\sqrt{MT}}\| \textbf{U}_{\bullet j \bullet} \|_2  =O_P\left(\frac{1}{\sqrt{M}} \vee \frac{1}{\sqrt{T}} \right),
\end{eqnarray*}
where the first inequality follows from $\|\bm{\Omega}\|_F \le \sqrt{\text{rank}(\bm{\Omega})} \cdot \|\bm{\Omega} \|_2$
, the second inequality follows from $\|\textbf{P}_{\widehat{\textbf{G}}} \|_2 = 1$
, and the last equality follows from a development similar to Lemma \ref{lemma2}.3 using Assumption \ref{ass4}.

\medskip

Based on the above development, we obtain that

\begin{eqnarray*}
&&\frac{1}{\sqrt{MT}}\left\|\bm{\Gamma}_{I,j}\textbf{G}' - \widehat{\bm{\Gamma}}_{I,j} \widehat{\textbf{G}}'\right\|_F \\
&=&O_P\left( \frac{\sqrt{\ln(MN)}}{ \min \{\sqrt{M}, \sqrt{N}, \sqrt{T}\}}  \right)+O_P\left(\frac{\sqrt{\ln(MN)}}{\sqrt{M} \wedge \sqrt{T}} \right)+O_P\left( \frac{ \sqrt{\ln(MN)}}{ \min \{\sqrt{M}, \sqrt{N}, \sqrt{T}\}}  \right)\\
&&+O_P( T^{a_{I,j}} ) +O_P\left(\frac{1}{\sqrt{M}} \vee \frac{1}{\sqrt{T}} \right)\\
&=&O_P\left( \frac{\sqrt{\ln(MN)}}{ \min \{\sqrt{M}, \sqrt{N}, \sqrt{T}\}}+T^{a_{I,j}}   \right).
\end{eqnarray*}
The proof of the first result is then complete.

\medskip

(2). The second result can be proved in exactly the same way as the first result. Thus, omitted. \hspace*{\fill}{$\blacksquare$}

\bigskip

\noindent \textbf{Proof of Lemma \ref{lemma8}}:

(1). Write

\begin{eqnarray*}
&&\frac{1}{MT} \|(\bm{\Gamma}_{I,j} \textbf{G}' - \widehat{\bm{\Gamma}}_{I,j} \widehat{\textbf{G}}' )' \diag\{\bm{\Lambda}_{E,\bullet j}' \}'\textbf{F}_{E}' \|_2\\
&\leq &\frac{1}{MT} \|\bm{\Gamma}_{I,j} \textbf{G}' - \widehat{\bm{\Gamma}}_{I,j} \widehat{\textbf{G}}' \|_2 \cdot \|  \diag\{\bm{\Lambda}_{E,\bullet j}' \} \|_2 \cdot \| \textbf{F}_E \|_2 \\
&= &\frac{1}{MT} \|\bm{\Gamma}_{I,j} \textbf{G}' - \widehat{\bm{\Gamma}}_{I,j} \widehat{\textbf{G}}'  \|_2 \cdot \sqrt{ \max_{i\ge 1, j\ge 1}\| \bm{\lambda}_{E,ij}\|_{F}^2 } \cdot \| \textbf{F}_{E} \|_2 \\
&= &O_P(1)\frac{1}{\sqrt{MT}}\cdot \left( \frac{\sqrt{\ln(MN)}}{ \min \{\sqrt{M}, \sqrt{N}, \sqrt{T}\}}+T^{a_{I,j}}   \right) \cdot\sqrt{\ln (MN)} \cdot (\sqrt{T}\vee \sqrt{M})\\
&= &O_P(1)\frac{\sqrt{\ln(MN)}}{\sqrt{M} \wedge \sqrt{T}}\cdot\left( \frac{\sqrt{\ln(MN)}}{ \min \{\sqrt{M}, \sqrt{N}, \sqrt{T}\}}+T^{a_{I,j}}   \right)  ,
\end{eqnarray*}
where the second equality follows from Lemma \ref{lemma6}, Assumption \ref{ass1} and Assumption \ref{ass2}.

\medskip

(2). Write

\begin{eqnarray*}
&& \frac{1}{MT}\| (\bm{\Gamma}_{I,j} \textbf{G}' - \widehat{\bm{\Gamma}}_{I,j} \widehat{\textbf{G}}' )' \bm{\Lambda}_{I,\bullet j} \textbf{F}_{I,j}' \|_2\\
&\le &\frac{1}{MT} \| \bm{\Gamma}_{I,j} \textbf{G}' - \widehat{\bm{\Gamma}}_{I,j} \widehat{\textbf{G}}'\|_2 \cdot \|  \bm{\Lambda}_{I,\bullet j}  \|_2 \cdot \| \textbf{F}_{I,j} \|_2\\
&= &O_P(1)\frac{1}{\sqrt{MT}} \cdot  \left( \frac{\sqrt{\ln(MN)}}{ \min \{\sqrt{M}, \sqrt{N}, \sqrt{T}\}}+T^{a_{I,j}}   \right)\cdot \sqrt{M}\cdot \sqrt{T}\\
&=& O_P\left( \frac{\sqrt{\ln(MN)}}{ \min \{\sqrt{M}, \sqrt{N}, \sqrt{T}\}}+T^{a_{I,j}}   \right),
\end{eqnarray*}
where the first equality follows from Lemma \ref{lemma6}.

\medskip

(3). The proof is similar to the second result of this lemma, the result follows from $\| \textbf{U}_{\bullet j \bullet} \|_2 = O_P(\sqrt{M} \vee \sqrt{T})$.

\medskip

(4). Write

\begin{eqnarray*}
&&\frac{1}{MT}\|  \textbf{F}_{E} \diag\{\bm{\Lambda}_{E,\bullet j}' \} \diag\{\bm{\Lambda}_{E,\bullet j}' \}'\textbf{F}_{E}' \|_2 \\
&\le &\frac{1}{MT}\|\textbf{F}_E \|_2^2 \cdot \lambda_{\max} \{  \diag\{\bm{\Lambda}_{E,\bullet j}' \}'  \diag\{\bm{\Lambda}_{E,\bullet j}' \} \} \\
&\le &\frac{1}{MT} \| \textbf{F}_{E} \|_{2}^{2} \cdot \max_{i\ge 1, j\ge 1}\| \bm{\lambda}_{E,ij}\|_F^2 \\
&=&O_P(1)\frac{1}{MT}\cdot (T\vee M)\cdot \ln (MN)=O_P\left(\frac{\ln (MN)}{M\wedge T} \right),
\end{eqnarray*}
where the first equality follows from Assumption \ref{ass1} and Assumption \ref{ass2}.

\medskip

(5). Write

\begin{eqnarray*}
&&\frac{1}{MT} \|\textbf{F}_{E} \diag\{\bm{\Lambda}_{E,\bullet j}' \} \bm{\Lambda}_{I,\bullet j} \textbf{F}_{I,j}'\|_2\\
&\le &\frac{1}{MT} \|\textbf{F}_{E}\|_2 \cdot \|\diag\{\bm{\Lambda}_{E,\bullet j}' \}\|_2\cdot\| \bm{\Lambda}_{I,\bullet j}\|_2\cdot\| \textbf{F}_{I,j}'\|_2\\
&\le & \frac{1}{MT} \|\textbf{F}_{E}\|_2 \cdot \sqrt{\max_{i\ge 1, j\ge 1}\| \bm{\lambda}_{E,ij}\|_F^2}\cdot\| \bm{\Lambda}_{I,\bullet j}\|_2\cdot\| \textbf{F}_{I,j}'\|_2\\
&= &O_P(1)\frac{1}{MT}\cdot (\sqrt{T}\vee \sqrt{M})\cdot \sqrt{\ln (MN)} \cdot \sqrt{M}\cdot \sqrt{T}\\
&=&O_P\left(\frac{\sqrt{\ln (MN)}}{ \sqrt{M} \wedge \sqrt{T}} \right),
\end{eqnarray*}
where the second inequality follows from a procedure similar to the third result of this lemma, and the first equality follows from Assumption \ref{ass1} and Assumption \ref{ass2}.
 
\medskip

(6). Similar to the proof of the fifth result, the result follows from $\| \textbf{U}_{\bullet j \bullet} \|_2 = O_P(\sqrt{M} \vee \sqrt{T})$.

\medskip

(7). Write

\begin{eqnarray*}
&&\frac{1}{MT}\| (\bm{\Gamma}_{I,j} \textbf{G}' - \widehat{\bm{\Gamma}}_{I,j} \widehat{\textbf{G}}' )' (\bm{\Gamma}_{I,j} \textbf{G}' - \widehat{\bm{\Gamma}}_{I,j} \widehat{\textbf{G}}' )\|_2\le \frac{1}{MT} \| \bm{\Gamma}_{I,j} \textbf{G}' - \widehat{\bm{\Gamma}}_{I,j} \widehat{\textbf{G}}' \|_2^2\\
&= &O_P\left( \frac{\ln(MN)}{ \min \{M, N, T\}}+T^{2a_{I,j}}   \right),
\end{eqnarray*}
where the first equality follows from Lemma \ref{lemma6}.

\medskip

(8). Write

\begin{eqnarray*}
&&\frac{1}{MT} \|\textbf{F}_{I,j} \bm{\Lambda}_{I,\bullet j}'  \textbf{U}_{\bullet j \bullet}\|_2\le \frac{1}{MT} \|\textbf{F}_{I,j}\|_2 \cdot \| \bm{\Lambda}_{I,\bullet j} \textbf{U}_{\bullet j \bullet}\|_F\\
&= &O_P(1)\frac{1}{MT}\cdot \sqrt{T} \cdot \sqrt{MT} = O_P\left( \frac{1}{\sqrt{M}} \right),
\end{eqnarray*}
where the first equality follows from a development similar to Lemma \ref{lemma2}.6 using Assumption \ref{ass3} and Assumption \ref{ass4}.

\medskip

(9). Write

\begin{eqnarray*}
\frac{1}{MT} \| \textbf{U}_{\bullet j \bullet}^{\prime} \textbf{U}_{\bullet j \bullet} \|_2 =O_P\left(\frac{1}{\sqrt{M}} \vee \frac{1}{\sqrt{T}} \right),
\end{eqnarray*}
where the result follows from a development similar to Lemma \ref{lemma2}.3 using Assumption \ref{ass3}. \hspace*{\fill}{$\blacksquare$}

\end{document}